\documentclass[aps, pra, groupedaddress, twocolumn, floatfix]{revtex4-1}
\usepackage{color}
\usepackage{marvosym}
\usepackage{amssymb}
\usepackage{amsmath}
\usepackage{graphicx}
\usepackage{natbib}
\usepackage[caption=false]{subfig}
\usepackage{dcolumn}
\usepackage{booktabs}
\usepackage{longtable}
\usepackage{hyperref}

\AtBeginDocument{
	\heavyrulewidth=.08em
	\lightrulewidth=.05em
	\cmidrulewidth=.03em
	\belowrulesep=.65ex
	\belowbottomsep=0pt
	\aboverulesep=.4ex
	\abovetopsep=0pt
	\cmidrulesep=\doublerulesep
	\cmidrulekern=.5em
	\defaultaddspace=.5em
}

\begin{document}

\title{Uncovering the Dynamics of Correlation Structures Relative to
       the Collective Market Motion}

\author{Anton J. Heckens}
\email{anton.heckens@uni-due.de}
\author{Sebastian M. Krause}
\email{sebastian.krause@uni-due.de}
\author{Thomas Guhr}
\email{thomas.guhr@uni-due.de}
\affiliation{
	Fakult\"at f\"ur Physik, Universit\"at Duisburg--Essen, Duisburg, Germany		
}

\begin{abstract}
  The measured correlations of financial time series in subsequent
  epochs change considerably as a function of time. When studying the
  whole correlation matrices, quasi--stationary patterns, referred to
  as market states, are seen by applying clustering methods. They
  emerge, disappear or reemerge, but they are dominated by the
  collective motion of all stocks. In the jargon, one speaks of the
  market motion, it is always associated with the largest eigenvalue
  of the correlation matrices. Thus the question arises, if one can
  extract more refined information on the system by subtracting the
  dominating market motion in a proper way. To this end we introduce a
  new approach by clustering reduced--rank correlation matrices which
  are obtained by subtracting the dyadic matrix belonging to the
  largest eigenvalue from the standard correlation matrices. We
  analyze daily data of 262 companies of the S\&P~500 index over a
  period of almost 15 years from 2002 to 2016.  The resulting dynamics
  is remarkably different, and the corresponding market states are
  quasi--stationary over a long period of time. Our approach adds to the
  attempts to separate endogenous from exogenous effects.
\end{abstract}

\maketitle

\section{\label{sec:Introduction}Introduction}

In the financial media, one often hears phrases such as ``the S\&P~500
breaks out", ``the markets stabilized" or ``the current state of the
market".  The term ``state" is widely used in physics.  It is
therefore interesting and challenging for physicists working on
complex systems to explore whether such qualitative statements can be
substantiated by devising meaningful quantitative procedures.
Economists often also speak of
``regimes"~\cite{campbellEconometricsFinancialMarkets1997,
  hamiltonNewApproachEconomic1989, hamiltonAnalysisTimeSeries1990}
instead of states, sometimes these regimes are related to business
cycles.

Extending the work of
Ref.~\cite{munnixIdentifyingStatesFinancial2012}, we identify and
characterize the market states by clustering a set of correlation
matrices for subsequent epochs. The resulting clusters are then viewed
as the market states. The important new ingredient here is a new
approach to do this relative to the collective motion of the market,
\textit{i.e.}~to the coherent motion of all stocks.  We obtain a
considerable reduction of the information contained in these
correlation matrices to a trajectory of the financial market in the
space of the market states which are quasi-stationary,
\textit{i.e.}~they emerge, disintegrate, reemerge and eventually
disappear. Clustering~\cite{jainDataClustering502010,
  ronanAvoidingCommonPitfalls2016,
  KaufmanRousseeuw1990:FindingGroupinData} is an often applied method
in complex systems, \textit{i.e.}~one tries to find groups or typical
representatives of these groups in the data sets.  In econophysics,
clustering is most often used for unsupervised sector
classification~\cite{martiReviewTwoDecades2017}.  However, instead of
comparing the return time series with each other, one can compare the
returns of the market from one trading day with those of other trading
days~\cite{marsiliDissectingFinancialMarkets2002}.

It is also possible to cluster correlation matrices. The approach
emphasizes the time-dependence of the interactions of stocks.  A first
application for correlation matrices was put forward
in~\cite{munnixIdentifyingStatesFinancial2012}.  Since then, however,
there have also been new related investigations of correlation
matrices.  For example, it was found that the jumps from one market
state to another are caused by strong changes in the mean correlation,
while within the market state one moves around this mean correlation,
which can be described by noise
models~\cite{stepanovStabilityHierarchyQuasistationary2015,
  rinnDynamicsQuasistationarySystems2015,
  chetalovaZoomingMarketStates2015, pharasi2020market}.  This is 
supported by the finding that the market remains in one and the same state 
for a relative long time after a jump. Furthermore, market states can 
be used as long-time indicators (``precursors``) for other market states 
which is particularly interesting for estimating the transition probabilities 
into crises states~\cite{pharasiIdentifyingLongtermPrecursors2018,pharasi2020market}.
Relationships between financial crises were compared and
characterized~\cite{qiu2018state}.  In addition to the equity markets,
the market states of the futures markets have also been analyzed using
correlation matrices~\cite{papenbrockHandlingRiskonRiskoff2015}.

Here, we propose a modification of the clustering method of
Ref.~\cite{munnixIdentifyingStatesFinancial2012} which turns out
rather substantial as it facilitates a refined analysis of the
correlation structure and characterizes the system more precisely.
The clusters resulting of the standard correlation matrix and standard
covariance matrix is dominated by marketwide
correlations~{\cite{song2011evolution,stepanovStabilityHierarchyQuasistationary2015},
\textit{i.e.}~the time-dependence of the correlation structures in and
between the different industry sectors is blurred. The marketwide
correlation is mostly captured by the dyadic matrix corresponding to
the largest eigenvalue of the spectral decomposition.  The largest
eigenvalue and the corresponding eigenvector can be assigned to the
``market'', where, as already emphasized, ``market'' means here the
coherent, collective motion of all
stocks~\cite{gopikrishnan2001quantifying,
  plerouRandomMatrixApproach2002, benzaquen2017dissecting}.  The next
largest eigenvalues can be assigned to the industry sectors.  To
investigate the dynamics of the dyadic matrices of the sectors and
thus the correlation structure, we subtract the dyadic matrix from the
largest eigenvalue. Hence we carry out our analysis in a ``moving
frame'', as one might say in analogy to dynamical problems in
traditional physics, defined by the collective motion of the market as
a whole. Our new approach might help to shed new light on the
perpetual challenge of how to distinguish, from the measured data,
exogenous and endogenous effects. Although these effects are always
likely to mutually affect each others, it is at least plausible to
view the collective market motion as particularly strongly influenced
by exogenous effects. Hence, our analysis relative to the market
motion should leave us with correlation structures in which the
endogenous effects are better seen than previously, but
  one has to be aware that the intrinsic structure of the collective
  market motion might indicate endogenous effects.

On the technical side it is important that the corresponding reduced
matrices are well-defined covariance matrices.  The variances on the
main diagonals can thus be used to construct new and, once more,
well-defined correlation matrices.  These correlation matrices are
called \emph{reduced-rank correlation
  matrices}~\cite{brigoNoteCorrelationRank,rebonatoMostGeneralMethodology1999,
  PieterszGroenen2004:RankReductionOfCorrelationMatricesByMajorization,
  GrubivsicPietersz2007:EfficientRankReduction}. It is also important
to notice that we in our approach remove the largest dyadic matrix
whilst, in contrast, removal of the small ones, subjected to purely
statistical behavior, defines the filtering
method~\cite{Laloux1999:NoiseDressingFinancialCorrelationMatrices,
  alterSingularValueDecomposition2000, kimSystematicAnalysisGroup2005,
  macmahonCommunityDetectionCorrelation2015} for noise reduction.
Filtering and a variety of other techniques 
are important tools for estimating the correlation matrix 
elements for portfolio optimization~\cite{laloux2000random,bouchaudfinancial,
	potters2005financial,tola2008cluster,pantaleo2011improved,
	bun2017cleaning,bongiorno2020nonparametric,bongiorno2020covariance}.
Reduced-rank correlation matrices calculated from the filtered
standard correlation matrix were analyzed
in~\cite{miceliUltrametricityFundFunds2004, tumminello2007kullback,
  tumminello2007shrinkage}.  We want to point out that there are other
techniques to remove the ``market" from the standard correlation
matrices~\cite{plerouRandomMatrixApproach2002,
  borghesiEmergenceTimehorizonInvariant2007, shapira2009index,
  kenett2009rmt, meng2014systemic, bommarito2018spectral},
particularly the ``center of mass" approach, linear regression methods
and partial correlations.  Furthermore, the reduced-rank correlation
matrices should not be confused with the much more common ``reduced
correlation matrices" from factor
analysis~\cite{wiki:2019:FactorAnalysis}.

These are our goals: we compare the non-stationarity of the standard
correlation matrices with the one of two types of reduced-rank
correlation matrices, one calculated from the covariance matrices, the
other one evaluated from the correlation matrices.  First, we define
via singular value decomposition (SVD) reduced data matrices from
which we can calculate the reduced-rank correlation matrices.  Second,
the temporal evolution of market states is calculated via hierarchal
$k$-means (bisecting $k$-means) clustering.  Third, the averaged
correlation matrices of a market state -- the typical market state --
as a measure for the averaged correlation structure of the respective
market state is determined.  It is known that the dynamics of the
largest eigenvectors of the standard correlation matrix corresponding
to the ``market" and the sectors are known to be quite stable in time
for intraday data and even more for daily
data~\cite{gopikrishnan2001quantifying,
  plerouRandomMatrixApproach2002}.  We go one step further.  We
investigate the quasi-stationarity of the sum of dyadic matrices
without the ``market" part -- that means we analyze additionally to
just eigenvector dynamics the time evolution of the combinations of
eigenvectors and their ``weights", the eigenvalues -- in a time period
of massive financial crises.  Forth, we calculate the mean correlation
of the standard correlation matrices and the reduced ones and compare
this with historical events of financial crises.  Various questions
arise with regard to market states: How does the quasi-stationarity of
the market states of the reduced-rank correlation matrix change?  When
and how often do jumps between these market states occur and how does
the result compare with the market states of the standard correlation
matrices?

In Sec.~\ref{sec:DataSet}, we present the data.  We briefly sketch for
the convenience of the reader the approach of
Ref.~\cite{munnixIdentifyingStatesFinancial2012}, \textit{i.e.} the
clustering of standard correlation matrices in
Sec.~\ref{sec:ClusteringStandCorrMat}.  We show in
Sec.~\ref{sec:ResultClusteringReducedRankCorrelationMatrices2} the
relationship between reduced-rank correlation matrices and their
corresponding data matrices.  The reduced-rank correlation matrices
are clustered and the cluster results are compared with those of the
standard correlation matrix.  We conclude the paper with
Sec.~\ref{sec:Conclusion}.

\section{\label{sec:DataSet}Data set and conventions}

The data were collected from QuoteMedia~\cite{DynamicStockMarket}
and acquired by us from Quandl~\cite{Quandl}.
The investigation period is January 02, 2002 to July 08, 2016.
We analyze the daily data of $K = 262$ stocks in the S\&P~500 index 
(see Appx.~\ref{sec:ListStocks} and~\cite{wiki:2019:List500Companies}).
From the downloaded OHLCV data (Open{$\vert$}High{$\vert$}Low{$\vert$}Close{$\vert$}Volume), 
we take the closing prices $S_i(t)$ of company $i$, which are split and dividend 
adjusted.
A time series $S_i(t)$ has 3655 trading days.
The time series of the logarithmic returns are calculated from the daily closing 
prices,
\begin{equation} \label{eqn:LogReturn}
G_i(t) = \log \frac{S_i(t+\Delta t)}{S_i(t)} , \hspace{0.5cm} i = 1, 
\ldots, K 
\,.
\end{equation}
We set $\Delta t$~=~1~\text{day} for calculating daily trading day returns.
So we have $T_{\text{tot}}$~=~3654~\text{trading days} for the return time series 
of company $i$.
The data matrix resulting from the returns is
\begin{equation} \label{eqn:DatamatrixG}
G = \begin{bmatrix}  G_1(1) & \dots & G_1(T_{\text{tot}})  \\
\vdots & & \vdots \\
G_i(1) & \dots & G_i(T_{\text{tot}}) \\
\vdots & & \vdots \\
G_K(1) & \dots & G_K(T_{\text{tot}}) 
\end{bmatrix} \,.
\end{equation}
The rows of the data matrix $G$ contain $K$ time series of length $T_{\text{tot}}$.
These rows
and therefore later the rows of the correlation matrices are arranged according to 
the sectors in Tab.~\ref{tab:GICS} (cf. Ref.~\cite{wiki:2019:GlobalIndustryClassificationStandard}).
In addition, the sub-sectors of the sectors were taken into account and sorted alphabetically 
within the sectors.
Particularly noteworthy here are the Real Estate sector, which has been added in 
recent years and was formerly part of the Financials sector, and the Communication 
Services sector, which has emerged in part from the former Telecommunications Services 
sector, from the Consumer Discretionary sector and Information Technology 
sector.

For the technical part of our analysis like principal component analysis (PCA) 
and $k$-means clustering (see Sec.~\ref{sec:DeterminationMarketStates}, Appx.~\ref{sec:FacilClust} and 
Appx.~\ref{sec:StandKMeans}), we use 
the 
implementations
in the R-stats package~\cite{RProject}.
\begin{table}[!htb]
	\centering
	\caption{\label{tab:GICS}
		Global Industry Classification Standard (GICS) 
		(see~\cite{wiki:2019:GlobalIndustryClassificationStandard}).
	}
		\begin{tabular}{l@{\hspace{5em}}l@{\hspace{2em}}d}
			\toprule
			\multicolumn{1}{l}{Abbreviation} &
			\multicolumn{1}{l}{Sector}{}  &
			\multicolumn{1}{c}{Number of} \\
			\multicolumn{1}{l}{} &
			\multicolumn{1}{l}{} &
			\multicolumn{1}{c}{companies}
			\\
			\midrule
			E & Energy & 18 \\
			M & Materials & 14 \\
			I & Industrials & 46 \\
			CD & Consumer Discretionary & 29 \\
			CST & Consumer Staples & 24 \\
			HC & Health Care & 27 \\
			F & Financials & 37 \\
			RE & Real Estate & 8 \\
			I & Information Technology & 29 \\
			CSE & Communication Services & 9 \\
			U & Utilities & 21 \\
		\bottomrule
		\end{tabular}
\end{table}

\section{\label{sec:ClusteringStandCorrMat}Clustering standard correlation matrices}

In Sec.~\ref{sec:CorrMatricesEpoches}, we define the standard correlation 
matrix. 
A short overview 
of the clustering method, the determination of the number of clusters and the
definition of a typical market state are 
introduced in 
Sec.~\ref{sec:DeterminationMarketStates}. The results of the clustering, \textit{i.e}. 
the time evolution of the market states and the typical market states are 
presented in 
Sec.~\ref{sec:ResultsClusteringStandCorrMat} as well as a comparison between 
the mean correlation of the standard correlation matrix and the historical 
events of financial crises.

\subsection{\label{sec:CorrMatricesEpoches}Correlation matrices and epochs}

The time series of the returns $G_i(t)$ are divided into disjoint intervals of equal 
length, so-called epochs, of $T = 42$ trading days (2 trading months or one sixth 
of a trading year). 
Thus there are 87 data matrices (or correlation matrices) or $N_{\text{ep}} = T_{\text{tot}}/T 
= 87$ epochs.
We denote the individual 87 epochs by the number $n_{\text{ep}} = 1, \ldots, N_{\text{ep}}$.
The resulting correlation matrices do not have full rank. As $T \ll K$, the rank 
is $T-1$ and the number of vanishing eigenvalues is $K-(T-1)$.
The additional disappearing eigenvalue is caused by the normalization to zero mean 
value of the return time series (see Sec.~\ref{sec:DefinitionRedRankCorrMat}).

In contrast to~\cite{munnixIdentifyingStatesFinancial2012}, we calculate the correlation 
matrices over
2 trading months, not the trading days of 2 calendar months. The total number of 
trading days within 2 calendar months intervals
have not always the same length.
Each data matrix or correlation matrix receives a time stamp which corresponds to 
the center of the epoch and which is used to assign the 42-day interval as a black 
dot in later figures (cf. Fig.~\ref{fig:StandCorrMeanCorr}). 
\begin{figure}[t]
	\centering
	\includegraphics[width=1.0\columnwidth]{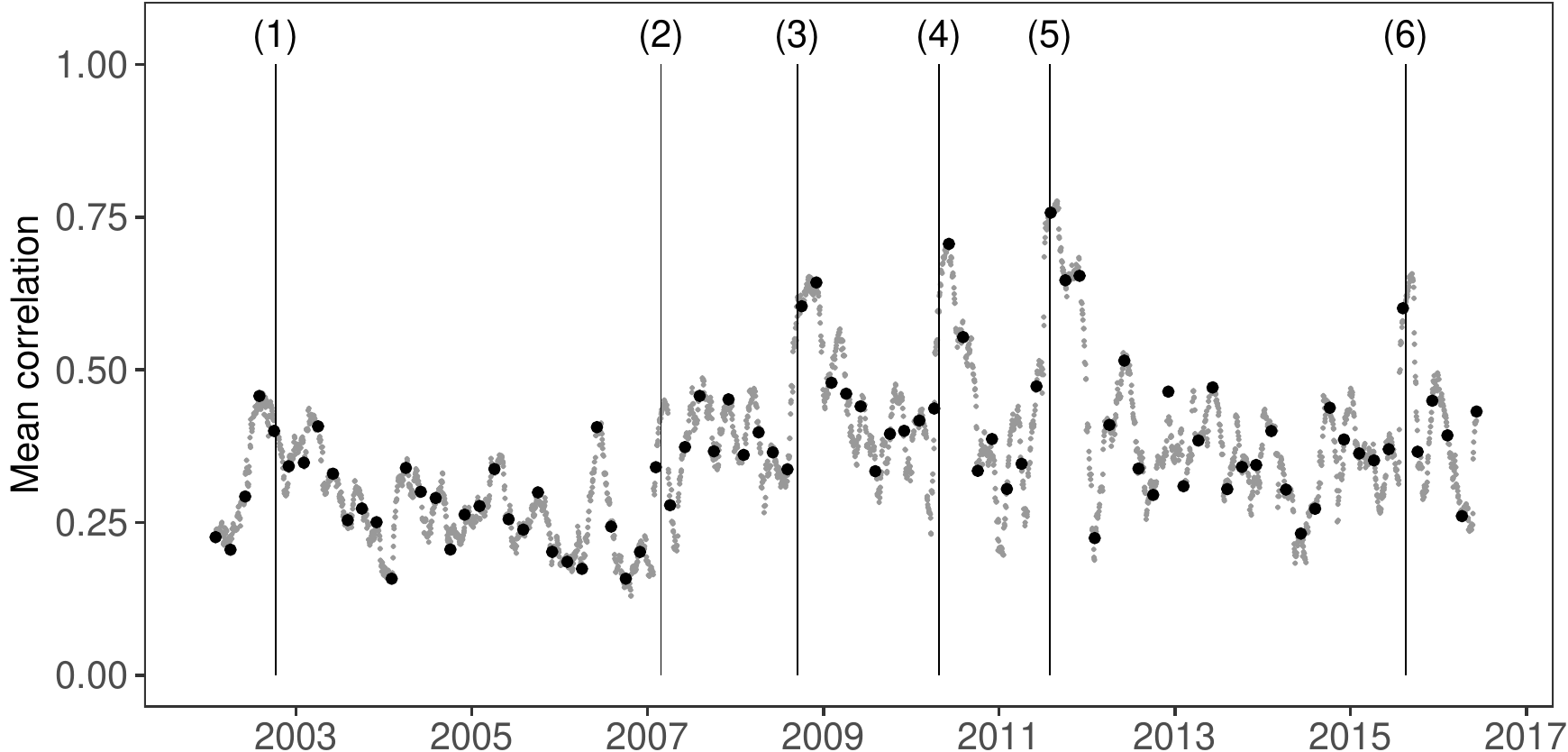}
	\caption{Mean correlation of the standard correlation matrix (Eq.~(\ref{eqn:StandCorrMat})). 
The larger black dots belong to the middle of the epochs for which the corresponding 
correlation matrices are clustered. The smaller grayish dots belong to the middle 
of 42 trading day epochs calculated of overlapping intervals (1~trading day sliding 
windows) in order to show the relation to crises in Tab.~\ref{tab:FinancialCrises} 
(\href{https://www.quandl.com/}{Data from QuoteMedia via Quandl}).}
	\label{fig:StandCorrMeanCorr}
\end{figure}
\begin{table}[t]
	\centering
	\caption{\label{tab:FinancialCrises}
		Financial crises events taken 
		from~\cite{wiki:2019:LISTSTOCKMARKETCRASHESBEARMARKETS}.
	}
		\begin{tabular}{cl@{\hspace{1em}}c}
			\toprule
			Number &
			Crisis &
			\multicolumn{1}{c}{Date}
			\\
			 &
			 &
			\multicolumn{1}{c}{(Year-Month-Day)} \\
			\midrule
			(1) & Stock market downturn of 2002 & 2002-10-09 \\
			(2) & Chinese stock bubble & 2007-02-27 \\
			(3) & Lehman Brothers crisis & 2008-09-16 \\
			(4) & European debt crisis & 2010-04-27 \\
			(5) & August 2011 stock markets fall & 2011-08-01 \\
			(6) & The Great Fall of China & 2015-08-18 \\
			\bottomrule
		\end{tabular}
\end{table}
The 42 trading days represent a compromise between the noise,
which increases for smaller trading intervals, and the dynamics, which change
the correlations due to true economic relations between the companies.

First, the return time series $M_i(t)$ -- standardized to mean zero and standard 
deviation one -- are calculated for an epoch $n_{\text{ep}}$ with time indices $t \in [\left(n_{\text{ep}}-1\right)T + 1,\; n_{\text{ep}} T ]$ from the time series of the returns 
$G_i(t)$ by
\begin{equation} \label{eqn:TimeSeriesM}
M_i(t) = \frac{G_i(t) - \mu_i( n_{\text{ep}} )}{\sigma_i( n_{\text{ep}} )} , \hspace{0.5cm} 
i = 1, \ldots, K \,
\end{equation}
with the mean value (drift) of the returns
\begin{equation} \label{eqn:AverageTimeSeries}
\mu_i(n_{\text{ep}}) = \frac{1}{T} \sum_{t=1}^T G_i \left( 
\left( n_{\text{ep}}-1 \right) T + t \right)
\end{equation}
and the standard deviation (volatility) of the returns
\begin{equation} \label{eqn:StandardDeviationTimeSeries}
\sigma_i(n_{\text{ep}}) = \sqrt{ \frac{1}{T} \sum_{t=1}^T \left[ G_i\left( 
	\left(n_{\text{ep}}-1\right)T + t \right) - 
\mu_i(n_{\text{ep}}) \right]^2 } \,.
\end{equation}
This results in the
Pearson correlation matrix 
\begin{equation} \label{eqn:StandCorrMat}
C(n_{\text{ep}}) = \frac{1}{T} \, M(n_{\text{ep}}) M^{\dagger}(n_{\text{ep}}) 
\end{equation} 
calculated from the normalized data matrix $M(n_{\text{ep}})$, where
$M^{\dagger}$ is the transposed matrix of $M$.

\subsection{\label{sec:DeterminationMarketStates}Clustering method and determination of number of market 
states}

For the $N_{\text{ep}} = 87$ epochs, 87 correlation matrices $C(n_{\text{ep}})$ were 
calculated.
The correlation matrices are now to be divided into groups, so-called clusters, using 
a cluster algorithm.
By clustering $87 \times 262^2$ correlation matrix elements are reduced to 87 integer numbers.
The market is in the cluster or market state
in which the correlation matrix $C(n_{\text{ep}})$ has been clustered.
That is why we call this procedure market state analysis.
The goal of the market state analysis will be to analyze the temporal behavior of 
the correlation dynamics and correlation structure more closely.

We choose the bisecting $k$-means algorithm \cite{Steinbach2000:AComparisonDocumentClusteringTechinques, 
	TanSteinbachKumar:IntroductionDataMining} to cluster the correlation matrices
(a possibility to speed up the clustering is to use 
PCA~\cite{Pearson1901:LinesPlanesCLosestFit, 
	Hotelling1933:AnalysisComplexStatisticalVariablesPrincipalComponents, 
	jolliffePrincipalComponentAnalysis2002}, see Appx.~\ref{sec:FacilClust}),
which in contrast to the standard $k$-means (vanilla 
$k$-means~\cite{Steinhaus1956:DivisionCorpMaterielEnParties,BallHall1965:ISODATA,MacQueen1967:SomeMethodsClassification,LLoyd1982:LeastSquaresQuantization}, see 
Appx.~\ref{sec:StandKMeans})
 has the advantage that the number of clusters $k$ can be determined by a 
geometric 
criterion.
To cluster the correlation matrices of two different epochs 
$n_\text{ep}$ and $n^{\prime}_{\text{ep}}$, we use the 
Euclidean distance
\begin{align} \label{eqn:EuclidDist}
d \left(n_\text{ep}, n^{\prime}_\text{ep} \right) &= \sqrt{ \sum_{i,j} 
	\left( C_{ij}(n_{\text{ep}}) - C_{ij}(n^{\prime}_{\text{ep}})    
	\right)^2 } \\  
	&=  || C(n_{\text{ep}}) - 
	C(n^{\prime}_{\text{ep}}) ||
 \,.
\end{align}
The bisecting $k$-means algorithm is
a hierarchical procedure (top-down approach). 
Within the procedure, whenever a cluster is split, the $k$-means algorithm is 
applied 
for $k=2$, \textit{i.e.} a cluster (parent cluster) is split into 2 clusters (child 
clusters).
For a specific $k$, the cluster algorithm divides the set 
$Z=\{ C(1), 
C(2), \ldots, C(87) \}$ of all correlation matrices
into $k$ subsets $Z = \{ z_1, z_2, \ldots, 
z_l,\ldots, z_k \}$.
Every subset $z_l$ is a cluster.
In order to get the cluster solution $Z$ for a cluster number $k$, a 
threshold is introduced
\begin{equation} \label{eqn:BisecThreshold}
\chi = p \, d^{\text{(max)}}_{\text{width}}
\end{equation}
with the average width of a cluster
\begin{equation} \label{eqn:Width}
d^{(l)}_{\text{width}} = \frac{1}{m_l}  
	\sum_{ n_{\text{ep} } \in z_l } || C(n_{\text{ep}}) - {\langle C 
	\rangle}^{(l)}    
	|| 
\end{equation}
and the so-called centroid of a cluster
\begin{equation} \label{eqn:Cendroid}
{\langle C \rangle}_{ij}^{(l)} = \frac{1}{m_l} \sum_{ n_{\text{ep} } \in z_l  
} 
C_{ij}({{n_\text{ep}}}) \,.
\end{equation}
The number of correlation matrices in a cluster is $m_l$. The cluster with the 
maximum average width $d^{\text{(max)}}_{\text{width}}$ is usually the largest 
cluster, \textit{i.e}.
the average width of all 87 correlation matrices.
But it is also possible that -- after splitting -- one of the child clusters 
is larger than the parent cluster. $p \in [0,1] $ is the parameter for 
the threshold $\chi$ in Eq.~(\ref{eqn:BisecThreshold}) which allows to adjust 
the 
cluster solution $Z$ for a specific $k$ due to the different average widths of 
the 
clusters in the bisecting $k$-means hierarchy.
The basic idea of the hierarchy is as follows.
For $p=1$, there is only one cluster ($k=1$). By decreasing the value for $p$, 
the parent cluster of all correlation matrices will split into two child clusters 
when the threshold $\chi$ is smaller than the average width of the parent cluster.
For the value $p=0$, there are $k=N_{\text{ep}}$ clusters as 
the width of the clusters containing only one correlation matrix is zero.

Our goal is to find the value $k^{\text{(opt)}}$ for the cluster number, 
which gives the best possible spatial separation of the clusters.
We calculate for every child cluster of a cluster solution $Z$ with number $k$ 
a quotient of the 
distance between the centroids of the child clusters which both emerged from splitting the same
parent cluster
$d_{\text{CtoC}}$ and their 
individual width $d_{\text{width}}^{(l)}$
\begin{equation} \label{eqn:Quotient}
\xi_{\text{child}}^{(l)} = \frac{ d_{\text{CtoC}} }{ 
d_{\text{width}}^{(l)} \,.
} 
\;.
\end{equation}  
Thus the mean quotient of the quotients $\xi_{\text{child}}^{(l)}$ reads
\begin{equation} \label{eqn:QuotientMean}
\langle \xi_{\text{child}} \rangle = \frac{1}{k} \sum_{l=1}^{k} 
\xi_{\text{child}}^{(l)} 
 \,.
\end{equation} 
It is possible that a cluster contains only one correlation matrix.
Since the width for this ``isolated" cluster is zero and thus the
quotient in Eq.~(\ref{eqn:Quotient}) is not defined, this cluster is
omitted when calculating the mean quotient.
Fig.~\ref{fig:StandCorrDetClusterNumber} illustrates the mean quotient
as function of the cluster number $k$ for the standard correlation
matrix, where $k^{\text{(opt)}} =2$. Nevertheless, we
choose $k^{\ast}=4$ due to the following reasons: In general, the
criteria for the $k^{\ast}$ should be, first, a larger mean quotient
$\xi_{\text{child}}$ as a purely geometrical criterion, second, a
$k^{\ast}$ not too small to facilitate capturing the temporal
evolution of the correlation matrices, including smaller changes,
and third, a cluster number not too large to reflect information on
the time evolution of the correlation matrices which is a major
goal of the market state analysis.  Thus, our choice results
from a variety of motivations.
 
In contrast to the correlation matrices of stationary epochs of length
$T=42$ trading days~\cite{schmitt2013non}, the correlation matrices
within the market states themselves behave
quasi-stationary~\cite{chetalovaZoomingMarketStates2015}.  The
correlation matrices change within a market state, \textit{i.e.} they
fluctuate around an average correlation matrix.  We call the centroid
-- the average correlation matrix of one of the clusters of the
cluster solution $Z^{\text{(opt)}}$ corresponding to
$k^{\text{(opt)}}$ or $Z^{\ast}$ corresponding to $k^{\ast}$ -- in
Eq.~(\ref{eqn:Cendroid}) in the language of market states a
\emph{typical market state}.  It is a representative of the
correlation structure of the respective market state.  The averaging
reduces the noise and shows a clearer picture of the correlation
structure.

\subsection{\label{sec:ResultsClusteringStandCorrMat}Results of clustering standard correlation 
matrices}

We cluster 87 Pearson correlation matrices using the standard correlation matrix 
defined in Eq.~(\ref{eqn:StandCorrMat}) and apply the bisecting $k$-means 
algorithm 
from Sec.~\ref{sec:DeterminationMarketStates}.
We determine the number of clusters.
\begin{figure}[!htb]
	\centering
	\includegraphics[width=1.0\columnwidth]{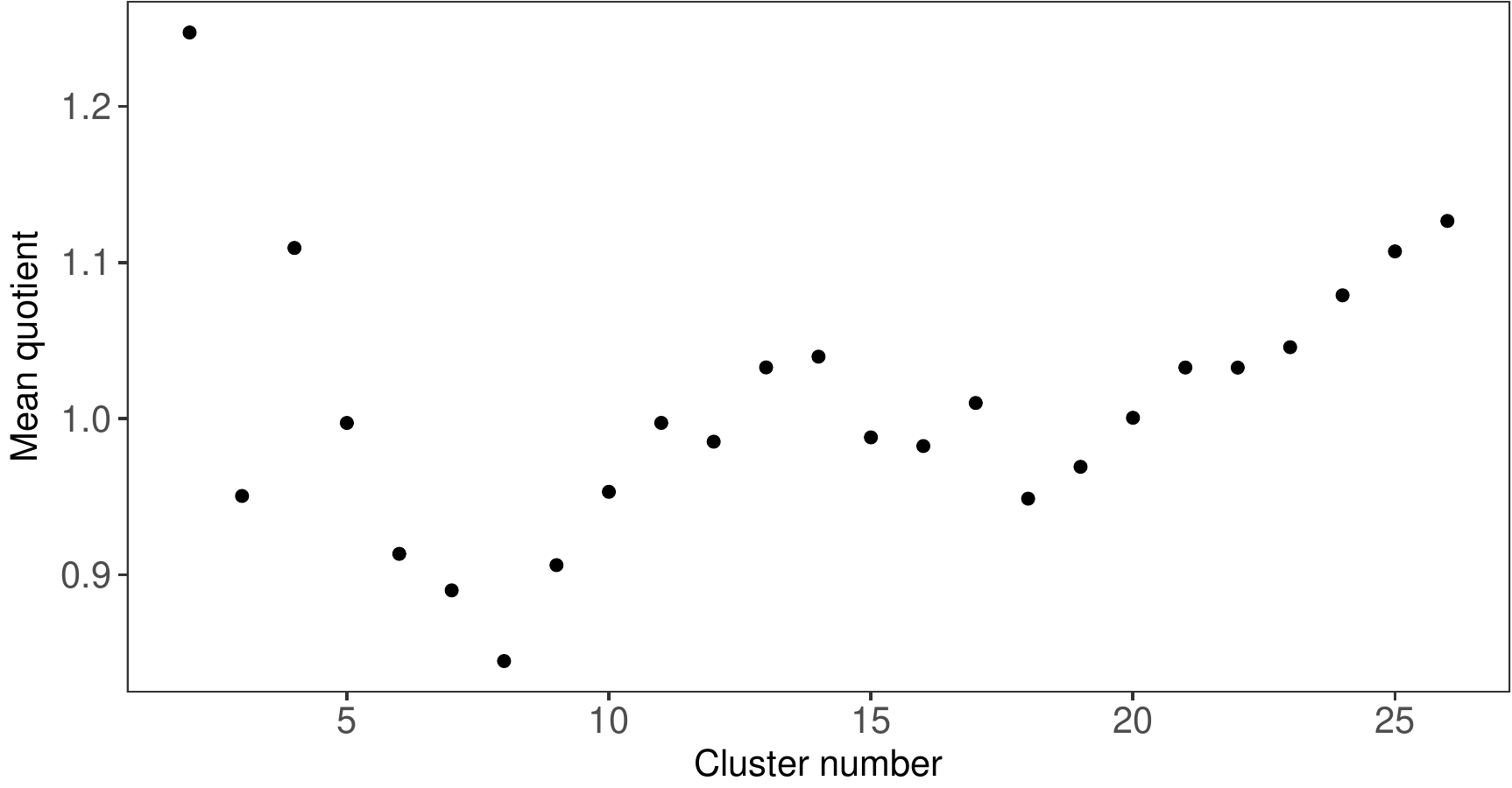}
	\caption{Determination of cluster number for the standard correlation matrix in 
Eq.~(\ref{eqn:StandCorrMat}) (\href{https://www.quandl.com/}{Data from QuoteMedia 
via Quandl}).}
	\label{fig:StandCorrDetClusterNumber}
\end{figure}
According to Fig.~\ref{fig:StandCorrDetClusterNumber}, there is an increase in 
the quotient for smaller $k$ at $k=4$.
We assume $k^{\ast}=4$ as the number of market states.
The temporal evolution of market states is shown in Fig.~\ref{fig:TimeEvolutionStandCorr}.
\begin{figure*}[!htb]
	\centering
	\includegraphics[width=1.0\textwidth]{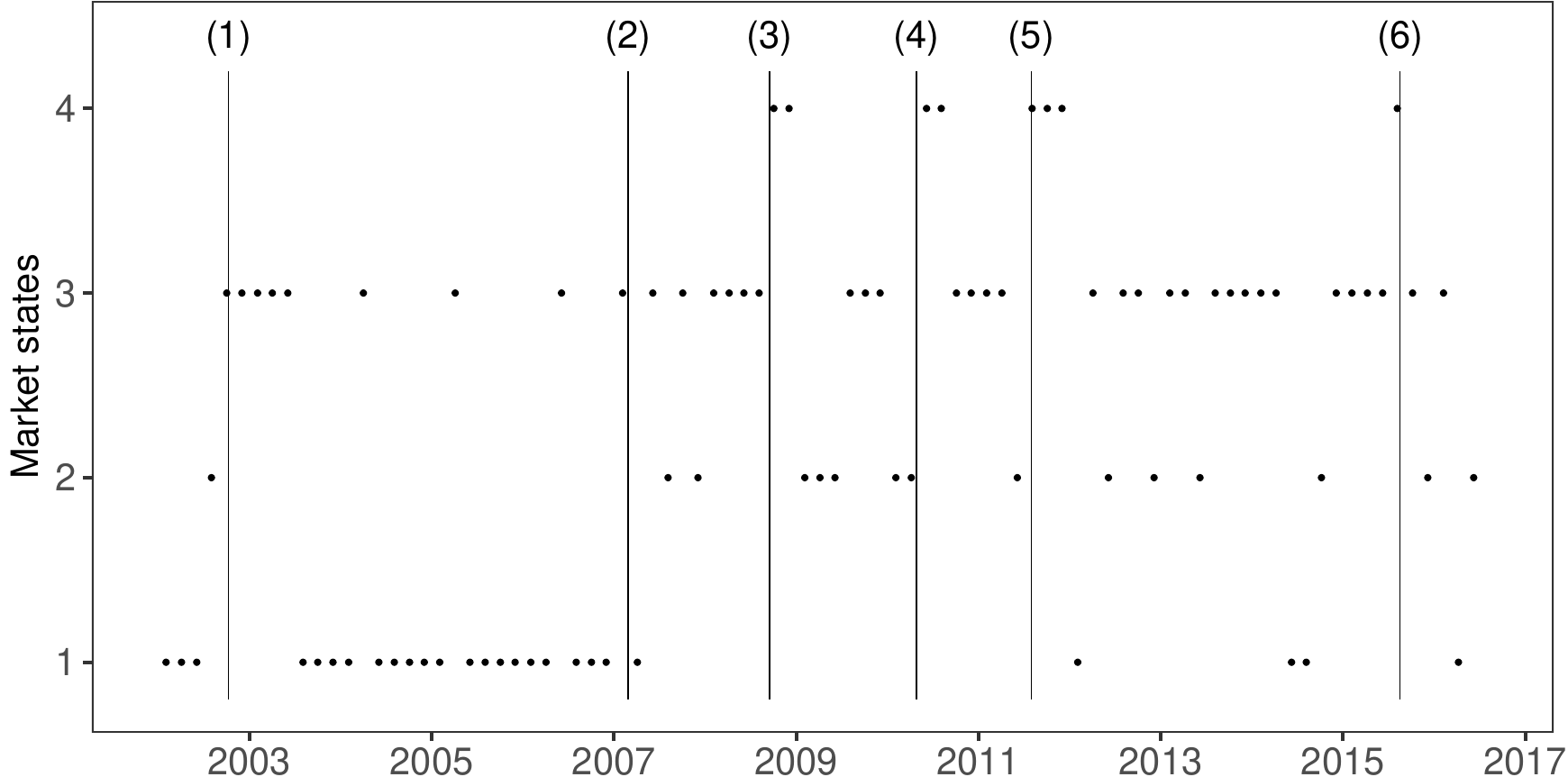}
	\caption{Temporal evolution of the standard correlation matrix in Eq.~(\ref{eqn:StandCorrMat}). 
The numbers in brackets are historical financial crises according to Tab.~\ref{tab:FinancialCrises} 
(\href{https://www.quandl.com/}{Data from QuoteMedia via Quandl}).}
	\label{fig:TimeEvolutionStandCorr}
\end{figure*}
The middle of the correlation intervals is depicted as black dots in the plots.
A number is assigned to the market states.
The state that occurs first in the period from 2002 to the middle of 2016 receives 
the number 1, the state that occurs second, number 2, etc.
In order to better understand market states, six historical events were added to 
the plots (cf. Tab.~\ref{tab:FinancialCrises}).
According to Sec.~\ref{sec:DeterminationMarketStates}, the correlation matrices assigned 
to a market state are used to form the typical market states shown in Fig.~\ref{subfig:StandCorrTypClustCent:main}.
\begin{figure*}[!htb]
	\vspace{-1cm}
	\centering
	\subfloat[\label{subfig:StandCorrTypClustCent:a}state 1]{
		\includegraphics[width=0.33\textwidth]{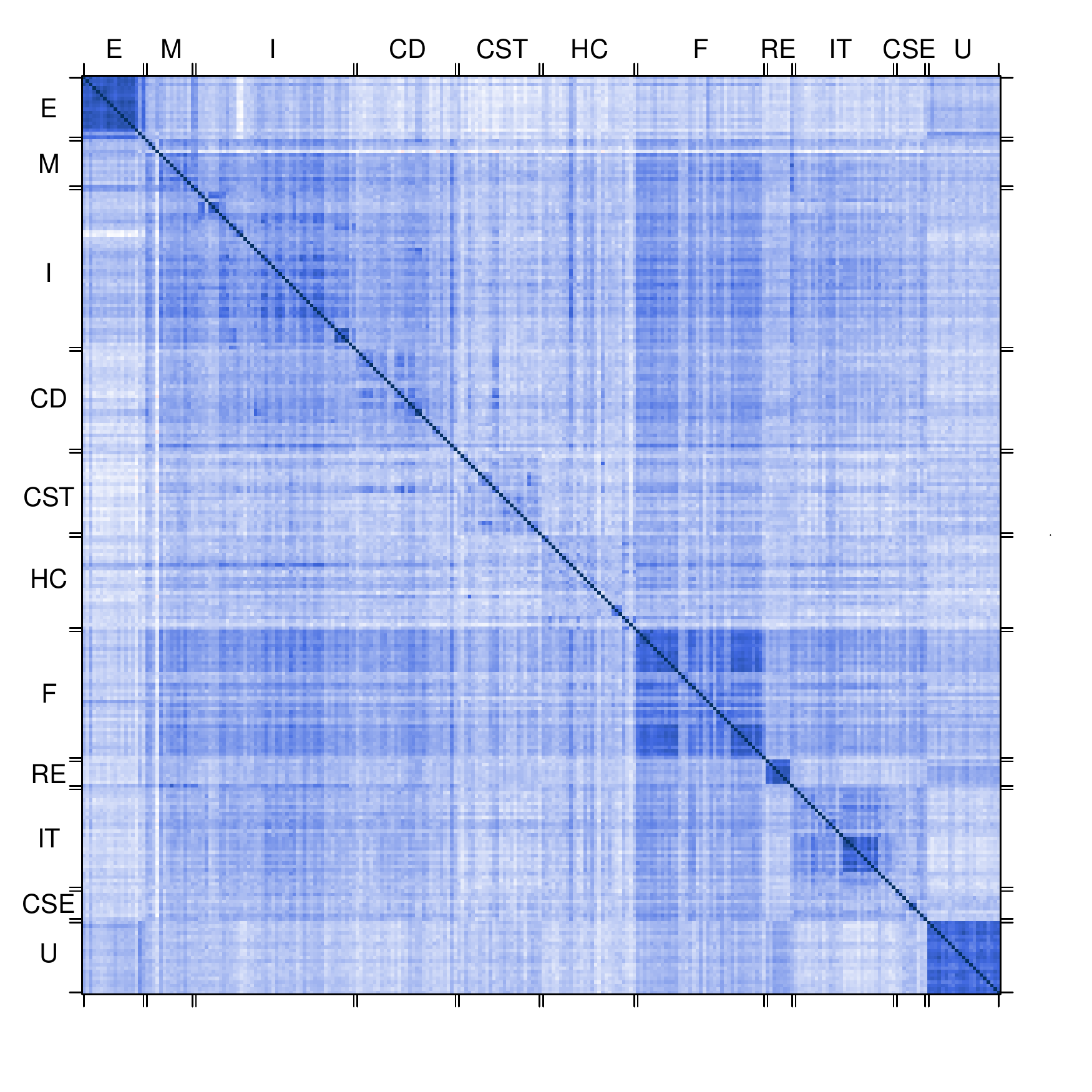}
	}
	\subfloat[\label{subfig:StandCorrTypClustCent:b}state 2]{
		\includegraphics[width=0.33\textwidth]{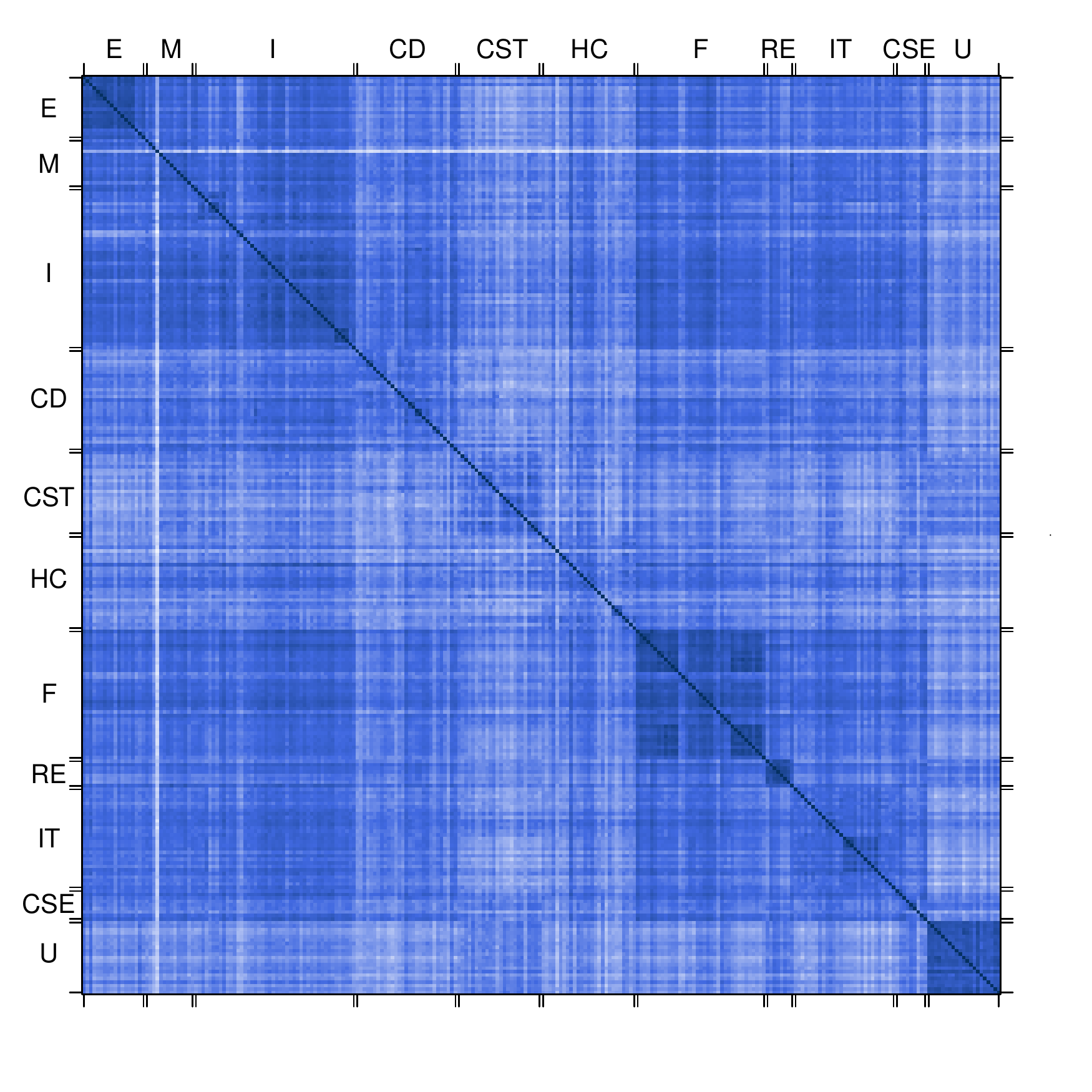}
	}\\
	\subfloat[\label{subfig:StandCorrTypClustCent:c}state 3]{
		\includegraphics[width=0.33\textwidth]{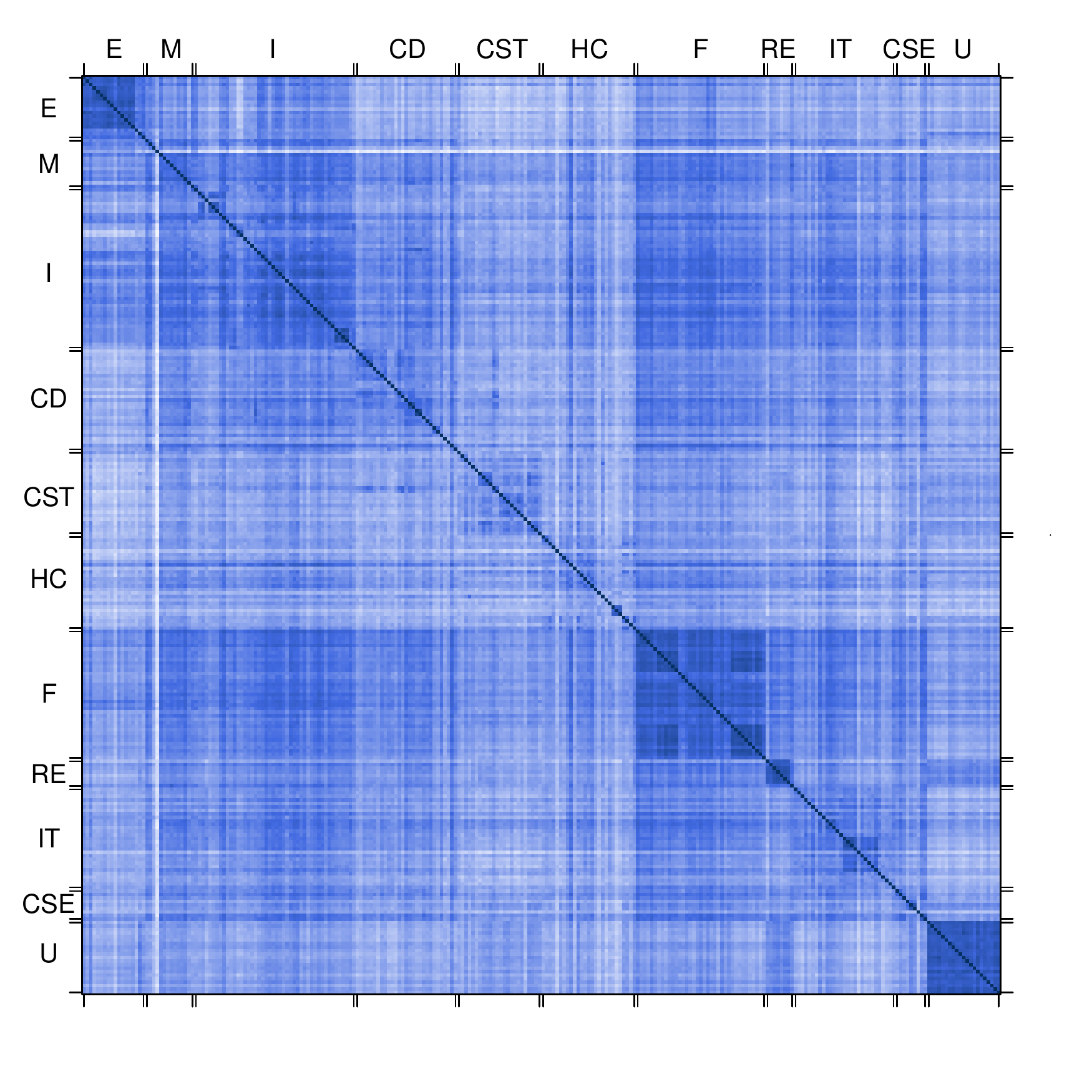}
	}
	\subfloat[\label{subfig:StandCorrTypClustCent:d}state 4]{
		\includegraphics[width=0.33\textwidth]{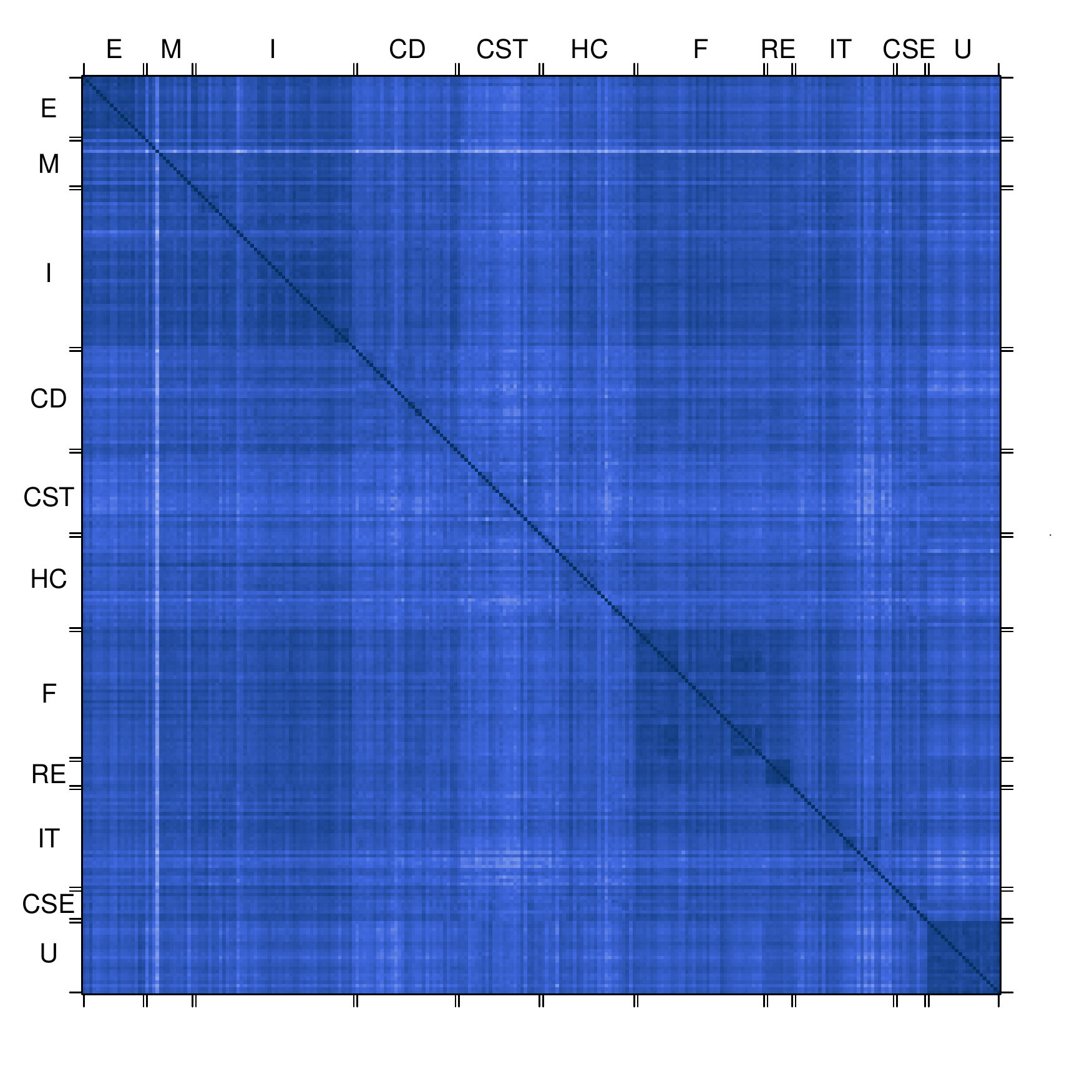}
	}\\
	\subfloat[\label{subfig:StandCorrTypClustCent:e}overall average correlation matrix 
(averaged over all 87 correlation matrices)]{%
		\includegraphics[width=0.58\textwidth]{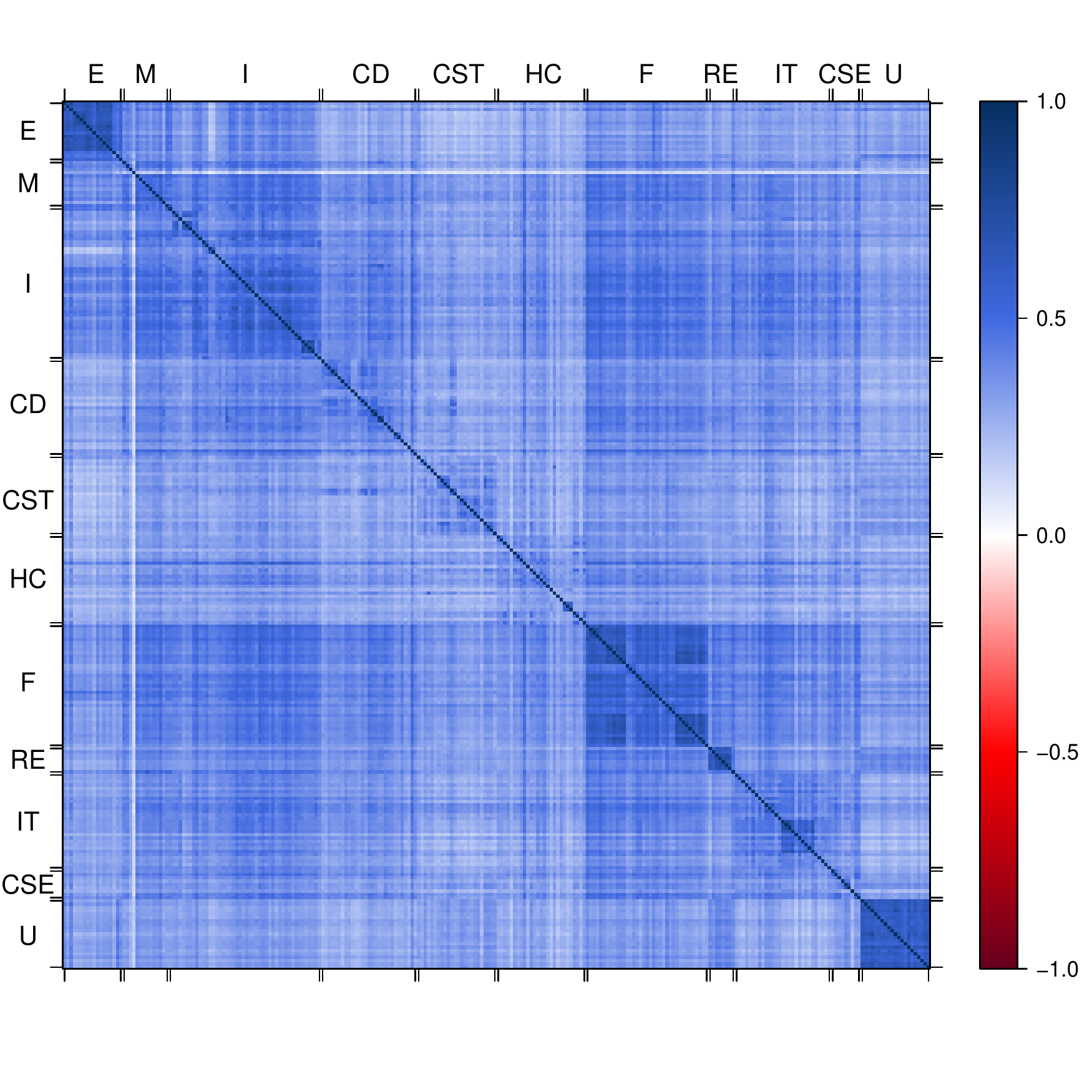}
	}
	\caption{\label{subfig:StandCorrTypClustCent:main}Typical market states of the 
standard correlation matrix in Eq.~(\ref{eqn:StandCorrMat}) calculated as element-wise 
average of the correlation matrices belonging to a market state (see Tab.~\ref{tab:NumberCorrMatMarketStates}). 
Sector legend: E: Energy; M: Materials; I: Industrials; CD: Consumer Discretionary; 
CST: Consumer Staples; HC: Health Care; F: Financials; RE: Real Estate; I: Information 
Technology; CSE: Communication Services; U: Utilities (\href{https://www.quandl.com/}{Data 
from QuoteMedia via Quandl}).}	
\end{figure*}
Tab.~\ref{tab:NumberCorrMatMarketStates} shows how many correlation matrices are 
in the respective market states.
\begin{table}[!htb]
	\centering
	\caption{\label{tab:NumberCorrMatMarketStates}
		Number of correlation matrices assigned to market states (\href{https://www.quandl.com/}{Data 
from QuoteMedia via Quandl}). 
	}
		\begin{tabular}{d@{\hspace{0em}}d@{\hspace{0em}}d@{\hspace{0em}}d}
			\toprule
			\multicolumn{1}{c}{State} &
			\multicolumn{1}{c}{Standard} &
			\multicolumn{1}{c}{Reduced-Rank}  &
			\multicolumn{1}{c}{Reduced-Rank}  \\
			 &
			 &
			\multicolumn{1}{c}{(Cov Approach)}  &
			\multicolumn{1}{c}{(Corr Approach)}
			\\
			\midrule
			1 & 26 & 73 & 44 \\
			2 & 15 & 5 & 7 \\
			3 & 38 & 1 & 17 \\
			4 & 8 & 8 & 11 \\
			5 &  &  & 8 \\
		\bottomrule
		\end{tabular}
\end{table}
According to Fig.~\ref{fig:TimeEvolutionStandCorr}, the first three market states 
are already emerging before 2003.
Fig.~\ref{subfig:StandCorrTypClustCent:main} shows that the associated typical 
market states are very similar.
The sectors CST and HC (cf. Tab.~\ref{tab:GICS}) appear as a lighter cross due 
to their lower 
correlation to the rest of the market.
The mean correlation (cf. Tab.~\ref{tab:MeanCorrTypMarkStates}) is essentially the 
difference between the first four market states.
\begin{table}[!htb]
	\centering
	\caption{\label{tab:MeanCorrTypMarkStates}%
		Mean correlation of typical market states (\href{https://www.quandl.com/}{Data 
from QuoteMedia via Quandl}).
	}
		\begin{tabular}{dddd}
			\toprule
			\multicolumn{1}{c}{State} &
			\multicolumn{1}{c}{Standard} &
			\multicolumn{1}{c}{Reduced-Rank} &
			\multicolumn{1}{c}{Reduced-Rank}
			\\
			 &
			 &
			\multicolumn{1}{c}{(Cov Approach)}  &
			\multicolumn{1}{c}{(Corr Approach)}
			\\
			\midrule
			1 & 0.23908 & 0.01818 & 0.00077 \\
			2 & 0.45601 & 0.06624 & 0.00196 \\
			3 & 0.35807 & 0.16216 & 0.00026 \\
			4 & 0.64563 & 0.04145 & 0.00153 \\
			5 &  &  & 0.00211 \\
			\bottomrule
		\end{tabular}
\end{table}
Market state 4 is characterized by its high mean correlation.
The crises events (3) to (6) as highlighted by vertical lines in Fig.~\ref{fig:TimeEvolutionStandCorr} show that this market 
state can be referred to as
``crisis state".
Fig.~\ref{fig:StandCorrMeanCorr} also illustrates the relationship between mean 
correlation and the crisis state.
Interestingly, the vertical lines (3), (4) and (5) are located at the beginning of 
the fourth market state.

The quasi-stationary sections are interrupted due to many jumps.
The market states 1, 2, and 3 also reappear at the end of the investigated data set.
A development towards new market states is hardly discernible.
The bottom line is that one actually clusters the mean correlation rather than the 
correlation structure.

We emphasize that due to the dominating behavior of the
largest eigenvalue (or the mean correlation) the Euclidean distance
in Eq.~(\ref{eqn:EuclidDist}) is governed by the differences of the
largest eigenvalues (or the mean correlation)
\cite{song2011evolution,stepanovStabilityHierarchyQuasistationary2015}.
For comparison, we also carry out a market state analysis of the
matrix structure corresponding to the leading eigenvector in
Appx.~\ref{sec:DeMeanMatrices}. The differences are substantial,
demonstrating the fundamental disparity to our main line of study,
particularly since the collective market motion itself has a matrix
structure, Appx.~\ref{sec:DeMeanMatrices}.  Nevertheless, it is
interesting to see that, \textit{e.g.}, more quasi--stationary
periods result in the analysis corresponding to the leading
eigenvector.

\section{\label{sec:ResultClusteringReducedRankCorrelationMatrices2}Clustering reduced-rank 
correlation matrices}

In Sec.~\ref{sec:MotDataMat}, we want to motivate the definition of
formal data matrices corresponding to reduced-rank correlation
matrices. Their definition and implications are given in
Sec.~\ref{sec:DefinitionRedRankCorrMat}.  In
Sec.~\ref{sec:ResultClusteringReducedRankCorrelationMatrices}, we
analyze the time evolution of market states of the reduced-rank
correlation matrices and compute their typical market states. We
compare these results with the results of the standard correlation
matrix in detail. The influence of major historical events of
financial crises on the market states and on the mean correlation of
all three correlation matrices is scrutinized as well.

\subsection{\label{sec:MotDataMat}Motivation for formal data matrices 
corresponding 
to reduced-rank correlation matrices}

Our goal is to define formal $K \times T$ data matrices for 
the 
calculation of reduced-rank correlation matrices analogously to the definition 
of standard correlation matrices (cf. Eq.~\ref{eqn:StandCorrMat}).
It is sufficient to consider the reduced-rank covariance matrices.
For each epoch $n_{\text{ep}}$ with time indices, $t \in [\left(n_{\text{ep}}-1\right)T + 1,\; n_{\text{ep}} T ]$, we normalize the rows (time series) of the data matrix $G$ in Eq.~(\ref{eqn:DatamatrixG}) 
to mean value zero 
\begin{equation} \label{eqn:TimeSeriesA}
A_i(t) = G_i(t) - \mu_i(n_{\text{ep}}) , \hspace{0.5cm} i = 1, \ldots, K \,
\end{equation}
and define the standard covariance matrix as
\begin{equation} \label{eqn:CovarianceMatAADagger}
\Sigma = \frac{1}{T} A A^{\dagger} \,.
\end{equation}
The spectral decomposition of the standard covariance matrix reads
\begin{equation} \label{eqn:SpectralDecompositionCovarianceMat}
\Sigma = \frac{1}{T} U \Lambda U^{\dagger} \,,
\end{equation}
where $U$ is an orthogonal $K \times K$ matrix whose columns are eigenvectors 
of $T \Sigma$.
$\Lambda$ is the diagonal matrix of the eigenvalues of $T \Sigma$ where the 
eigenvalues 
are ordered descendingly.
In Sec.~\ref{sec:DefinitionRedRankCorrMat}, it will become clear why we included 
the prefactor $1/T$ into the definition.
Eq.~(\ref{eqn:SpectralDecompositionCovarianceMat}) can be written as
\begin{equation} \label{eqn:WWDagger}
\Sigma = W W^{\dagger}
\end{equation}
with
\begin{equation} \label{eqn:ULambda}
W = \frac{1}{\sqrt{T}} \, U \Lambda^{1/2} \,.
\end{equation}
If we set eigenvalues in the eigenvalue matrix $\Lambda$ in 
Eq.~(\ref{eqn:ULambda}) to zero, we are able to define reduced-rank covariance 
matrices.
In Sec.~\ref{sec:DefinitionRedRankCorrMat}, we will see that we can construct formal $ K \times T $ data matrices that can be used to 
compute covariance and correlation 
matrices where the rows are formal time series of the companies without the 
largest 
eigenvalue.
Furthermore, we also want to investigate whether correction terms may exist, that 
means whether a
reduced-rank covariance
matrix calculated from a 
formal data matrix actually correspond to that
obtained by setting the largest eigenvalue in Eq.~(\ref{eqn:ULambda}) to zero.

\subsection{\label{sec:DefinitionRedRankCorrMat}Definition of reduced-rank correlation 
matrices}

In Sec.~\ref{sec:ResultsClusteringStandCorrMat}, we saw that
essentially the market states are dominated by the mean value of the
standard correlation matrix. The mean correlation and the largest
eigenvalue show the same
dynamics~\cite{stepanovStabilityHierarchyQuasistationary2015,song2011evolution}
and the largest eigenvalue corresponds to the marketwide collective
behavior~\cite{plerouRandomMatrixApproach2002}.  Therefore, it is
useful to analyze what happens if this effect is removed from the
standard correlation matrices.  The correlation matrices free of this
dominating effect are called reduced-rank correlation matrices.  To
better understand the concept of reduced-rank correlation matrices, we
use the singular value decomposition (SVD) of the data matrix.
Analogous to the definition of the Pearson correlation matrix in
Eq.~(\ref{eqn:StandCorrMat}), it makes sense to define the
reduced-rank correlation matrices via a data matrix as well.  We are
going to derive two different approaches for the reduced-rank
correlation matrices.

A data matrix $A$, which is calculated from the data matrix $G$ in Eq.~(\ref{eqn:DatamatrixG}), 
has time series normalized to mean value zero, as introduced 
in Eq.~(\ref{eqn:TimeSeriesA}).
The normalization for all time series $A_i$ ($i = 1, \ldots, K$) or rows of the data 
matrix $A$ can also be formulated
as
\begin{equation} \label{eqn:DataMatrixANormalization}
\begin{pmatrix}
\langle A_1(t) \rangle, \ldots, \langle A_i(t) \rangle, \ldots, \langle 
 A_K(t) \rangle 
\end{pmatrix}
=
\frac{1}{T} A \, e =
\text{\MVZero} 
 \,,
 \end{equation}
where 
\begin{equation} \label{eqn:AverageValueA}
\langle A_i(t) \rangle =
\frac{1}{T} \sum_{t=1}^{T} A_i(t)
\end{equation}
is the mean value for the time series $A_i(t)$,
\begin{equation} \label{eqn:eVector}
e = \left(1, \ldots, 1 \right)
\end{equation}
is the $T$-dimensional column vector with ones in all entries and
\begin{equation} \label{eqn:NullVec}
\text{\MVZero} = \left( 0,\ldots,0 \right)
\end{equation}
is the zero column vector.

The singular value decomposition of the data matrix $A$ reads
\begin{equation} \label{eqn:SingValDecomp}
A = U \alpha V^{\dagger} \,,
\end{equation}
where $U$ is the orthogonal $K \times K$ matrix introduced in 
Eq.~(\ref{eqn:SpectralDecompositionCovarianceMat}) and 
$V$ an orthogonal $T \times T$ matrix, where the columns of these matrices are the 
eigenvectors to $A A^{\dagger}$ or $A^{\dagger} A$ respectively.
For $AA^{\dagger}$ and $A^{\dagger}A$ we arrive at
\begin{align} \label{eqn:AADagger}
A A^{\dagger} &= U \alpha \alpha^{\dagger} U^{\dagger}
\\ 
\label{eqn:ADaggerA}
A^{\dagger} A &= V \alpha^{\dagger} \alpha V^{\dagger} \,,
\end{align}
where
$\alpha$ is a $K \times T$ matrix.
For $T < K$ it is 
structured 
as
\begin{equation} \label{eqn:SingValMatAlpha}
\alpha = \begin{bmatrix}  \alpha_1 & \dots & 0 \\
\vdots & \ddots & \vdots \\
0 & \dots & \alpha_T \\
0 & \dots & 0 \\
\vdots & \ddots & \vdots \\
0 & \dots & 0 \\ 
\end{bmatrix} \,.
\end{equation}
$\alpha_t$, $t = 1, \ldots, T$ are the singular values which are non-negative real 
numbers.
Additionally, the zero block of the matrix $\alpha$ is a $(K-T) \times T$ 
matrix.
For $T \geq K$ the singular value matrix looks like
\begin{equation} \label{eqn:SingValMatAlphaTgeqK}
\alpha = \begin{bmatrix}  \alpha_1 & \dots & 0 & 0 & \dots & 0 \\
\vdots & \ddots & \vdots & \vdots & \ddots & \vdots \\
0 & \dots & \alpha_K & 0 & \dots & 0 \\
\end{bmatrix} \,,
\end{equation}
where the diagonal part has $\alpha_i$, $i=1,\ldots, K$ singular values.
The zero block is a $K \times (T-K)$ matrix.
Explicitly writing the columns of $U$ and $V$ as eigenvectors $u_t$ and $v_t$, 
it 
follows from Eq.~(\ref{eqn:SingValDecomp})
\begin{align} \label{eqn:SingValDecompAsSum}
A = U \alpha V^{\dagger} &= \sum_{t=1}^{\text{min}(T,K)} \alpha_t u_t 
v_t^{\dagger}
\,,
\end{align}
where
\begin{align}
\text{min}(T,K) =
\begin{cases}
	T, & \text{for } T<K \\
	K, & \text{for } T \geq K  \,.
\end{cases}
\end{align}

Henceforth, we only discuss the case $T<K$.
Using Eq.~(\ref{eqn:SingValMatAlpha}) and the normalization in Eq.~(\ref{eqn:DataMatrixANormalization}), 
we obtain the eigenvalue equations
\begin{align} 
A A^{\dagger} u_i &= \alpha_i^2 u_i
				   \hspace{0.1cm}&&\textrm{for $i = 1, \ldots, t-1 , t+1 , 
				   	\ldots, 
				   	T$}  
				   \label{eqn:EigenEquTSmallK_1} 
\\
A A^{\dagger} u_i &= 0 \; u_i \hspace{0.1cm}&&\textrm{for $i = T+1, \ldots, 
K$}  
\label{eqn:EigenEquTSmallK_2} 
\\
A A^{\dagger} u_i &=  0 \; u_i \hspace{0.1cm}&&\textrm{for one value $i<T$}
\\
A^{\dagger} A v_t &= \alpha_t^2 v_t
				   \hspace{0.1cm}&&\textrm{for $t = 1, \ldots, t-1 , t+1 , 
				   \ldots, 
T$} \label{eqn:EigenEquTSmallK_3} \\
A^{\dagger} A v_t &=  0 \; v_t \hspace{0.1cm}&&\textrm{for one value $t$} 
\label{eqn:EigenEquTSmallK_4} 
\,.
\end{align}
Eq.~(\ref{eqn:EigenEquTSmallK_3}) does not have $T$, but only $T-1$ %
non-zero eigenvalues.
Due to the normalization (\ref{eqn:DataMatrixANormalization}) the following holds:
\begin{equation} \label{eqn:EigenEquNorm}
A{^\dagger} A e = A{^\dagger} (A e) = %
\text{\MVZero} \,.
\end{equation}
That means one of the $T$ eigenvalues
of $A{^\dagger} A$ is set to zero by normalization.
Therefore, Eq.~(\ref{eqn:EigenEquTSmallK_1}) has only non-zero $T-1$ eigenvalues as well.
The standard correlation matrix in Eq.~(\ref{eqn:StandCorrMat}) then has
$(K-T)+1$ zero eigenvalues for $T<K$.
We should keep in mind that %
for $T<K$ one eigenvalue of Eq.~(\ref{eqn:SingValDecompAsSum}) is zero. 
The normalization in Eq.~(\ref{eqn:DataMatrixANormalization}) now makes it possible 
to find the following relationship:
\begin{equation} \label{eqn:SingValDecompAsSumNormCondition}
A e = \sum_{t=1}^{T} \alpha_t u_t v_t^{\dagger} e  = \text{\MVZero}
\,.
\end{equation}
The orthogonalization condition is only
fulfilled if
\begin{equation} \label{eqn:NormConditionVectorV}
v_t^{\dagger} e  = 0  \hspace{0.5cm}\textrm{for $t = 1 \ldots t-1 , t+1 \ldots T$} 
\,.
\end{equation}
This means that the dyadic matrices and any combinations of dyadic matrices of the 
data matrix $A$ in Eq.~(\ref{eqn:SingValDecompAsSum}) are always automatically normalized 
to mean value zero if the entire data matrix $A$ has been normalized to mean value 
zero.

To define a correlation matrix without the largest eigenvalue $\alpha_T^2$, we 
define 
the following data matrix:
\begin{equation} \label{eqn:DatamatrixB}
B = A - \alpha_T u_T v_T^{\dagger} = \sum_{t=1}^{T-1} 
\alpha_t u_t 
v_t^{\dagger}  \,.
\end{equation}
According to Eq.~($\ref{eqn:NormConditionVectorV}$) $B$ is already normalized to 
mean value zero.
So we can now define a formal covariance matrix without the largest eigenvalue 
and without 
additional normalization to zero by
\begin{equation} \label{eqn:CovMatToDatamatrixB}
\Sigma_{B} = \frac{1}{T} B 
B^{\dagger}  \,.
\end{equation}
In addition, the rows of $B$ can be set to standard deviation one through
\begin{equation} \label{eqn:DatamatrixBAstNormVolat}
B^{\ast} =  \left( \sigma^B \right)^{-1} B
\end{equation}
with the formal volatility matrix (cf. 
Eq.~(\ref{eqn:StandardDeviationTimeSeries}))
\begin{equation} \label{eqn:VolatilityMatrixB}
\sigma^B =
\textrm{diag} \left( \sigma_1^{B} , \ldots , \sigma_K^{B} \right)  \,.
\end{equation}
This allows us to use Eq.~(\ref{eqn:CovMatToDatamatrixB}) to define a formal 
correlation 
matrix
\begin{equation} \label{eqn:CorrMatToDatamatrixB}
C_B = \left( \sigma^B \right)^{-1} \, \Sigma_B \, \left( \sigma^B \right)^{-1} = 
\frac{1}{T} B^{\ast} \left( B^{\ast} \right)^{\dagger} \,.
\end{equation}
This is one way to define a correlation matrix without the first eigenvalue.

Another possibility is to normalize the rows of data matrix $A$ not only to the mean 
value zero, but also to standard deviation one and then to repeat the procedure described 
above.
Like in Eq.~(\ref{eqn:DatamatrixBAstNormVolat}) the normalization to standard deviation 
one can be written as 
\begin{equation} \label{eqn:DatamatrixMNormVolat}
M =  \sigma^{-1} A \,.
\end{equation}
$M$ is a data matrix whose rows are normalized to mean zero and standard deviation 
one.
The additional normalization to standard deviation one does not change the single 
normalization of the dyadic matrices to mean value zero.
The eigenvalue equations for $MM^{\dagger}$ and $M^{\dagger}M$ read
\begin{align} 
M M^{\dagger} x_i &= \mu_i^2 x_i
\hspace{0.1cm}&&\textrm{for $i = 1, \ldots, t-1 , t+1 , 
	\ldots, 
	T$}   
\label{eqn:EigenEquTSmallK_1_MM} \\
M M^{\dagger} x_i &= 0 \; x_i \hspace{0.1cm}&&\textrm{for $i = T+1, \ldots, K$}  
\label{eqn:EigenEquTSmallK_2_MM} \\
M^{\dagger} M x_i &=  0 \; x_i \hspace{0.1cm}&&\textrm{for one value $i<T$}
\\
M^{\dagger} M y_t &=\mu_t^2 y_t \hspace{0.1cm}&&\textrm{for $t = 1, \ldots, t-1, 
t+1, 
\ldots, T$} \label{eqn:EigenEquTSmallK_3_MM} \\
M^{\dagger} M y_t &=  0 \; y_t \hspace{0.1cm}&&\textrm{for one value $t$} 
\label{eqn:EigenEquTSmallK_4_MM} 
\,.
\end{align}
In order to obtain the reduced-rank correlation matrix without the largest 
eigenvalue $\mu_T^2$, we can 
define another formal
data matrix:
\begin{equation} \label{eqn:DatamatrixN}
L =  M - \mu_T x_T y_T^{\dagger} = \sum_{t=1}^{T-1} \mu_t x_t 
y_t^{\dagger} 
\,.
\end{equation}
The rows of $L$ are now set to mean value zero but not to standard deviation 
one.
The formal covariance matrix is written as
\begin{equation} \label{eqn:CovMatToDatamatrixN}
\Sigma_L = \frac{1}{T} L L^{\dagger}  \,.
\end{equation}
Analogous to Eq.~(\ref{eqn:DatamatrixBAstNormVolat}), a data matrix is defined
whose rows are also normalized to standard deviation one
\begin{equation} \label{eqn:DatamatrixNAstNormVolat}
L^{\ast} =  \left( \sigma^L \right)^{-1} L
\end{equation}
with the formal volatility matrix
\begin{equation} \label{eqn:VolatilityMatrixN}
\sigma^L =
\textrm{diag} \left( \sigma_1^{L} , \ldots , \sigma_K^{L} 
\right) \,.
\end{equation}
The second definition of a reduced-rank correlation matrix is then
\begin{equation} \label{eqn:CorrMatToDatamatrixN}
C_L = \left( \sigma^L \right)^{-1} \, \Sigma_L \, \left( \sigma^L \right)^{-1} 
= 
\frac{1}{T} L^{\ast} \left( L^{\ast} \right)^{\dagger} \,.
\end{equation}
The two approaches for the reduced-rank correlation matrices can also be
expressed differently using Eqs.~(\ref{eqn:DatamatrixB}) and (\ref{eqn:DatamatrixN}).
Employing $v_t^{\dagger} v_{t^{\prime}} = \delta_{t, t^{\prime}}$
for the first approach of the covariance matrix $\Sigma$ corresponding to the data 
matrix $A$ yields 
\begin{align} 
C_B &=  \left( \sigma^B \right)^{-1}  \left( \frac{1}{T} A  A^{\dagger} - 
\frac{1}{T} \alpha_T^2 
u_T u_T^{\dagger}  \right) \left( \sigma^B \right)^{-1} \label{eqn:CorrMatToDatamatrixBAlternat_1} 
\\
&= \left( \sigma^B \right)^{-1} \left( \Sigma -
\frac{1}{T}\alpha_T^2 u_T u_T^{\dagger} \right) \left( \sigma^B \right)^{-1} 
\label{eqn:CorrMatToDatamatrixBAlternat_2}
\,.
\end{align}
and employing $y_t^{\dagger} y_{t^{\prime}} = \delta_{t, t^{\prime}}$ for the second 
approach of the correlation matrix $C$ corresponding to the data matrix $M$ yields
\begin{align} 
C_L &=  \left( \sigma^L \right)^{-1}  \left( \frac{1}{T} M  M^{\dagger} - 
\frac{1}{T}\mu_T^2 
x_T x_T^{\dagger}  \right) \left( \sigma^L \right)^{-1} 
\label{eqn:CorrMatToDatamatrixNAlternat_1} 
\\
&= \left( \sigma^L \right)^{-1} \left( C -
\frac{1}{T}\mu_T^2 x_T x_T^{\dagger} \right) \left( \sigma^L \right)^{-1} 
\label{eqn:CorrMatToDatamatrixNAlternat_2}
\,.
\end{align}
Eq.~(\ref{eqn:CorrMatToDatamatrixBAlternat_2}) and Eq.~(\ref{eqn:CorrMatToDatamatrixNAlternat_2}) 
describe well-defined correlation matrices reduced by one rank in which the first 
dyadic matrices are subtracted from the covariance matrix $\Sigma$ and from the correlation 
matrix $C$, respectively, and the associated standard deviations on the main diagonal 
are taken from the respective resulting covariance matrices to form the reduced-rank 
correlation matrices. 
Eq.~(\ref{eqn:CorrMatToDatamatrixBAlternat_2}) yields the same result for the 
reduced-rank covariance matrix part as the 
result obtained by spectral decomposition of 
the covariance matrix $\Sigma$ described in Sec.~\ref{sec:MotDataMat} 
setting 
the largest 
eigenvalue to zero. In general, there are no correction terms for the 
reduced-rank covariance matrices and for the 
reduced-rank correlation matrices.
In this Sec.~\ref{sec:DefinitionRedRankCorrMat}, we used additionally the $A^{\dagger}A$ 
matrix and the eigenvector matrix $V$ which is important for the normalization of 
the data matrix due to the normalization condition in Eq.~(\ref{eqn:NormConditionVectorV}). 
That we receive no correction term to force the normalization to mean value zero 
is not trivial.
In addition, any combinations of dyadic matrices without the noisy bulk can be 
used
for filtering before calculating reduced-rank correlation 
matrices~\cite{miceliUltrametricityFundFunds2004, tumminello2007kullback, 
	tumminello2007shrinkage}. 

The shown considerations lead for $T \geq K$ to the same results as in Eq.~(\ref{eqn:CorrMatToDatamatrixBAlternat_2}) 
and Eq.~(\ref{eqn:CorrMatToDatamatrixNAlternat_2}).
The considerations differ in the choice of the matrix $\alpha$ in 
Eq.~(\ref{eqn:SingValMatAlphaTgeqK}) 
and the normalization, which resulted in Eq.~(\ref{eqn:EigenEquTSmallK_4}) for 
$T<K$ 
in an additional zero eigenvalue, which for $T>K$ -- the typical case of a full-rank correlation matrix -- does not force any additional zero 
eigenvalue anymore.

An optical impression between the standard correlation matrix and the reduced-rank 
correlation matrices is given in Fig.~\ref{subfig:EntTimeCorr:main}.
\begin{figure*}[t]
	\vspace{-1cm}
	\centering
	\subfloat[\label{subfig:EntTimeStandCorr}Standard correlation matrix]{
		\includegraphics[width=0.33\textwidth]{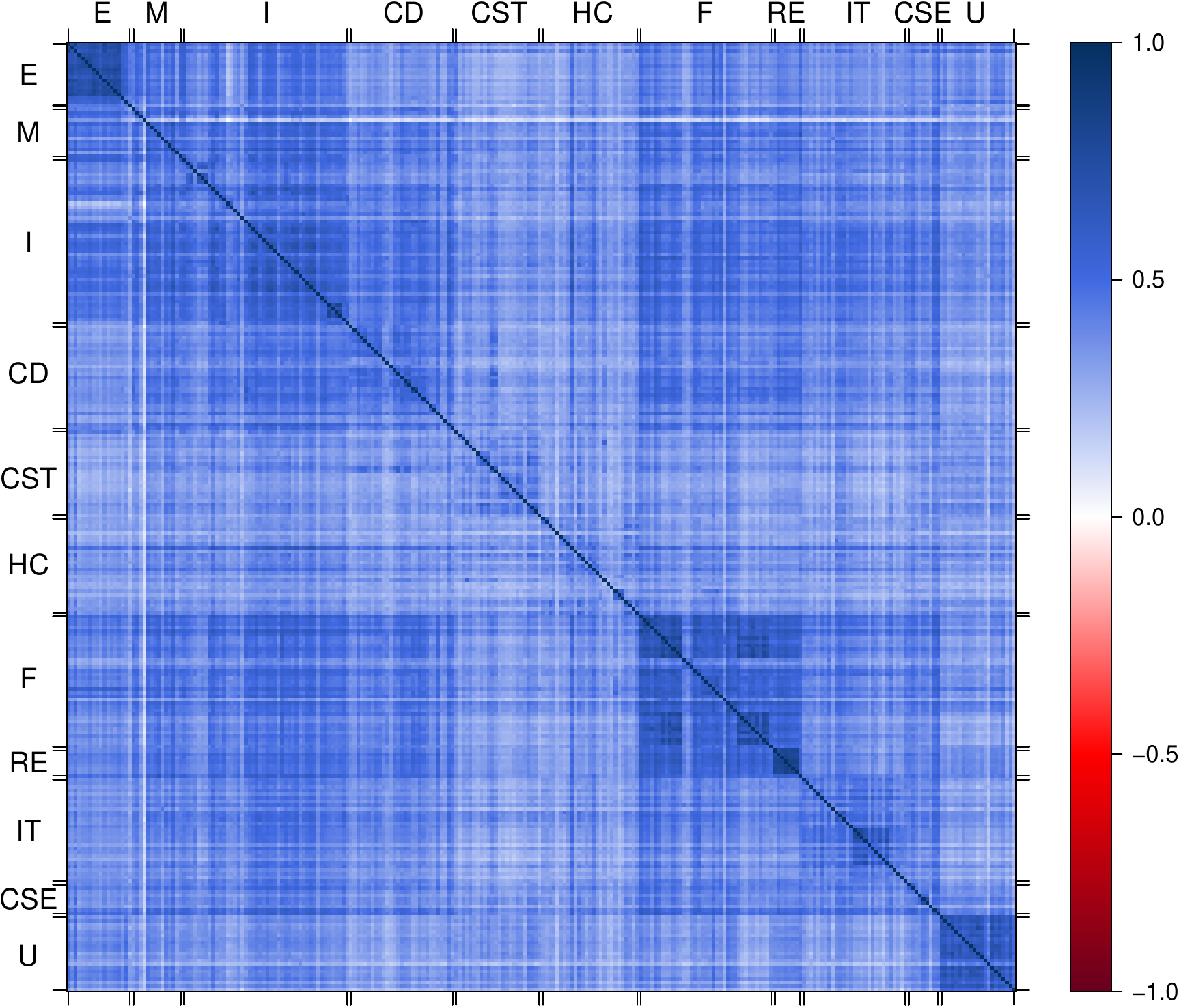}
	}
	\subfloat[\label{subfig:EntTimeRedRankCov:b}Reduced-rank correlation matrix (covariance 
approach)]{
		\includegraphics[width=0.33\textwidth]{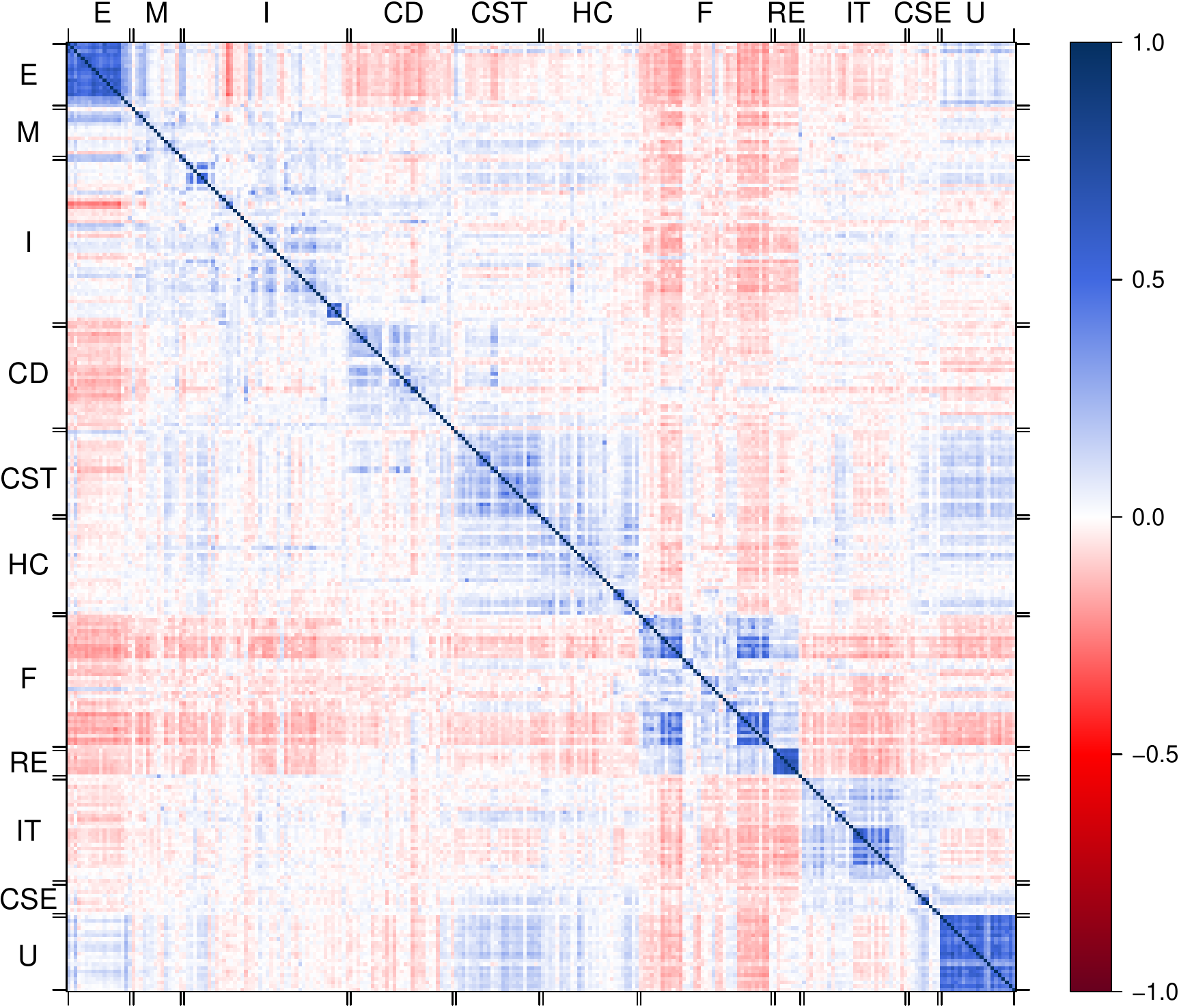}
	}
	\subfloat[\label{subfig:EntTimeRedRankCorr:c}Reduced-rank correlation matrix (correlation 
approach)]{
		\includegraphics[width=0.33\textwidth]{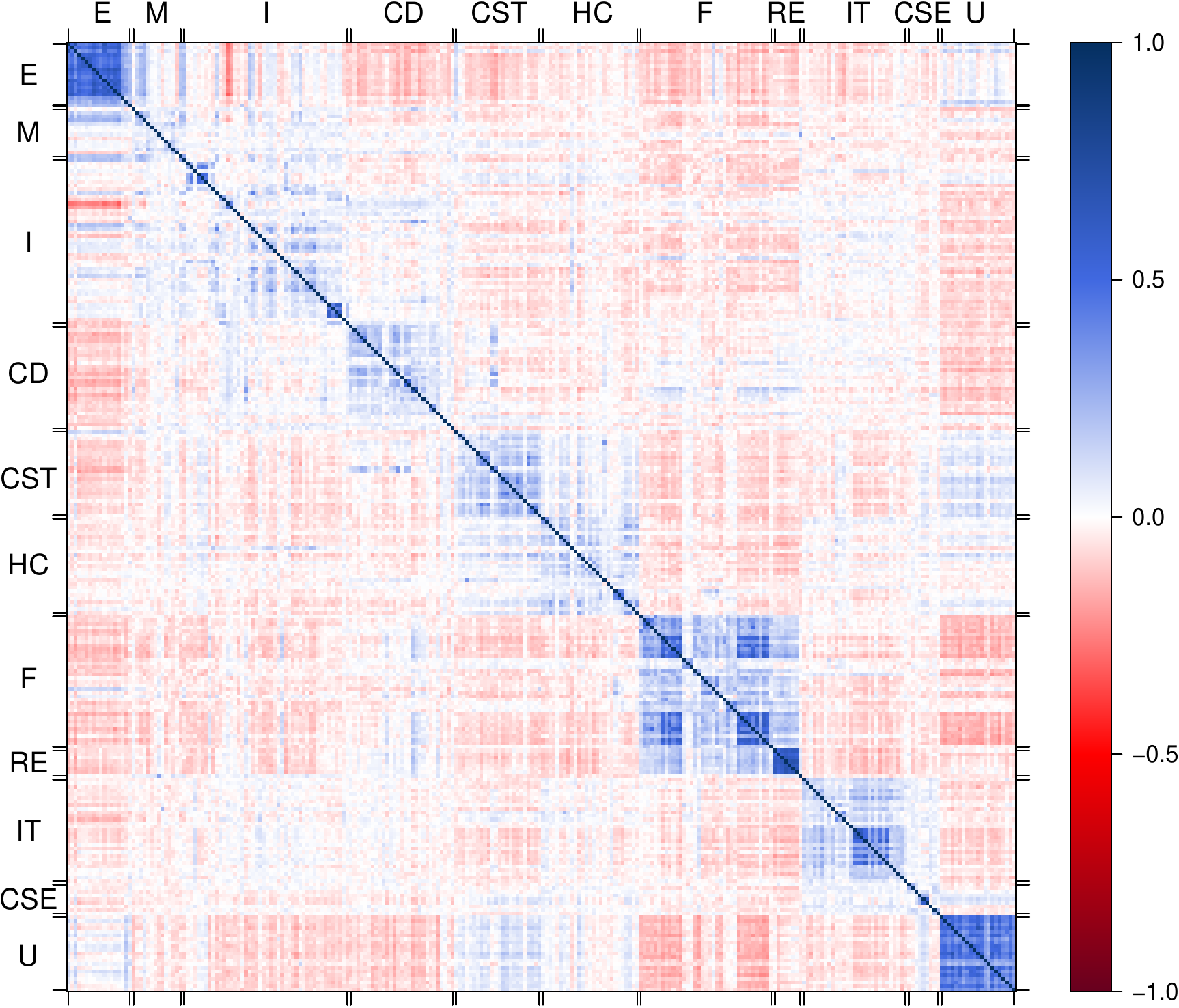}
	}
	\caption{\label{subfig:EntTimeCorr:main}The correlation matrices were calculated 
for the entire time period (t=1, \ldots, $T_{\text{tot}}$). Sector legend: E: Energy; 
M: Materials; I: Industrials; CD: Consumer Discretionary; CST: Consumer Staples; 
HC: Health Care; F: Financials; RE: Real Estate; I: Information Technology; CSE: 
Communication Services; U: Utilities (\href{https://www.quandl.com/}{Data from QuoteMedia 
via Quandl}).}	
\end{figure*}
They were calculated over the entire time period (t=1, \ldots, $T_{\text{tot}}$).
The correlation structure of the reduced-rank correlation matrices can be seen much 
more clearly.
Due to its positive intra-sector correlation structure, the block diagonal structure 
is more distinct from the rest of inter-sector correlations, which show significantly 
large negative correlations.
The reduced-rank correlation matrices hardly differ from each other for the entire 
time span.
As we will see later in 
Sec.~\ref{sec:ResultClusteringReducedRankCorrelationMatrices},
the use of correlation matrices of short time epochs allows us to
resolve large differences in the time evolution of the reduced-rank correlation 
matrices.

\subsection{\label{sec:ResultClusteringReducedRankCorrelationMatrices}Results of clustering 
reduced-rank correlation matrices}

Fig.~\ref{subfig:StandCorrTypClustCent:main} shows the typical market states of 
the standard correlation matrices.
All market states differ from each other in the sense of a first order effect only in 
the mean correlation.
First and foremost, the temporal evolution of the mean correlation is clustered.
However, we are not interested in
a global effect of the market here, but want to see exactly how the sectors change over time.
Therefore it is of interest to set the first dyadic matrix or the largest eigenvalue 
formally to zero.
In contrast to clustering the filtered covariance matrices without ``market", 
the 
additional normalization to standard deviation one offers the advantage, that one 
does not cluster the dominant variances but the correlation structure.    
\begin{figure}[b]
	\centering
	\includegraphics[width=1.0\columnwidth]{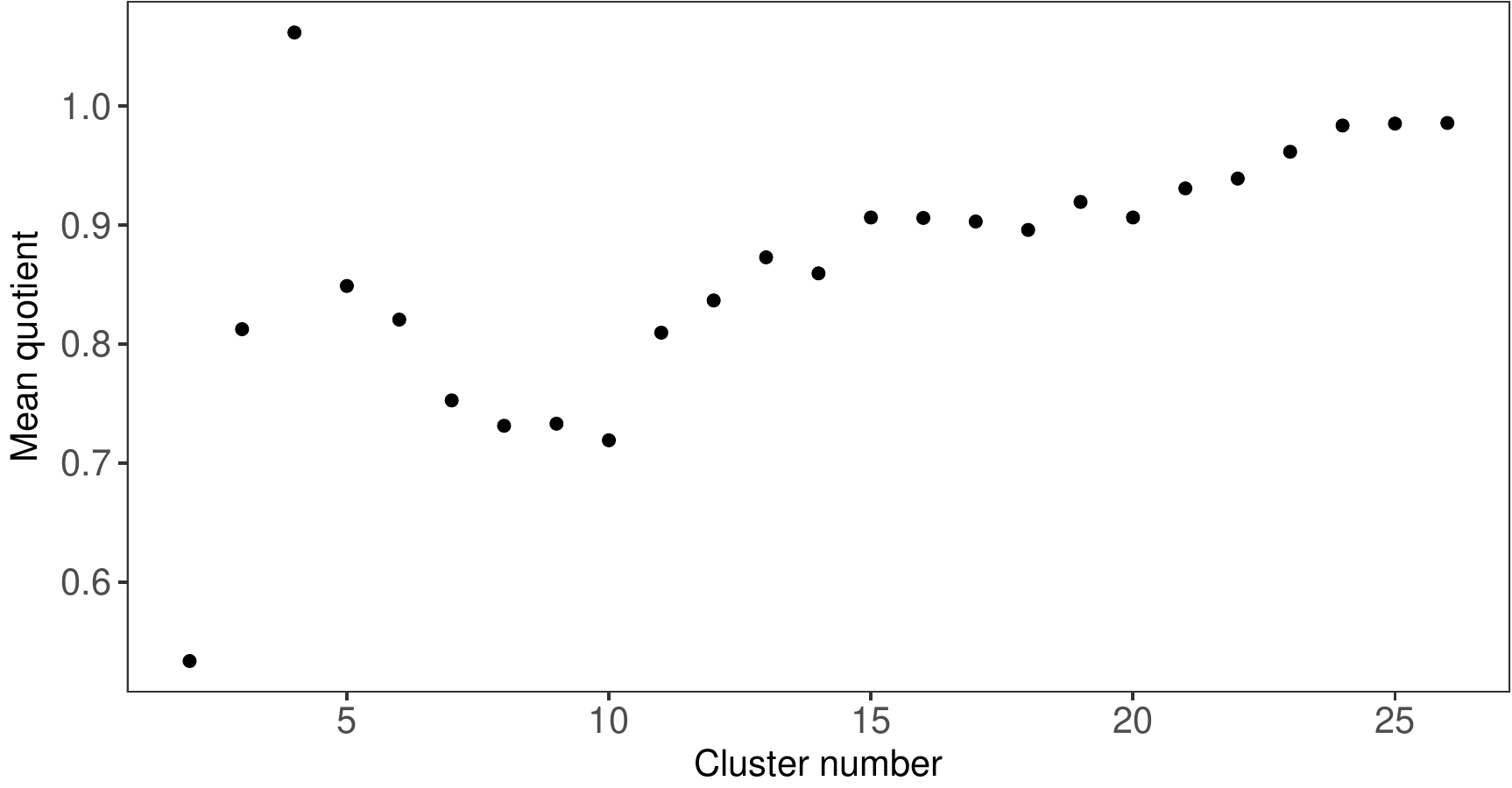}
	\caption{Determination of cluster number for the reduced-rank correlation matrix 
(covariance approach) in Eq.~(\ref{eqn:CorrMatToDatamatrixBAlternat_2}) (\href{https://www.quandl.com/}{Data 
from QuoteMedia via Quandl}).}
	\label{fig:RedRankCorrCovAnsatzDetClusterNumber}
\end{figure}
\begin{figure}[!htb]
	\centering
	\includegraphics[width=1.0\columnwidth]{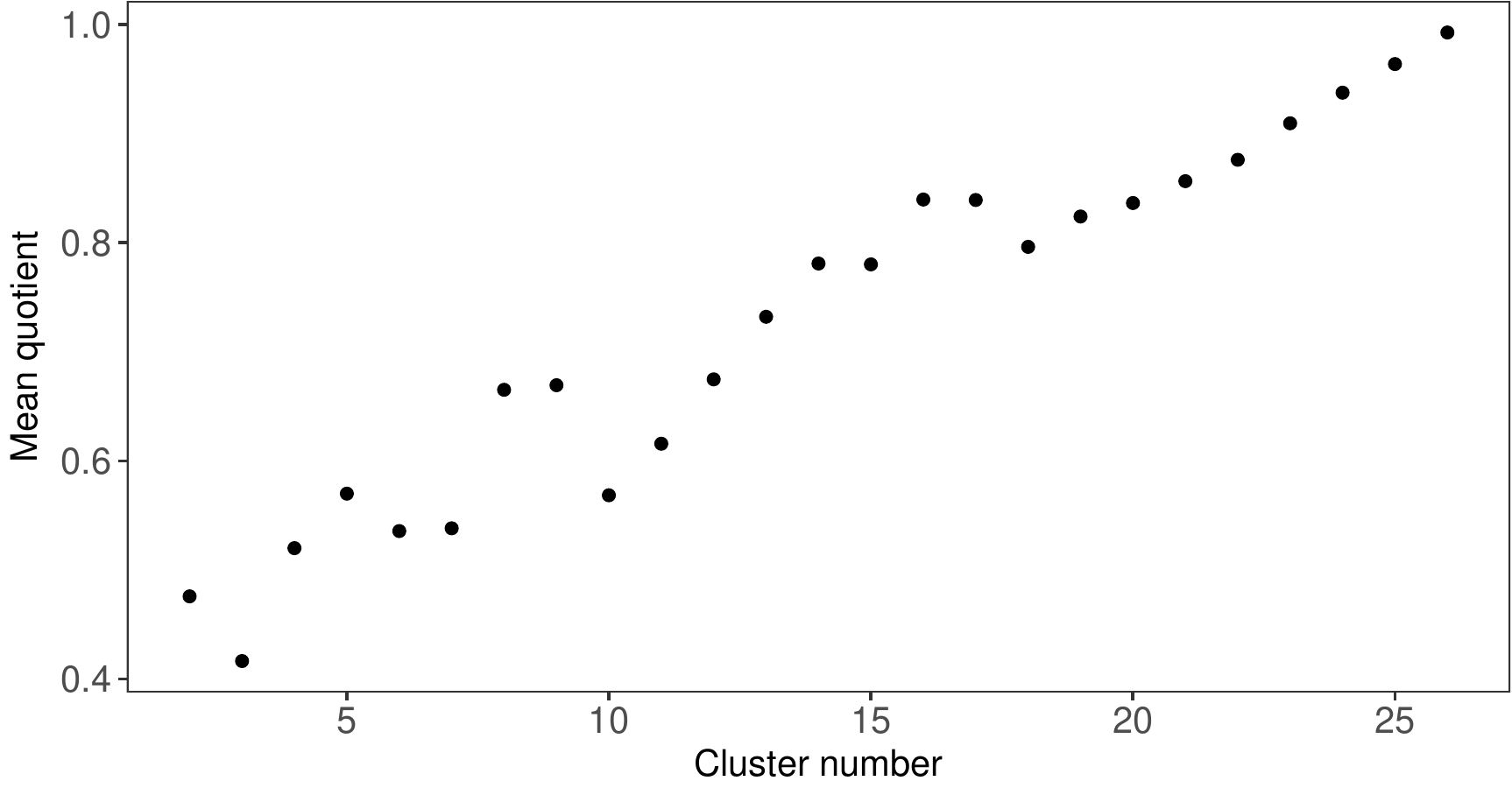}
	\caption{Determination of cluster number for the reduced-rank correlation matrix 
(correlation approach) in Eq.~(\ref{eqn:CorrMatToDatamatrixNAlternat_2}) (\href{https://www.quandl.com/}{Data 
from QuoteMedia via Quandl}).}
	\label{fig:RedRankCorrCorrAnsatzDetClusterNumber}
\end{figure}
\begin{table}[b]
	\centering
	\caption{\label{tab:MeasuredTurningPoints}
		Measured turning points (dashed lines in plots). (Cov) stays for the covariance 
approach in Eq.~(\ref{eqn:CorrMatToDatamatrixBAlternat_2}) and (Cor) for the correlation 
approach in Eq.~(\ref{eqn:CorrMatToDatamatrixNAlternat_2}) (\href{https://www.quandl.com/}{Data 
from QuoteMedia via Quandl}).
	}
		\begin{tabular}{c@{\hspace{3em}}l@{\hspace{2em}}l}
			\toprule
			\multicolumn{1}{c}{Number} &
			\multicolumn{1}{c}{(Cov)} &
			\multicolumn{1}{c}{(Cor)}
			\\
			\midrule
			1 & 2008-06-05/2008-08-05 & 2007-10-04/2007-12-04 \\
			2 & 2009-06-05/2009-08-05 & 2008-12-03/2009-02-04 \\
			3 & 2014-04-08/2014-06-09 & 2013-02-06/2013-04-09 \\
			4 & 2014-10-07/2014-12-05 & 2014-10-07/2014-12-05 \\
		\bottomrule
		\end{tabular}
\end{table}
\begin{figure}[b]
	\centering
	\includegraphics[width=1.0\columnwidth]{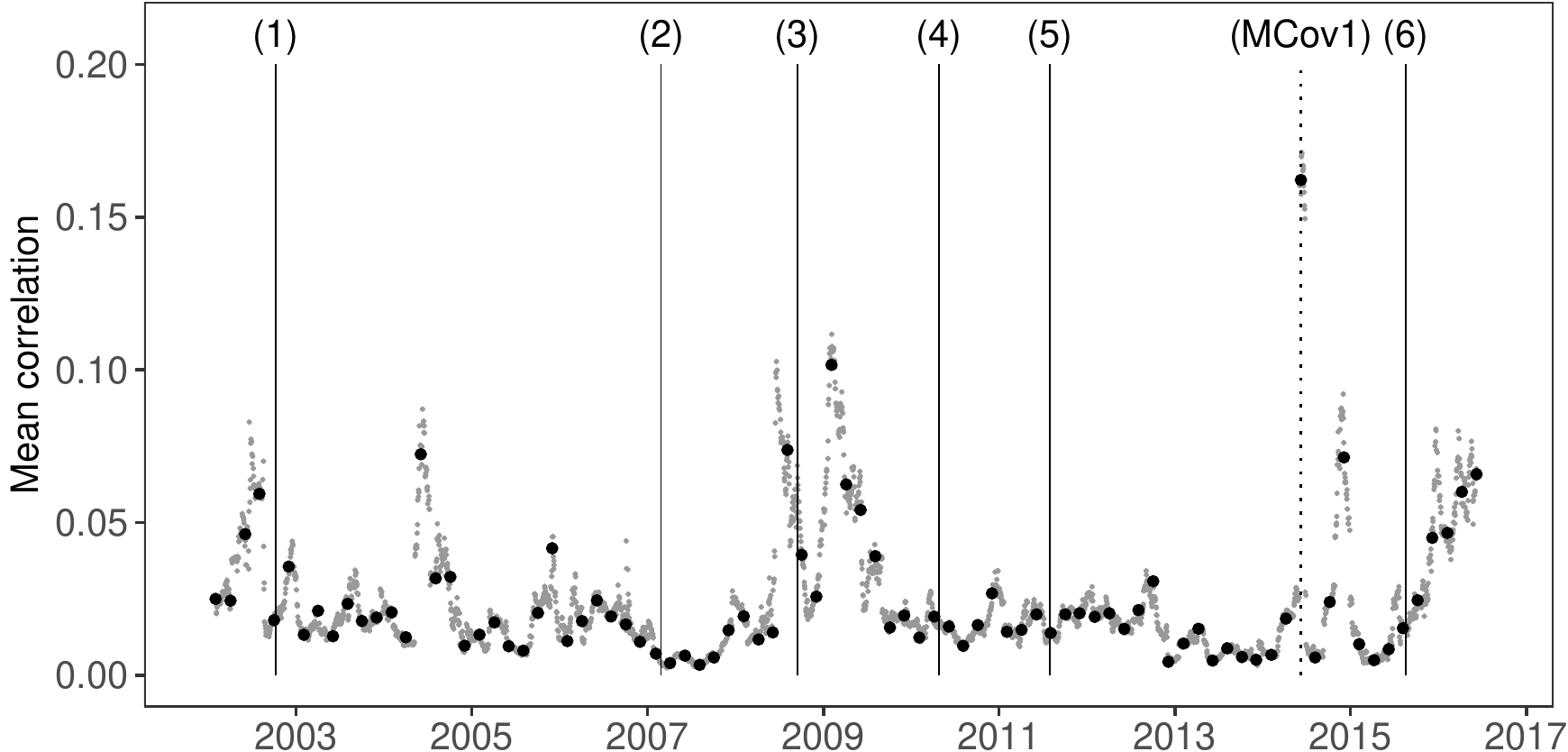}
	\caption{Mean correlation of the reduced-rank correlation matrix of the covariance 
approach (Eq.~(\ref{eqn:CorrMatToDatamatrixBAlternat_2})). The larger black dots 
belong to the middle of the epochs for which the corresponding correlation matrices 
are clustered. The smaller grayish dots belong to the middle of 42 trading day epochs 
calculated of overlapping intervals (1~trading day sliding windows) in order to show 
the relation to crises in Tab.~\ref{tab:FinancialCrises}. Measured event (MCov1) 
has the time stamp 2014-06-09 when the mean correlation is very high (\href{https://www.quandl.com/}{Data 
from QuoteMedia via Quandl}).}
	\label{fig:RedRankCorrMeanCovAnsatz}
\end{figure}
\begin{figure}[!htb]
	\centering
	\includegraphics[width=1.0\columnwidth]{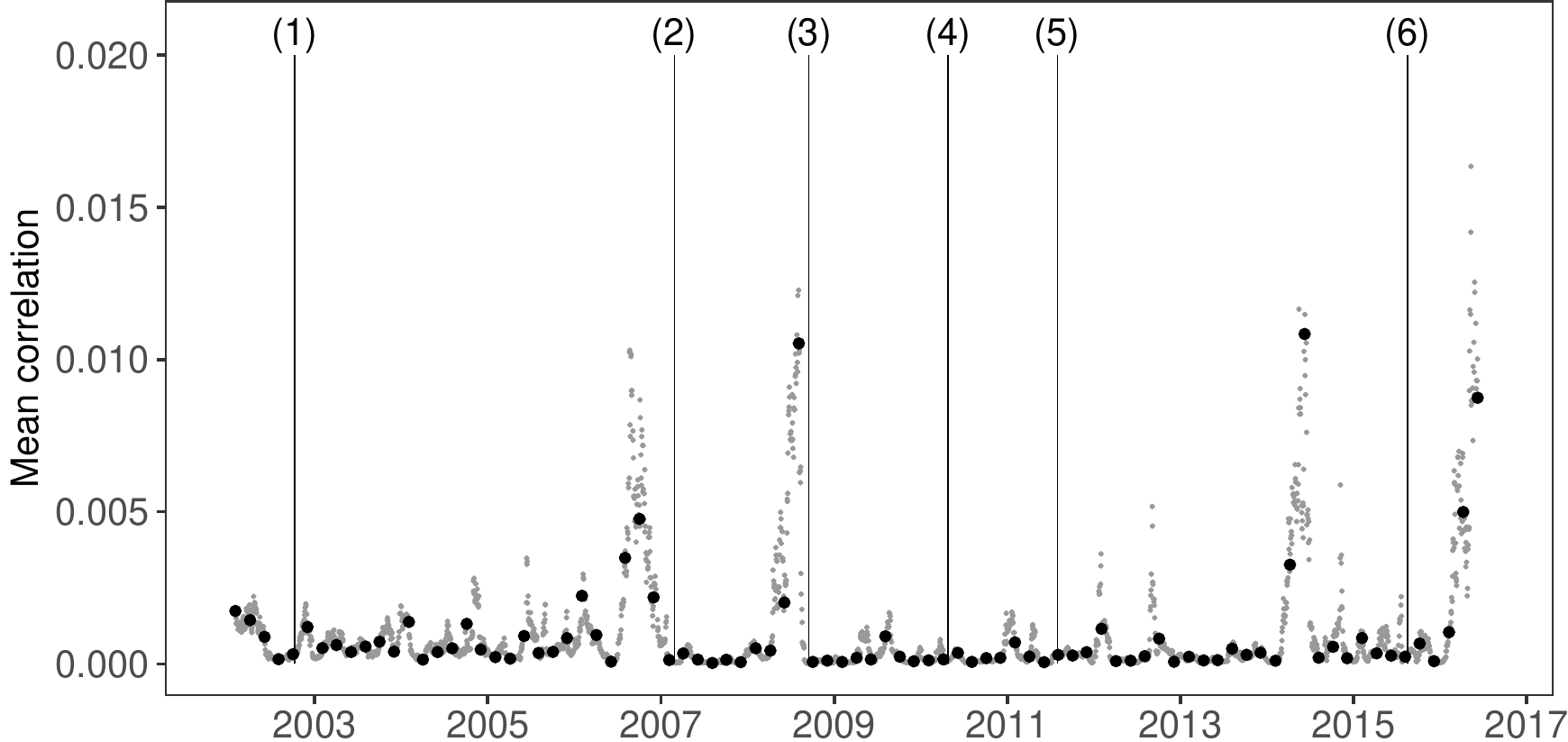}
	\caption{Mean correlation of the reduced-rank correlation matrix of the correlation 
approach (Eq.~(\ref{eqn:CorrMatToDatamatrixNAlternat_2})). The larger black dots 
belong to the middle of the epochs for which the corresponding correlation matrices 
are clustered. The smaller grayish dots belong to the middle of 42 trading day epochs 
calculated of overlapping intervals (1~trading day sliding windows) in order to show 
the relation to crises in Tab.~\ref{tab:FinancialCrises} (\href{https://www.quandl.com/}{Data 
from QuoteMedia via Quandl}).}
	\label{fig:RedRankCorrMeanCorrAnsatz}
\end{figure}
\begin{figure*}[!htb]
	\centering
	\includegraphics[width=1.0\textwidth]{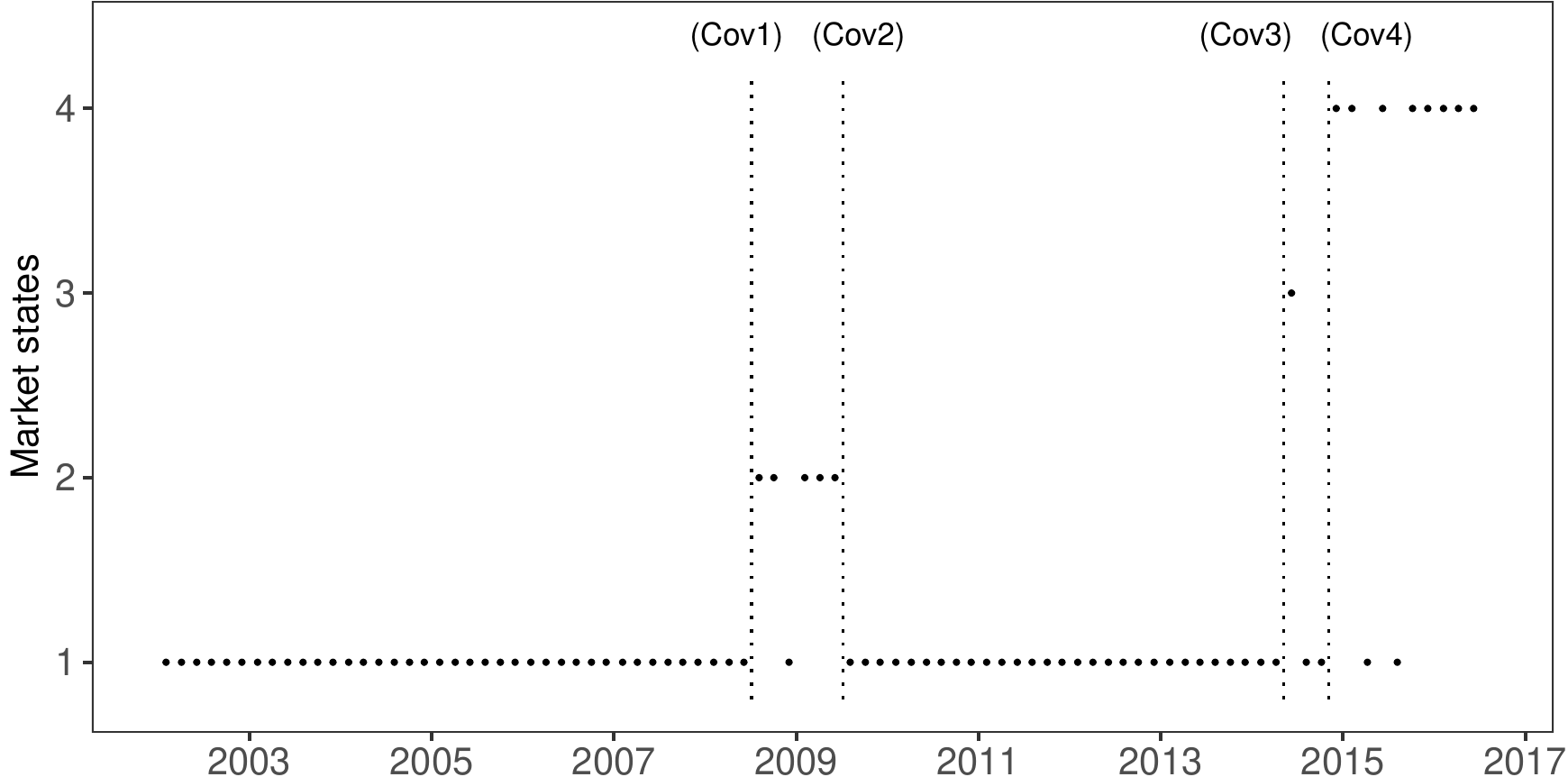}
	\caption{Temporal evolution of the reduced-rank correlation matrix of the covariance 
approach in Eq.~(\ref{eqn:CorrMatToDatamatrixBAlternat_2}). The dashed lines are 
marking the ``turning points" (see Tab.~\ref{tab:MeasuredTurningPoints}) of the 
market 
states when the market state changes fundamentally  (\href{https://www.quandl.com/}{Data 
from QuoteMedia via Quandl}).}
	\label{fig:TimeEvolutionRedRankCorrCovAnsatz}
\end{figure*}
\begin{figure*}[!htb]
	\vspace{-1cm}
	\centering
	\subfloat[\label{subfig:CorrCovAnsatzTypClustCent:a}state 1]{
		\includegraphics[width=0.33\textwidth]{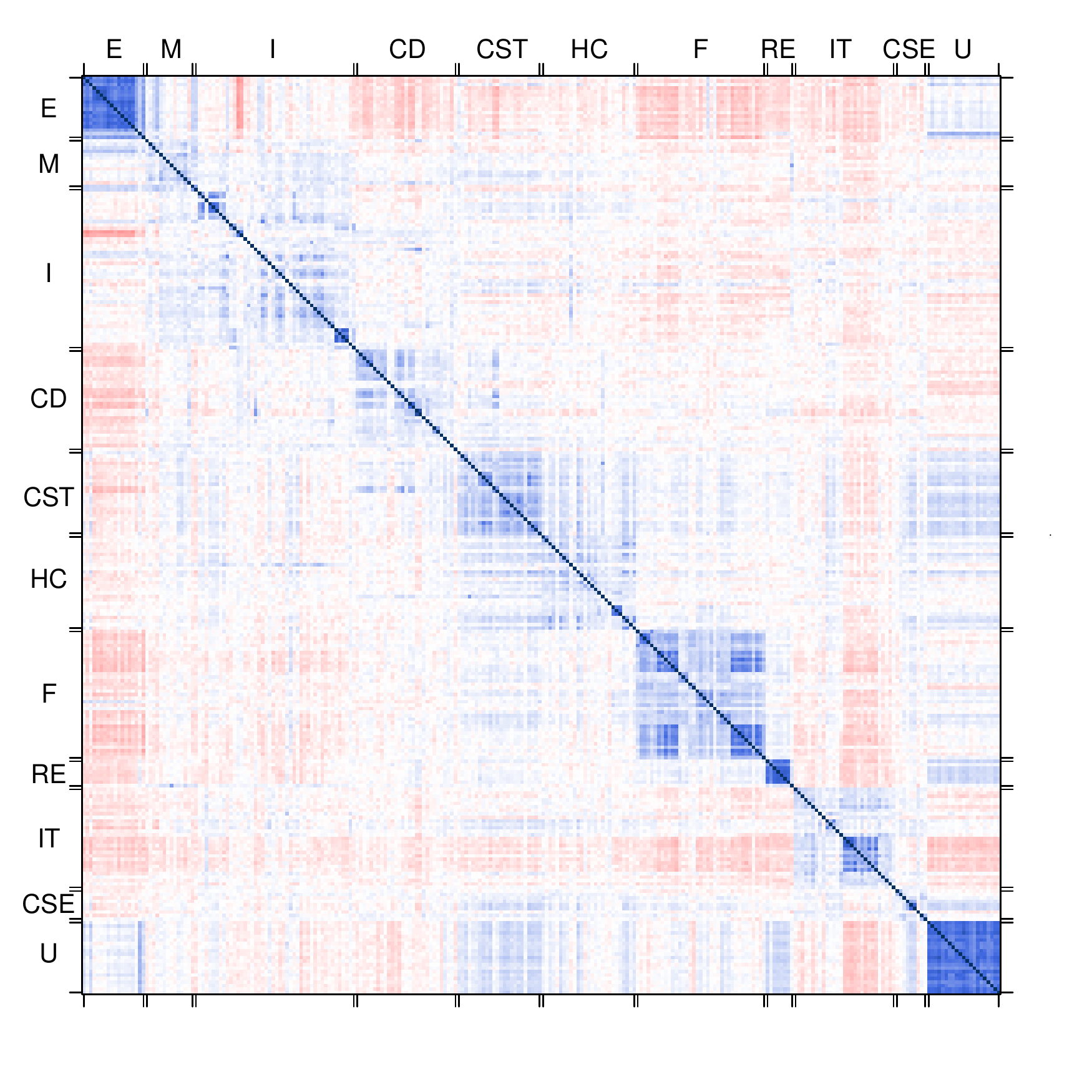}
	}
	\subfloat[\label{subfig:CorrCovAnsatzTypClustCent:b}state 2]{
		\includegraphics[width=0.33\textwidth]{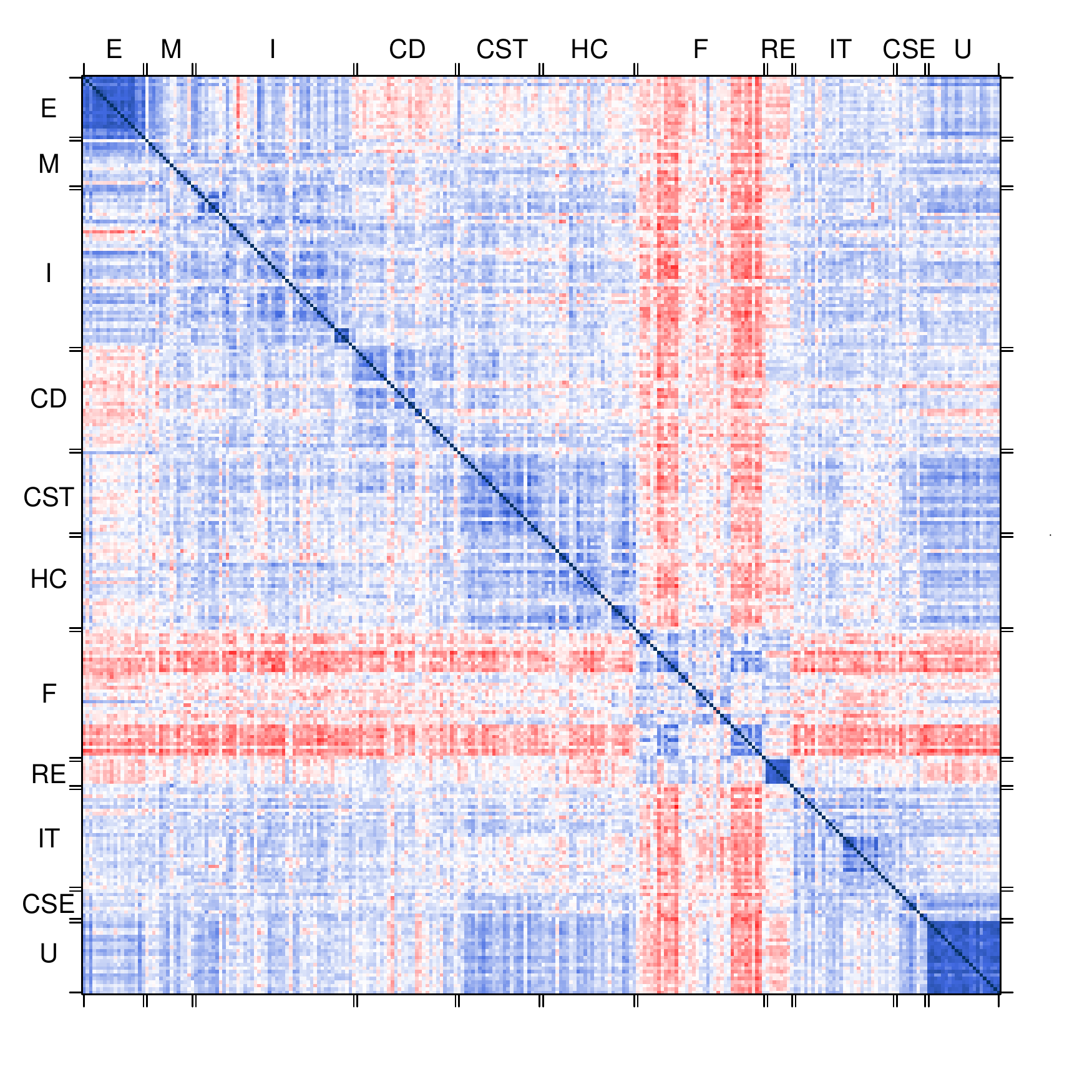}
	}\\
	\subfloat[\label{subfig:CorrCovAnsatzTypClustCent:c}state 3]{
		\includegraphics[width=0.33\textwidth]{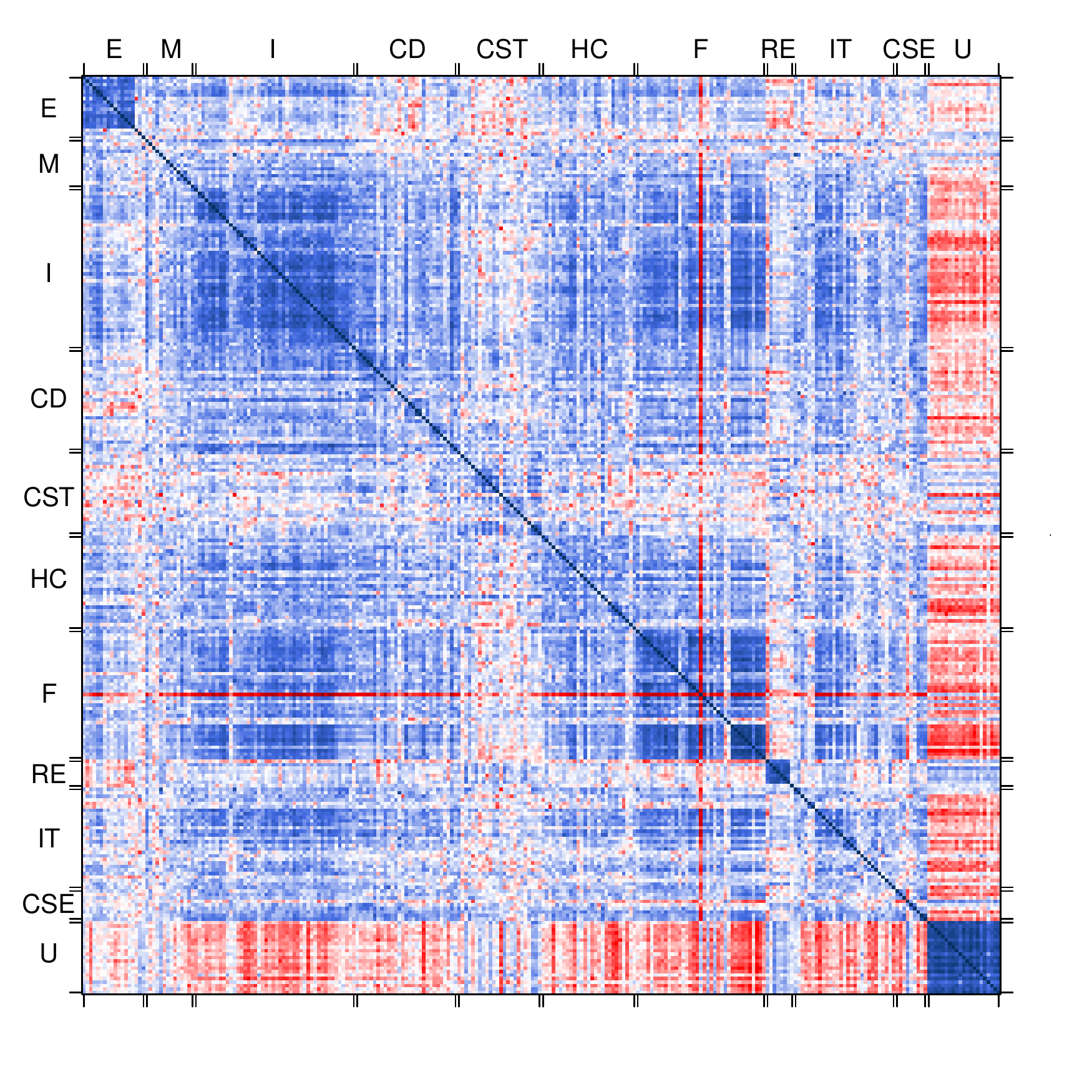}
	}
	\subfloat[\label{subfig:CorrCovAnsatzTypClustCent:d}state 4]{
		\includegraphics[width=0.33\textwidth]{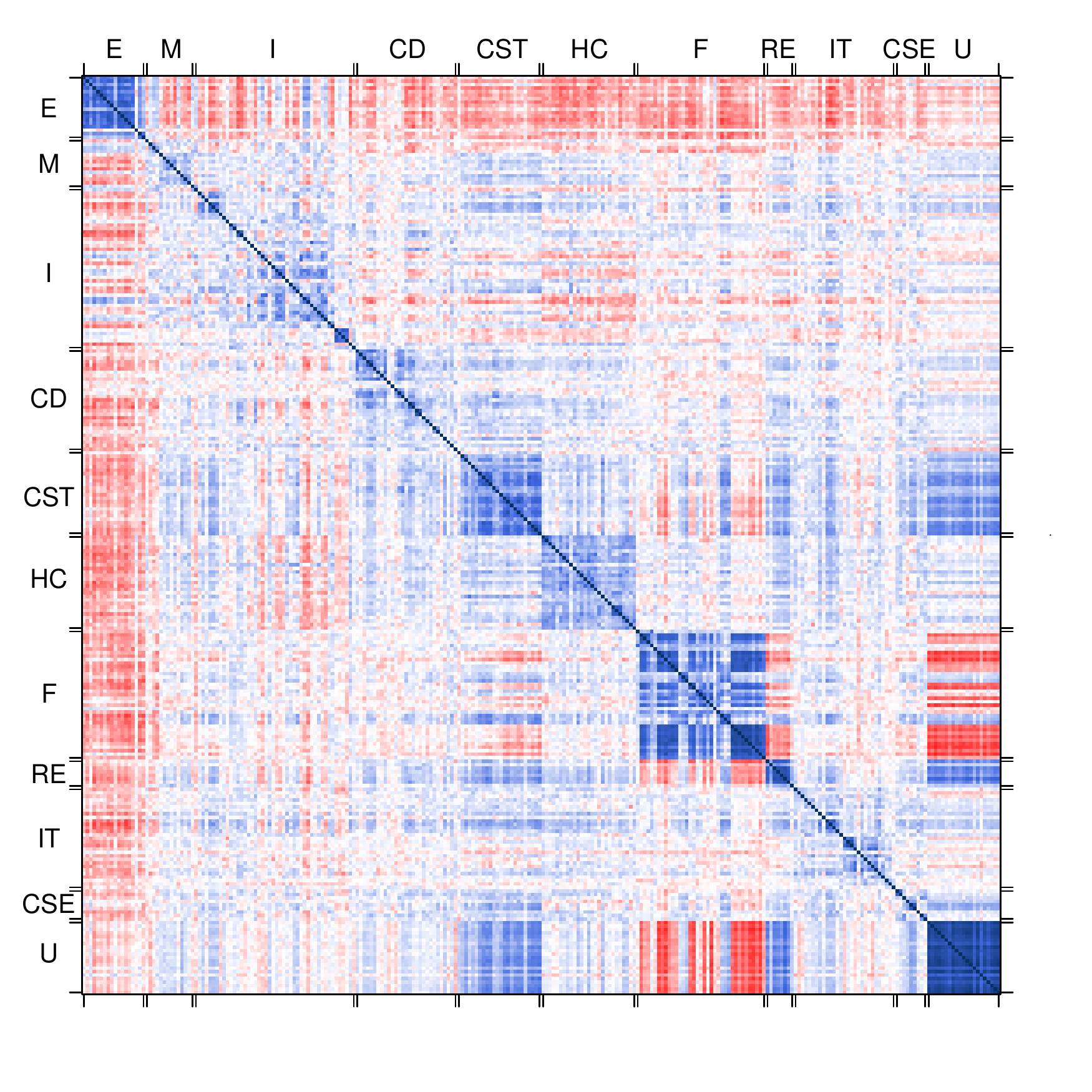}
	}\\
	\subfloat[\label{subfig:CorrCovAnsatzTypClustCent:e}overall average correlation 
matrix (averaged over all 87 correlation matrices)]{
		\includegraphics[width=0.58\textwidth]{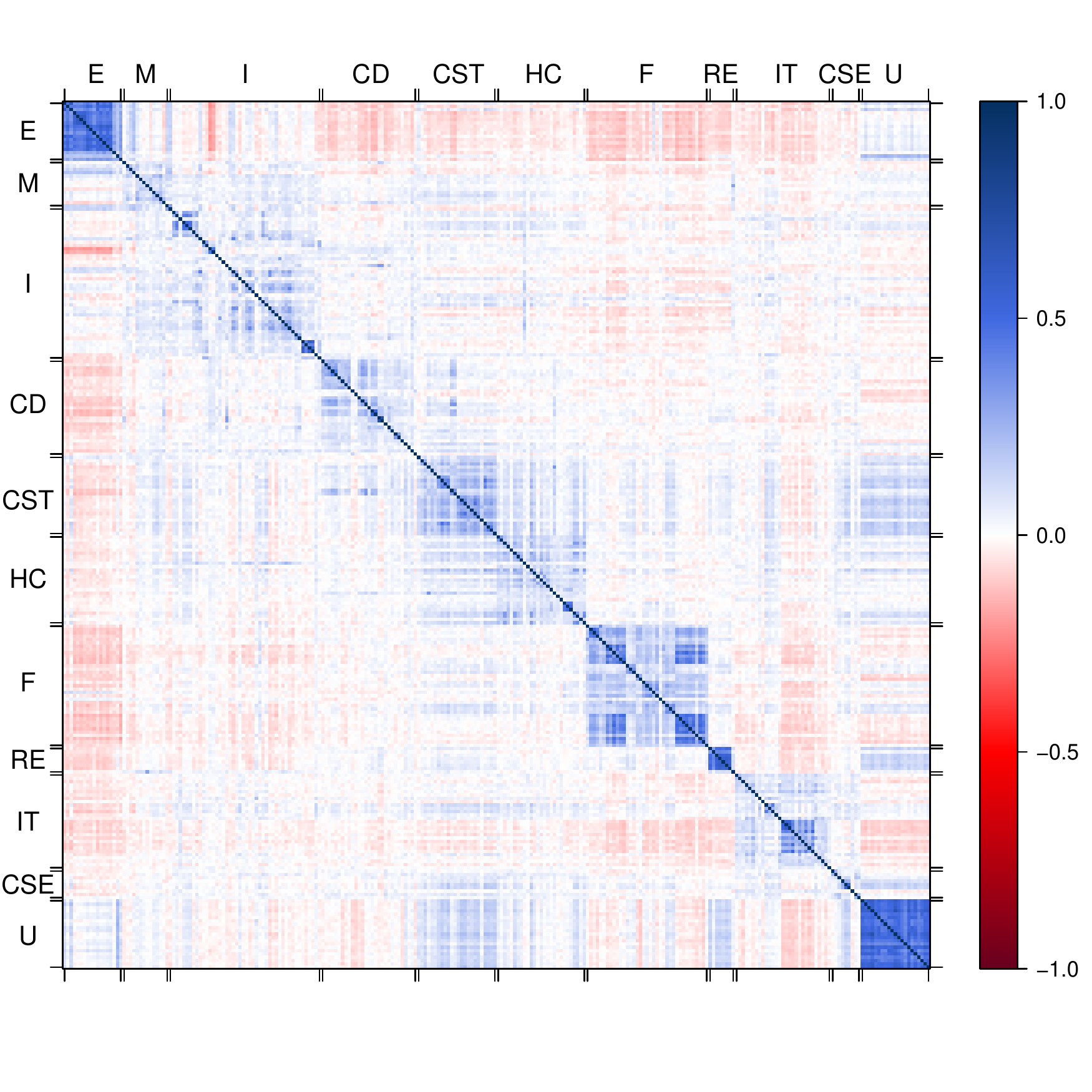}
	}
	\caption{\label{subfig:CorrCovAnsatzTypClustCent:main}Typical market states 
of the reduced-rank correlation matrix in Eq.~(\ref{eqn:CorrMatToDatamatrixBAlternat_2}) 
(covariance approach) calculated as element-wise average of the correlation matrices 
belonging to a market state (see Tab.~\ref{tab:NumberCorrMatMarketStates}). Sector 
legend: E: Energy; M: Materials; I: Industrials; CD: Consumer Discretionary; CST: 
Consumer Staples; HC: Health Care; F: Financials; RE: Real Estate; I: Information 
Technology; CSE: Communication Services; U: Utilities (\href{https://www.quandl.com/}{Data 
from QuoteMedia via Quandl}).}	
\end{figure*}
\begin{figure*}[!htb]
	\centering
	\includegraphics[width=1.0\textwidth]{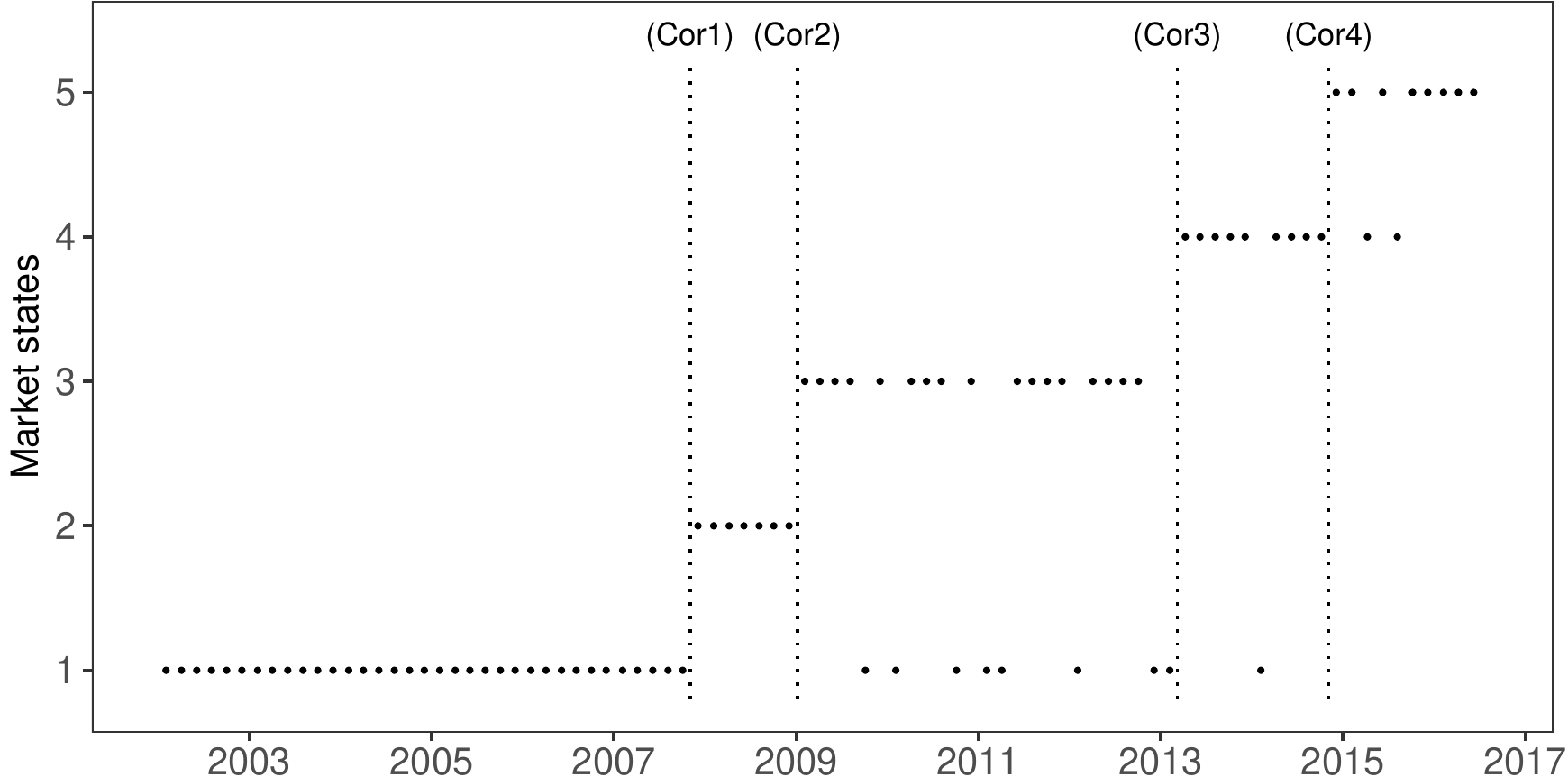}
	\caption{Temporal evolution of the reduced-rank correlation matrix of the correlation 
approach in Eq.~(\ref{eqn:CorrMatToDatamatrixNAlternat_2}). The dashed lines are 
marking the ``turning points" (see Tab.~\ref{tab:MeasuredTurningPoints}) of the 
market 
states when the market state changes fundamentally  (\href{https://www.quandl.com/}{Data 
from QuoteMedia via Quandl}).}
	\label{fig:TimeEvolutionRedRankCorrCorrAnsatz}
\end{figure*}
\begin{figure*}[!htb]
	\vspace{-1cm}
	\centering
	\subfloat[\label{subfig:CorrCorrAnsatzTypClustCent:a}state 1]{
		\includegraphics[width=0.33\textwidth]{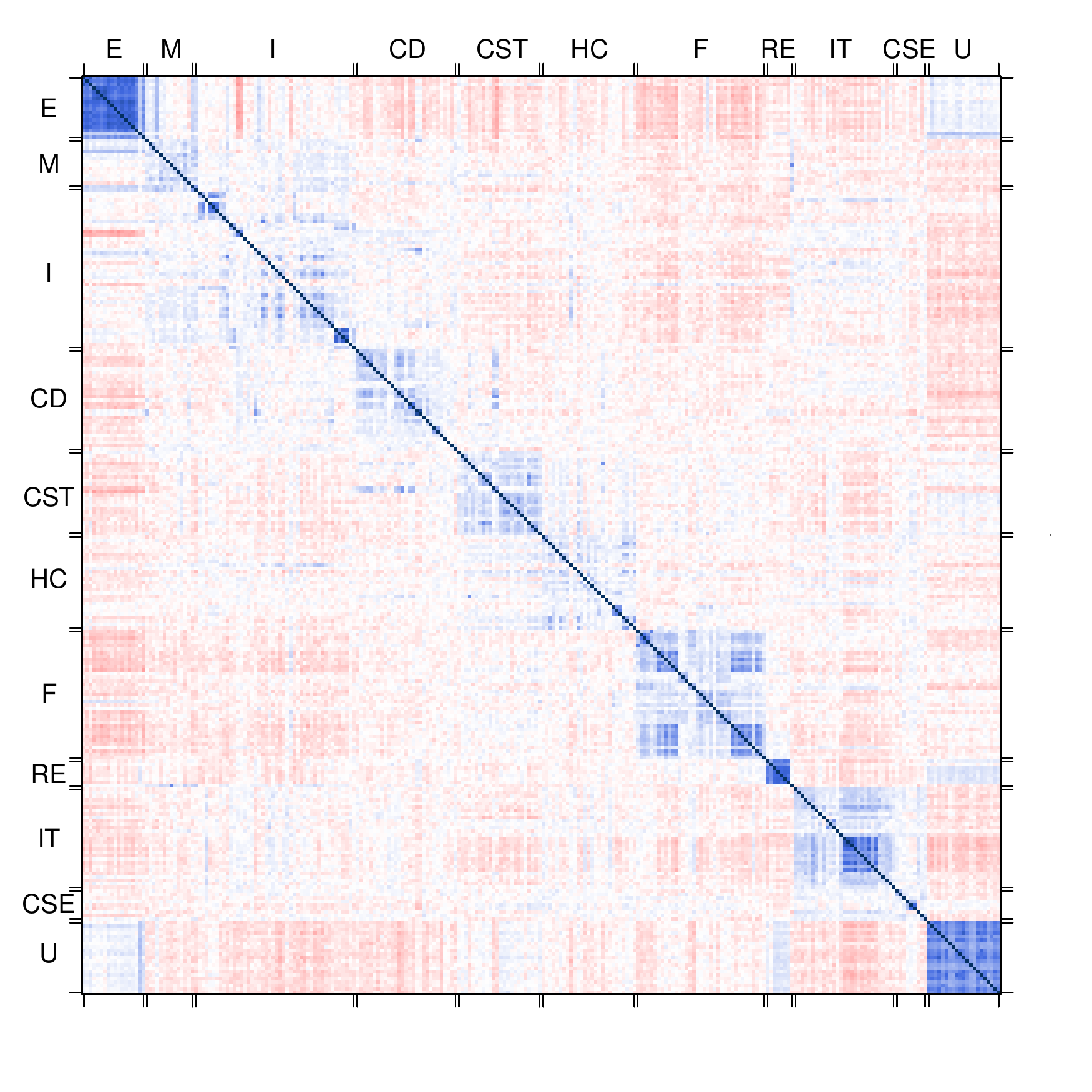}
	}
	\subfloat[\label{subfig:CorrCorrAnsatzTypClustCent:b}state 2]{
		\includegraphics[width=0.33\textwidth]{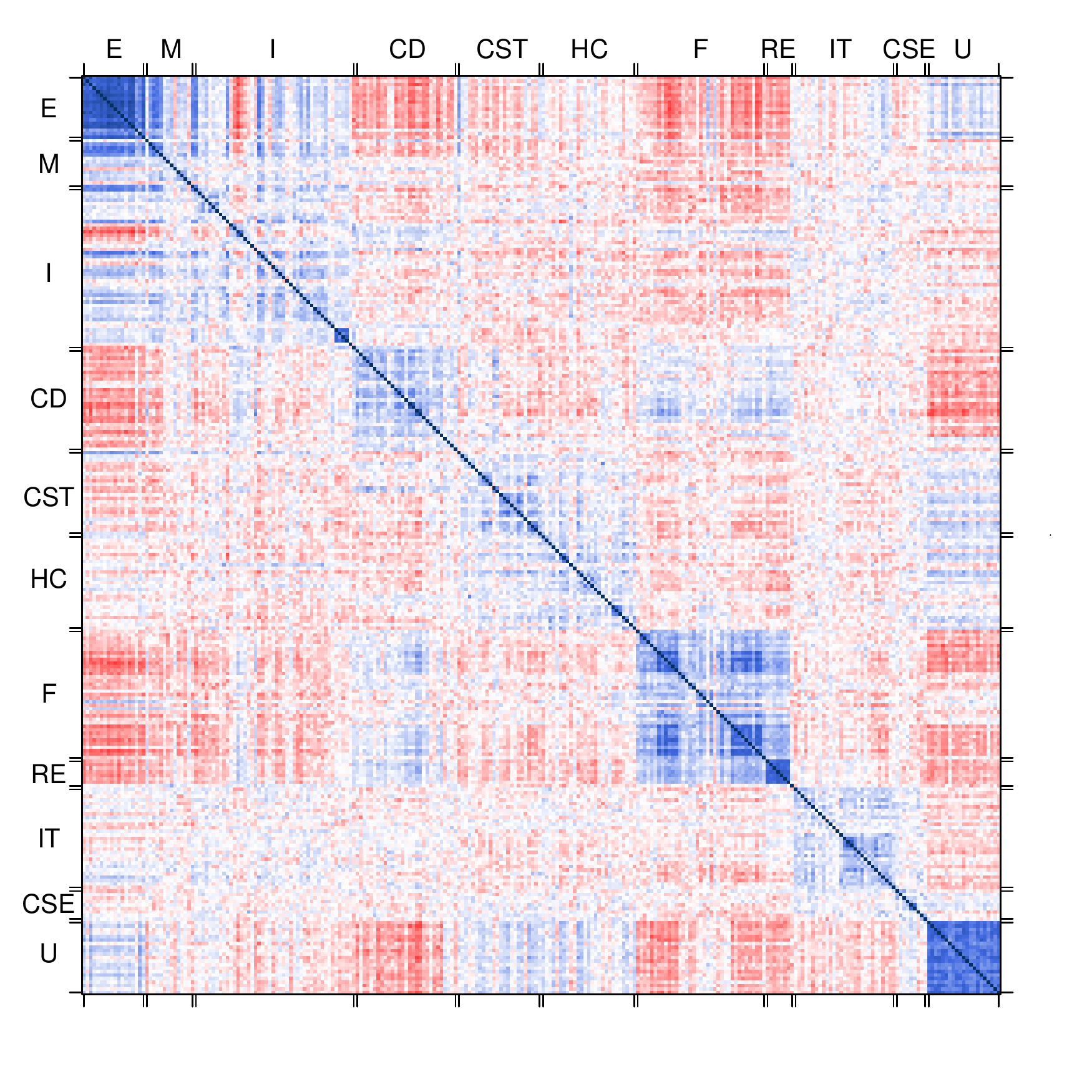}
	}
	\subfloat[\label{subfig:CorrCorrAnsatzTypClustCent:c}state 3]{
		\includegraphics[width=0.33\textwidth]{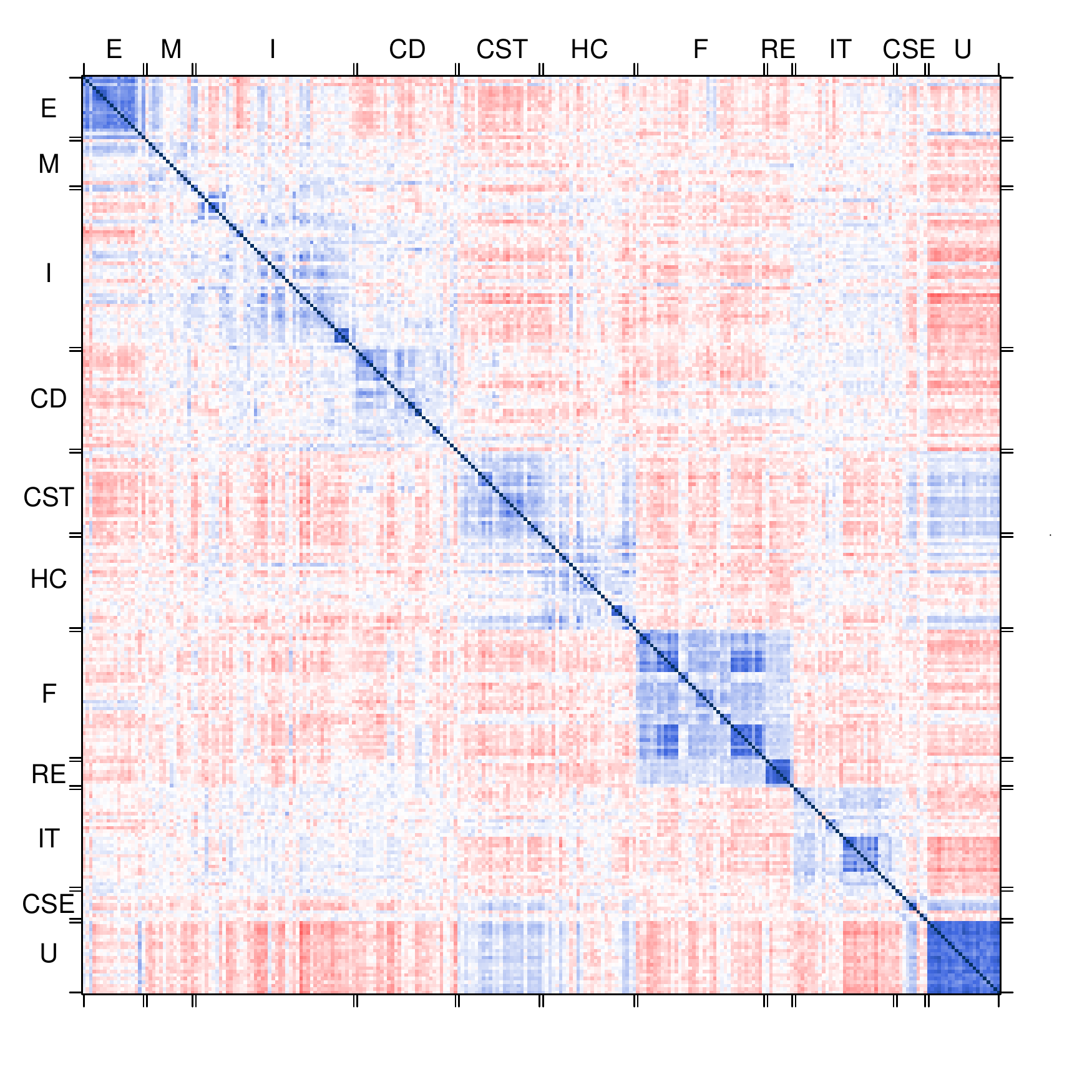}
	}\\
	\subfloat[\label{subfig:CorrCorrAnsatzTypClustCent:d}state 4]{
		\includegraphics[width=0.33\textwidth]{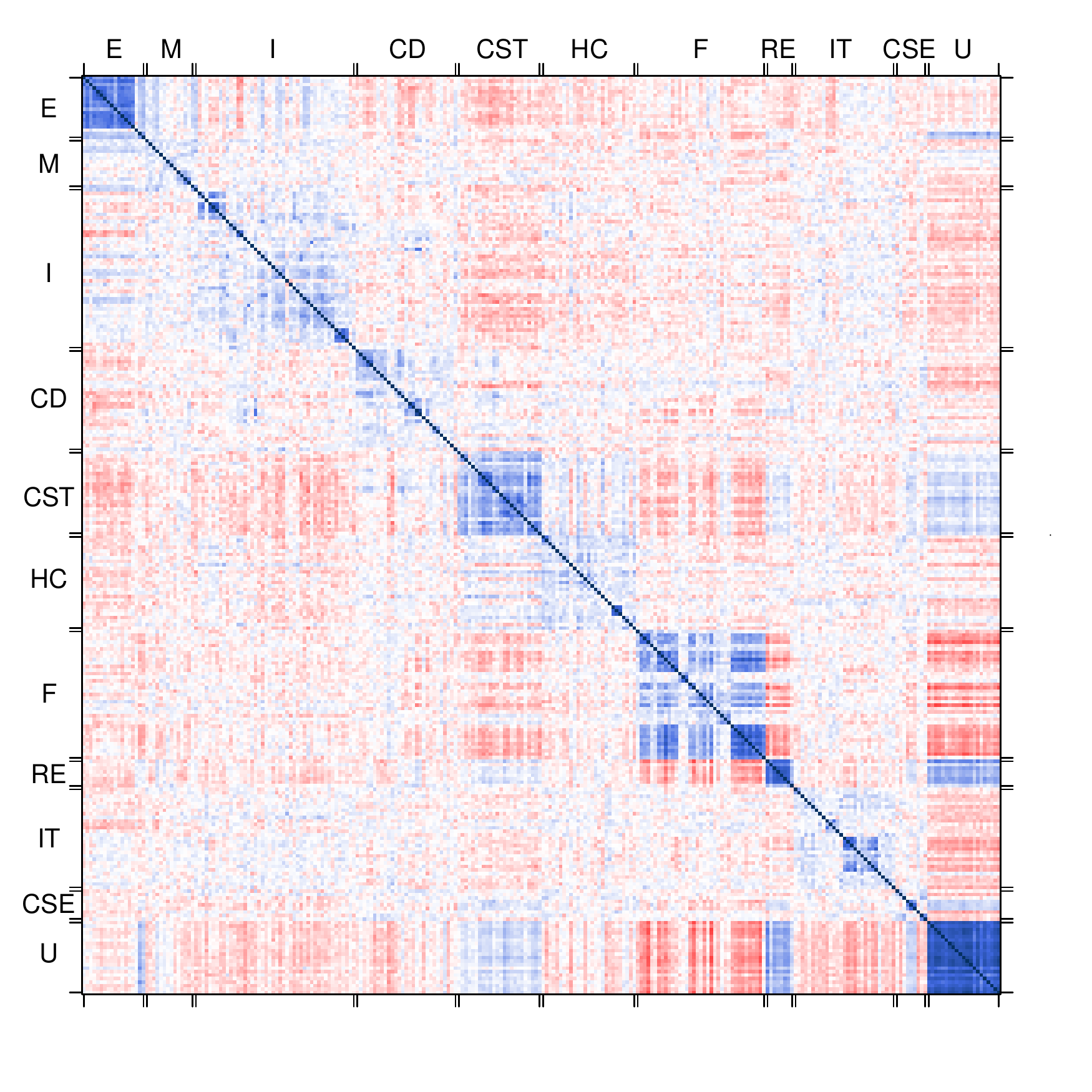}
	}
	\subfloat[\label{subfig:CorrCorrAnsatzTypClustCent:e}state 5]{
		\includegraphics[width=0.33\textwidth]{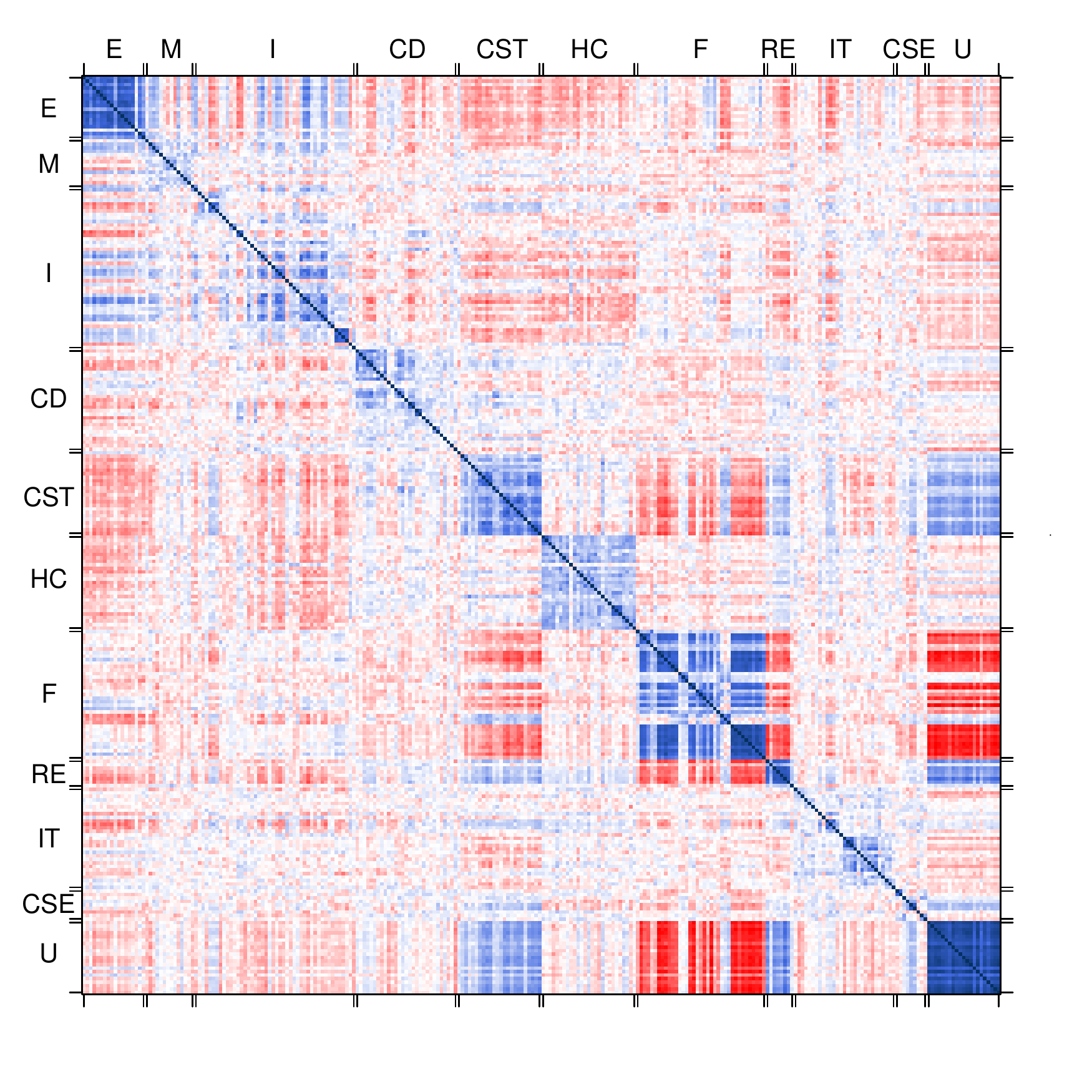}
	}\\
	\subfloat[\label{subfig:CorrCorrAnsatzTypClustCent:f}overall average correlation 
matrix (averaged over all 87 correlation matrices)]{
		\includegraphics[width=0.58\textwidth]{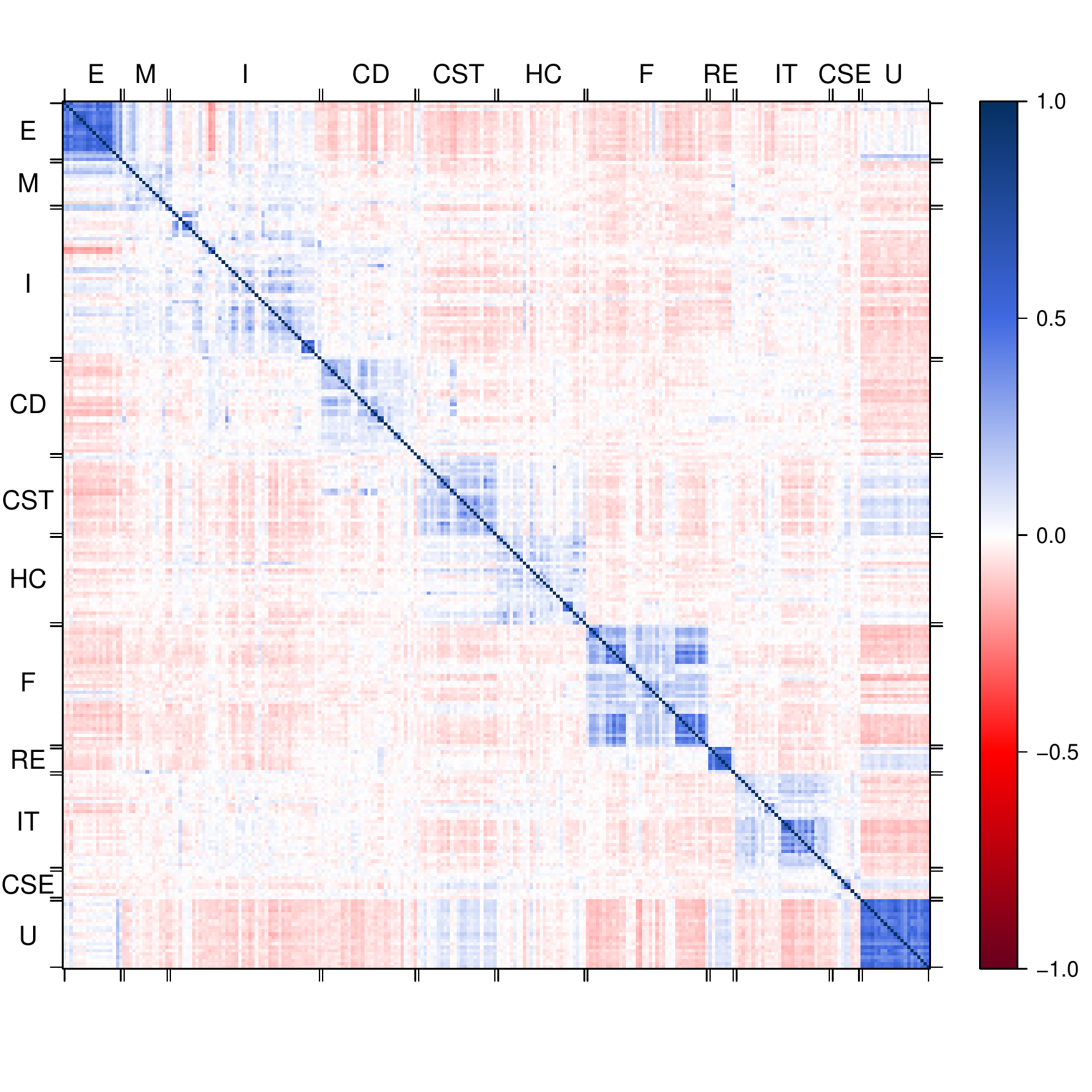}
	}
	\caption{\label{subfig:CorrCorrAnsatzTypClustCent:main}Typical market states 
of the reduced-rank correlation matrix in Eq.~(\ref{eqn:CorrMatToDatamatrixNAlternat_2}) 
(correlation approach) calculated as element-wise average of the correlation matrices 
belonging to a market state (see Tab.~\ref{tab:NumberCorrMatMarketStates}). Sector 
legend: E: Energy; M: Materials; I: Industrials; CD: Consumer Discretionary; CST: 
Consumer Staples; HC: Health Care; F: Financials; RE: Real Estate; I: Information 
Technology; CSE: Communication Services; U: Utilities (\href{https://www.quandl.com/}{Data 
from QuoteMedia via Quandl}).}
\end{figure*}

The number of market states in
Fig.~\ref{fig:RedRankCorrCovAnsatzDetClusterNumber} is chosen as
$k^{\text{(opt)}}=4$ for the covariance approach
Eq.~(\ref{eqn:CorrMatToDatamatrixBAlternat_2}) and as $k^{\ast}=5$
in Fig.~\ref{fig:RedRankCorrCorrAnsatzDetClusterNumber} for the
correlation approach Eq.~(\ref{eqn:CorrMatToDatamatrixNAlternat_2}).
As in Fig.~\ref{fig:StandCorrMeanCorr}, we plot in
Figs.~\ref{fig:RedRankCorrMeanCovAnsatz}
and~\ref{fig:RedRankCorrMeanCorrAnsatz} the mean correlation for
both approaches for later comparison with the temporal evolution of
the market states. We order the market states according to their
first temporal emergence as in
Fig.~\ref{fig:TimeEvolutionStandCorr}.  One sees in
Figs.~\ref{fig:TimeEvolutionRedRankCorrCovAnsatz}
and~\ref{fig:TimeEvolutionRedRankCorrCorrAnsatz} a substantially more
stable temporal evolution of the market states.  In
Figs.~\ref{subfig:CorrCovAnsatzTypClustCent:main}
and~\ref{subfig:CorrCorrAnsatzTypClustCent:main} the typical market
states are characterized by a much higher proportion of negative
correlations than in the case of standard correlation matrices.  The
overall correlation structure is more pronounced than with the latter,
\textit{i.e.}, the block structure of the industry sectors along the
diagonal.

Market state 3 for the covariance matrix ansatz consists of only one correlation 
matrix (see Fig.~\ref{fig:TimeEvolutionRedRankCorrCovAnsatz} and Tab.~\ref{tab:NumberCorrMatMarketStates}).
Fig.~\ref{fig:RedRankCorrMeanCovAnsatz}, Tab.~\ref{tab:MeanCorrTypMarkStates} and 
Tab.~\ref{tab:MeasuredTurningPoints} reveal that this is due to a strong mean correlation 
that dominates in Fig~\ref{fig:TimeEvolutionRedRankCorrCovAnsatz} at this time.
The mean correlations calculated using the covariance ansatz are much stronger (cf. 
Fig.~\ref{fig:RedRankCorrMeanCovAnsatz} and Fig.~\ref{fig:RedRankCorrMeanCorrAnsatz} 
and also Tab.~\ref{tab:MeanCorrTypMarkStates}).
We could not relate the peaks in Fig.~\ref{fig:RedRankCorrMeanCovAnsatz} and Fig.~\ref{fig:RedRankCorrMeanCorrAnsatz} 
to the historical events of Tab.~\ref{tab:FinancialCrises} like this is possible 
for the mean correlation of the standard correlation matrix in Fig.~\ref{fig:StandCorrMeanCorr}.
This means that the largest dyadic matrix describes crises.
Without the dominating dyadic matrix, this information is
lost in the reduced-rank correlation matrices.

The typical market state of state 1 is very similar in both reduced-rank correlation 
matrices.
One difference is that the correlation matrix elements from sectors CST to U 
are generally 
more positive for the covariance ansatz.
Tab.~\ref{tab:MeasuredTurningPoints} shows the "turning points" of the market states 
when for the first time the respective state changes significantly.
Until October 2007,
the ``market" stays in market state 1 and the typical market state is almost identical 
for both approaches.
The first temporal occurrence of the second market state is earlier for the correlation 
approach.
Within this state, the Lehman Brothers crisis took place (see Tab.~\ref{tab:FinancialCrises}).
So there is also some kind of ``crisis state" as in the case of the standard correlation 
matrices that existed before the crises.
In~\cite{musmeci2015risk} such a phenomenon is reported as pre-crisis 
structure by measuring a different quantity.
Here, however, we have automatically highlighted the phenomenon through the cluster 
procedure!
Market state 2 shows in the correlation matrix ansatz more negative inter-sector 
correlations.
For the covariance matrix approach, it is noticeable that two subbranches (Asset 
Management \& Custody Banks, Diversified Banks and Regional Banks (see Appx.~\ref{sec:ListStocks})) 
of the Financials sector show negative stripes to the other sectors.
This is interesting because the Financials sector was the starting point of the crisis.
Market state 2 even exists after the main phase of Lehman Brother crisis.

While in the covariance approach the market then returns to market state 1 (for $k=4$), 
in the correlation approach there are
two new market state.
Market state 3 in Fig.~\ref{subfig:CorrCorrAnsatzTypClustCent:c} is very 
similar to market state 1, as there are jumps between market state 1 and market state 
3.
Fig.~\ref{subfig:CorrCorrAnsatzTypClustCent:c} illustrates a typical market 
state that shows
larger positive and negative correlations between the sectors than market state 1.
Market state 4 in Fig~\ref{subfig:CorrCorrAnsatzTypClustCent:d} has
noticeable negative correlations between F and RE and F and U in comparison to market 
state 3.

The first appearance in time of market state 4 in the covariance approach and of 
market state 5 in the correlation approach is identical (see Tab.~\ref{tab:MeasuredTurningPoints}).
For the correlation ansatz, market state 5
and market state 4 share a similar correlation structure, but
the structure of market state 5 is even more pronounced
than the structure of market state 4.
For the covariance ansatz, market state 4 shows stronger positive correlations than 
market state 5 of the correlation ansatz.

The two correlation matrix approaches
show a further property in addition to their very quasi-stationary behavior.
The mean correlation according to Fig.~\ref{fig:StandCorrMeanCorr} can also become 
weaker for more recent times for the standard correlation matrices.
Since the mean correlation is mainly clustered, old market states may reappear in 
the correlation matrices with the largest dyadic matrix.
The market states of the
correlation approach die out after some time and no longer emerge, or the probability 
of emergence is very low.
This is comprehensible, because the economic relationships between the companies 
in a portfolio change over time as a result of changes in market regulations or technological 
developments, for example.

Hitherto, we included a set of 262 stocks from S\&P~500.
Obviously, the identification of market states
and the analysis of their time evolution must depend on the stocks
chosen. To quantify this effect, we carry out an additional analysis
of the market states depending on the choice and number of stocks in
Appx.~\ref{sec:MarketStatesDependNumberStocks}.  We compare the
cluster solutions $Z^{(\text{opt})}$ or $Z^{\ast}$
(cf. Sec.~\ref{sec:ResultsClusteringStandCorrMat}) in
Figs.~\ref{fig:TimeEvolutionStandCorr},
\ref{fig:TimeEvolutionRedRankCorrCovAnsatz}
and~\ref{fig:TimeEvolutionRedRankCorrCorrAnsatz} with the cluster
solutions $Z^{(K)}$ of randomly chosen companies from the set of all
$K = 262$ companies.  In the case of the standard correlation
matrices, the market states calculated from smaller samples ($K=50$
companies) are largely influenced by the collective behavior of all
stocks.  Therefore, the deviations from the cluster solutions
$Z^{\ast}$ is relatively small.  For both reduced-rank approaches,
the deviations of the cluster solutions $Z^{(K)}$ from
$Z^{(\text{opt})}$ or $Z^{\ast}$ are larger than in the case of the
standard correlation matrices. For each choice of $K$ companies, the
sector structure of the sample might change strongly.
Additionally, the time evolution of the 11 industry sectors is not
necessarily the same.

\section{\label{sec:Conclusion}Conclusion}

The dynamics of many complex systems is often, or at least for certain
periods of time, dominated by collective behavior. In the present
work on the dynamics of correlation structures in financial
markets, we capture this collective behavior using the dyadic matrix belonging
to the largest eigenvalue of the correlation  matrix.
Our goal was to cluster reduced-rank correlation matrices in which the influence 
of the largest eigenvalue or the corresponding dyadic matrix was removed.
In this way, we wanted to analyze the dynamics of the other non-dominant dyadic matrices 
assigned to the largest sectors,
more closely. 

We have shown by singular value decomposition that
we can construct mean-normalized data matrices that
lead in a simple manner to the reduced-rank correlation matrices.
There are no correction terms.
It is also possible to build data matrices to calculate reduced-rank correlation 
matrices which correspond to any combination of dyadic matrices.

The central result
is that we identify long-lasting quasi-stationary periods by clustering reduced-rank 
correlation matrices.
The single reduced-rank correlation matrices are characterized by a significantly differentiated 
correlation structure.
In addition, both reduced-rank correlation approaches show more negative correlations
than the standard correlation matrices, 
which lowers the mean correlation.
The market states develop
clearly visible for the correlation approach to new market states.
Overall, the correlation structure of the industry sectors show a high degree of quasi-stationarity 
over time.

In contrast to the long-lasting quasi-stationary periods of the reduced-rank approaches, 
the clustering of standard correlation matrices reveals a faster dynamics.
The mean correlation, which dominates the standard correlation matrices, causes
more jumps between market states.
Since the mean correlation of the standard correlation matrix is clustered in a first approximation, it is possible 
that old market states will reappear when the mean correlation decreases.
That is why we call the jumps of reduced-rank correlation matrices, which significantly 
change the market state, ``turning points".

One remarkable difference between the covariance approach and the correlation approach 
for the reduced-rank correlation matrices is that the covariance approach has much 
higher mean correlations.
If the mean correlation is understood as the systematic risk of a portfolio, then
standard correlation matrix has a high systematic risk, the reduced-rank correlation 
matrices show a much lower one.
The correlation matrix of the correlation approach shows by far the lowest systematic 
risk.
This reduced-rank correlation matrix only contains the diversification part.
The large mean correlation of the covariance approach
causes that one of the market states has only one correlation matrix, since the mean 
correlation is very high at this point in time and the cluster
algorithm perceives this point as an outlier.

As we could see from the comparison with historical events, the crisis
behavior is described by the mean correlations of the standard
correlation matrices.  The collective motion of the ``market" marks
crisis events.  Nevertheless, for both approaches we were able to
automatically identify a market state around the Lehman Brother crisis
by means of clustering, which differs from other typical market
states, in particular for the covariance matrix approach.  It is
interesting to note that even before the crisis, parts of the
Financials sector behaved differently in their correlation to other
sectors.

We have not investigated here what effect our approach of subtracting
the dyadic matrix to the largest eigenvalue has on the cluster result
for systems as in~\cite{meng2014systemic} where the spatiotemporal
dynamics of US housing markets was analyzed by looking at the
correlation dynamics using eigenvalues and eigenvectors.  A collective
behavior can then also be found in the other dyadic matrices
corresponding to the sectors.  This will cause some of the typical
market states to lose structure and these could look like typical
market states of the standard correlation matrix of our analyzed
system. That this effect can be observed in stock markets shows the
outlier in the case of the reduced-rank correlation matrix. This
effect however deserves further detailed discussion and we relegate it
to future publications.

From a conceptually viewpoint it is worth emphasizing that our
analysis in the ``moving frame'' defined by the collective motion of
the market as a whole also provides a new tool to help separating,
\textit{cum grano salis} and qualitatively, exogenous from endogenous
effects. Of course, financial crashes illustrate that exogenous
effects can prompt endogenous ones and the latter in turn trigger
reaction from politics, \textit{i.e.}~exogenous effects, and so
on. But notwithstanding the intertwining of these effects, the
collective market motion is commonly viewed as reflecting exogenous
effects stronger than other observables in the correlations.  Relative
to the market motion we thus obtain correlation structures in which
the endogenous effects are much better visible than before the
subtraction of the market impact. It is worth mentioning
that the collective market motion itself has a matrix structure.
One is tempted to speculate that the sector structure induced by the
leading eigenvector may indicate endogenous contributions to the
collective market motion. One might argue that endogenous effects
can act on some industry sectors stronger than on others. Indeed, but
this can be taken care of by iterating our approach further and
subtracting dyadic matrices corresponding to specific sectors. As this
would be a whole new project, we also leave it to future study.

\begin{acknowledgments}
	We thank Thomas H. Seligman for fruitful discussions and an
    unknown referee for useful suggestions.
\end{acknowledgments}

\clearpage
\bibliography{Lit.bib}

\clearpage
\onecolumngrid
\appendix

\section{\label{sec:FacilClust}Facilitations for clustering}

To speed up the clustering, we can perform a 
principal component analysis (PCA) 
before clustering~\cite{Pearson1901:LinesPlanesCLosestFit, 
Hotelling1933:AnalysisComplexStatisticalVariablesPrincipalComponents, 
	jolliffePrincipalComponentAnalysis2002}.
The PCA is not necessary to perform the cluster analysis, but it is an 
extremely 
time-saving operation since the problem scales
with $K^2$ without PCA.
To this end, the correlation matrix elements are arranged in rows of a $K^2 
\times 
N_{\text{ep}} $ matrix
\begin{equation} \label{eqn:DatamatrixCorrelationsPCA}
F_C = \begin{bmatrix}  1 & \dots & 1 \\
C_{12}(1) & \dots & C_{12}(N_{\text{ep}})  \\
\vdots & & \vdots \\
C_{i i-1}(1) & \dots & C_{i i-1}(N_{\text{ep}}) \\
1 & \dots & 1 \\
\vdots & & \vdots \\
C_{K K-1}(1) & \dots & C_{K K-1}(N_{\text{ep}}) \\
1 & \dots & 1
\end{bmatrix} \;.
\end{equation}
The rows of $F_C$ are then normalized to mean zero.
The matrix with the mean-normalized rows is referred to as $\widetilde{F}_C$.
The covariance matrix for the PCA is thus
\begin{equation} \label{eqn:CovarianceMatrixCorrelationsPCA}
\Sigma_C = \frac{1}{N_{\text{ep}}} \, \widetilde{F}_C \, 
\widetilde{F}_C^{\dagger} \;.
\end{equation} 
Since $\Sigma_C$ has $N_{\text{ep}}-1$ non-zero eigenvalues, we have 
drastically 
reduced the cluster problem using the PCA for $N_{\text{ep}} \ll K^2$.
The additional disappearing eigenvalue is caused by the normalization to zero 
mean 
value (see Sec.~\ref{sec:DefinitionRedRankCorrMat}).
The projections of the columns of Eq.~(\ref{eqn:DatamatrixCorrelationsPCA}) on 
the 
eigenvectors of $\Sigma_C$ then result in $N_{\text{ep}}$ vectors of length 
$N_{\text{ep}}-1$ 
that are clustered.

\section{\label{sec:StandKMeans}Standard \texorpdfstring{\textit{k}}{k}-means}

$k$-means assigns each correlation matrix exclusively to one cluster.
The input for $k$-means are the correlation matrices which are written in 
the transposed shape of
Eq.~(\ref{eqn:DatamatrixCorrelationsPCA}) in R \cite{RProject}. The alternative input for 
faster 
clustering are the projections on the eigenvectors of $\Sigma_C$  (cf. 
Appx.~\ref{sec:FacilClust}).
The clustering algorithm divides the set of correlation matrices $Z=\{ C(1), 
C(2), \ldots, C(87) \}$ into subsets,
\textit{i.e.} $Z = \{ z_1, z_2, \ldots, z_l,\ldots, z_k \}$. Every subset $z_l$ is 
a 
cluster. In total, we have $k$ clusters. The number $k$ has to be determined by 
other methods than standard $k$-means~\cite{jainDataClustering502010, 
	ronanAvoidingCommonPitfalls2016, KaufmanRousseeuw1990:FindingGroupinData}.
The $k$-means algorithm reads as follows
\begin{enumerate}
	\item Select $k$ correlation matrices as start centroids (see 
	Eq.~(\ref{eqn:Cendroid})). 
	\\
	Repeat (iterate) step 2. and 3. until the cluster assignment of the 
	correlation matrices
	no longer changes, in other words: the centroids no longer
	change.
	\item Form $k$ clusters by assigning each correlation matrix to its nearest 
	centroid.
	\item Calculate the centroid of each cluster.
\end{enumerate}
The start centroids are randomly selected correlation matrices.
The Euclidean distance of correlation matrices for different epochs 
Eq.~(\ref{eqn:EuclidDist})
was chosen as clustering distance measure.
For different start centroids we receive different clusterings $Z$.
Finally, we take the clustering $\widetilde{Z}^\text{(opt)}$ which minimizes the objective 
function
\begin{equation} \label{eqn:ObjFunction}
J(Z) = \sum_{l=1}^k \sum_{ n_{\text{ep} } \in z_l } 
\left|\left| 
C(n_{\text{ep}})  - 
{\langle 
C 
\rangle}^{(l)} 
\right|\right|^2 \,.
\end{equation}
The bisecting $k$-means algorithm uses the $k$-means cluster solution 
$\widetilde{Z}^\text{(opt)}$ for $k=2$, \textit{i.e.} for 
every splitting of a parent cluster into two child clusters (see Sec.~\ref{sec:DeterminationMarketStates}).

\section{\label{sec:DeMeanMatrices}Market states of de-meaned matrices}

\begin{figure}[!htb]
	\centering
	\includegraphics[width=1.0\textwidth]{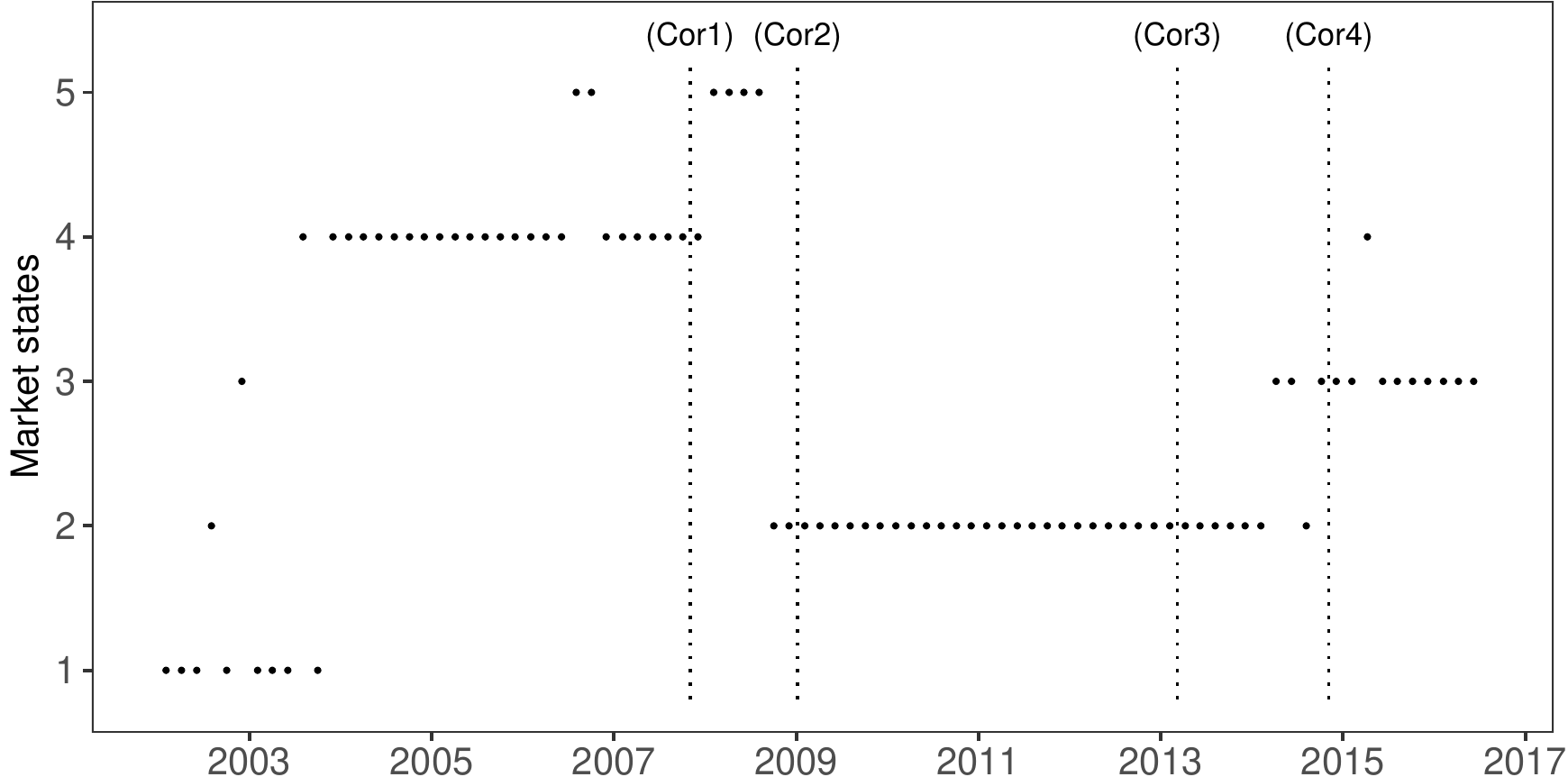}
	\caption{Temporal evolution of the de-meaned standard correlation matrices. The dashed lines are 
		marking the ``turning points" (see Tab.~\ref{tab:MeasuredTurningPoints}) of the 
		redcued-rank correlation matrices of the correlation approach  (\href{https://www.quandl.com/}{Data 
			from QuoteMedia via Quandl}).}
	\label{fig:TimeEvolutionStandSubstrMean-k5}
\end{figure}
\begin{figure}[!htb]
	\centering
	\subfloat[\label{subfig:SubtractMean-1:a}state 1]{
		\includegraphics[width=0.33\textwidth]{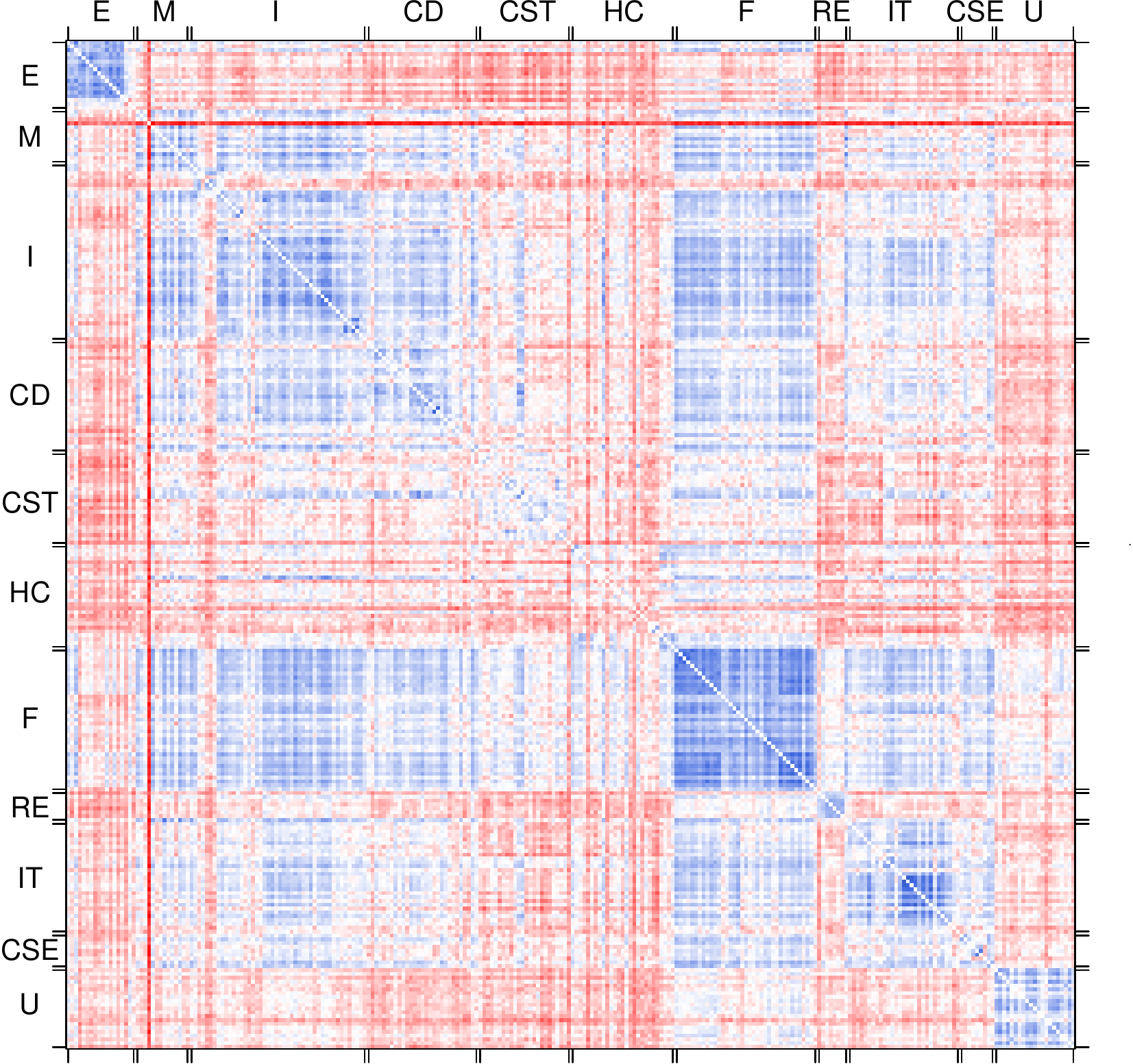}
	}
	\subfloat[\label{subfig:SubtractMean-2:b}state 2]{
		\includegraphics[width=0.33\textwidth]{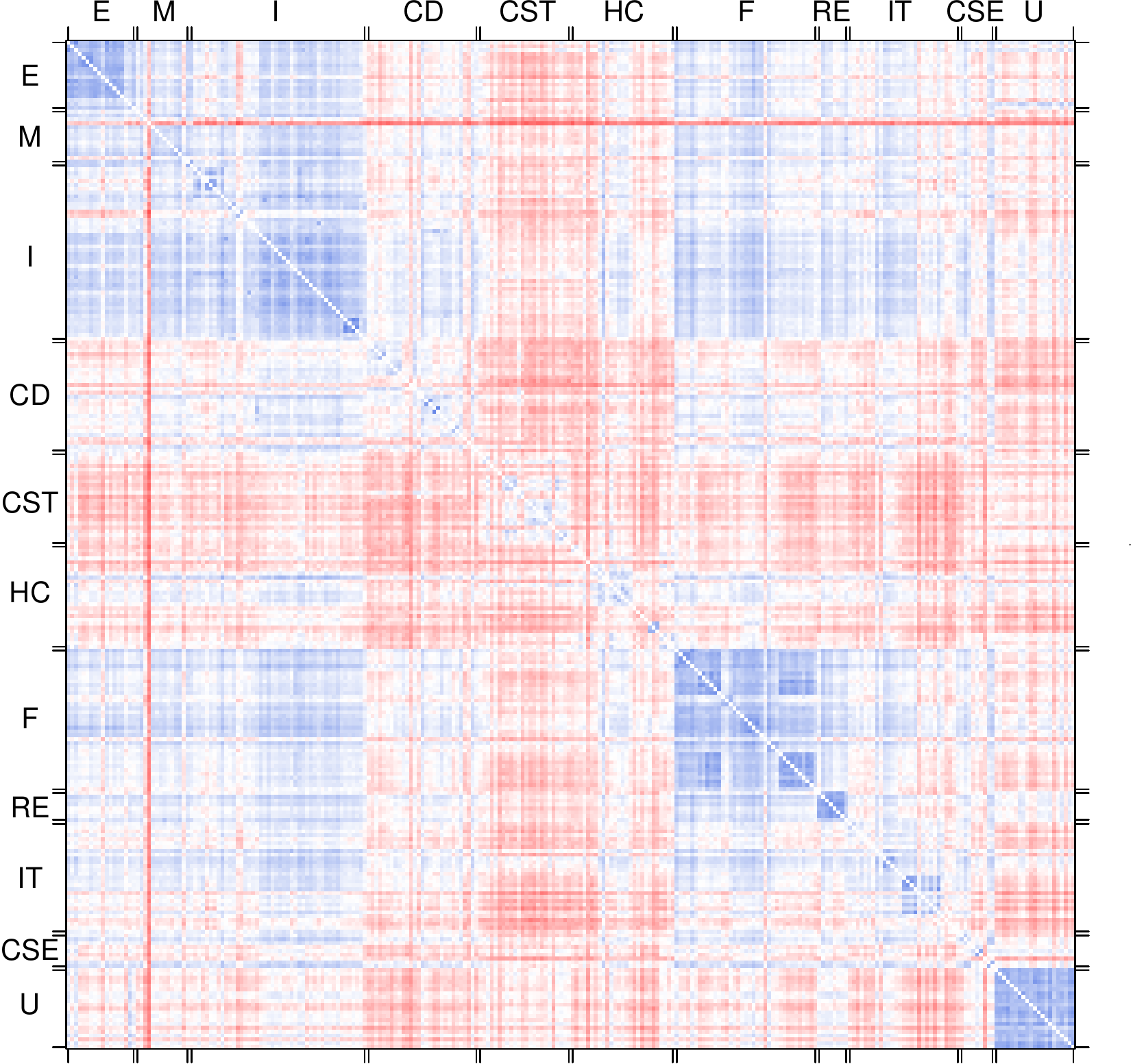}
	}
	\subfloat[\label{subfig:SubtractMean-3:c}state 3]{
		\includegraphics[width=0.33\textwidth]{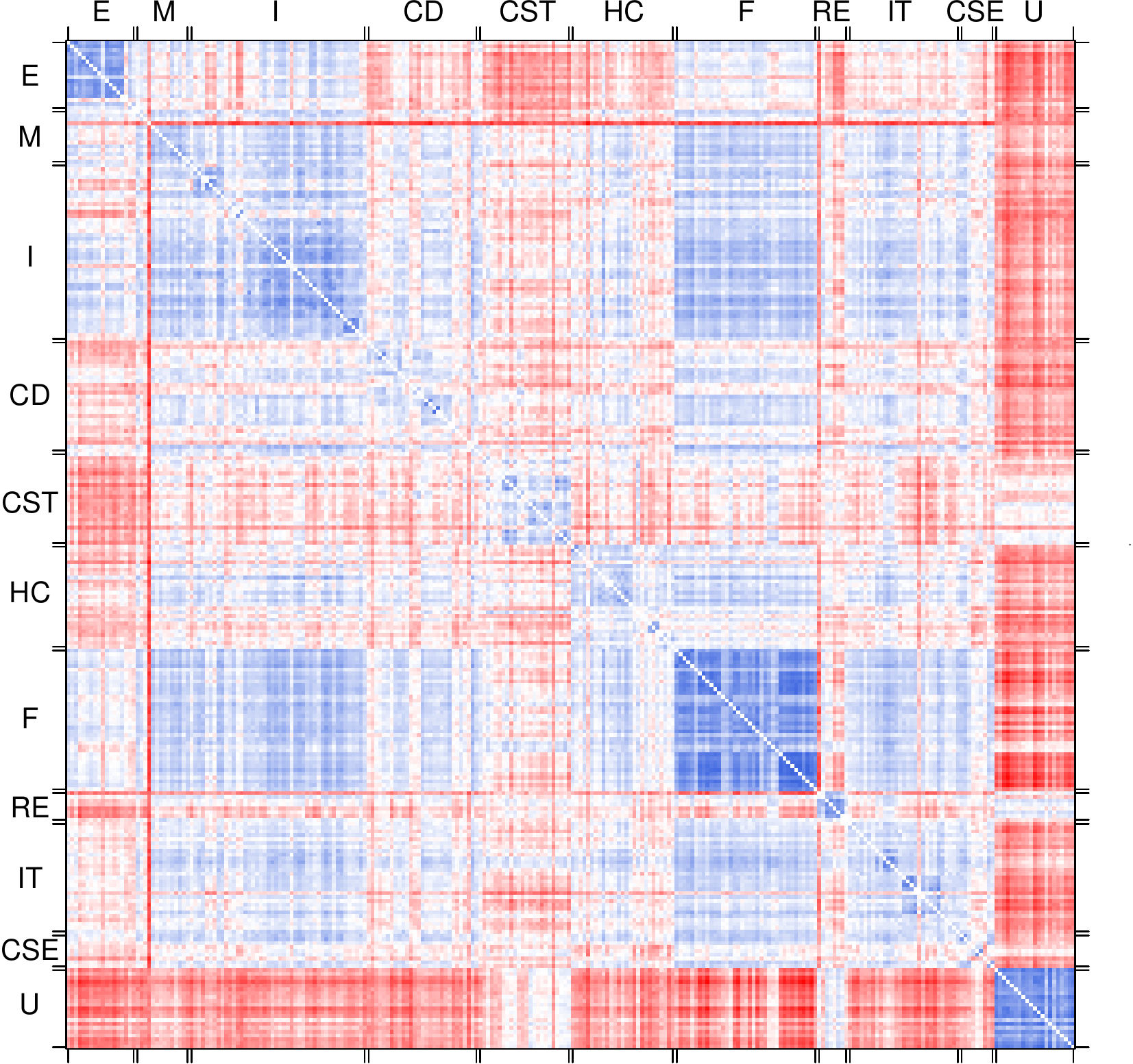}
	}\\
	\subfloat[\label{subfig:SubtractMean-4:d}state 4]{
		\includegraphics[width=0.33\textwidth]{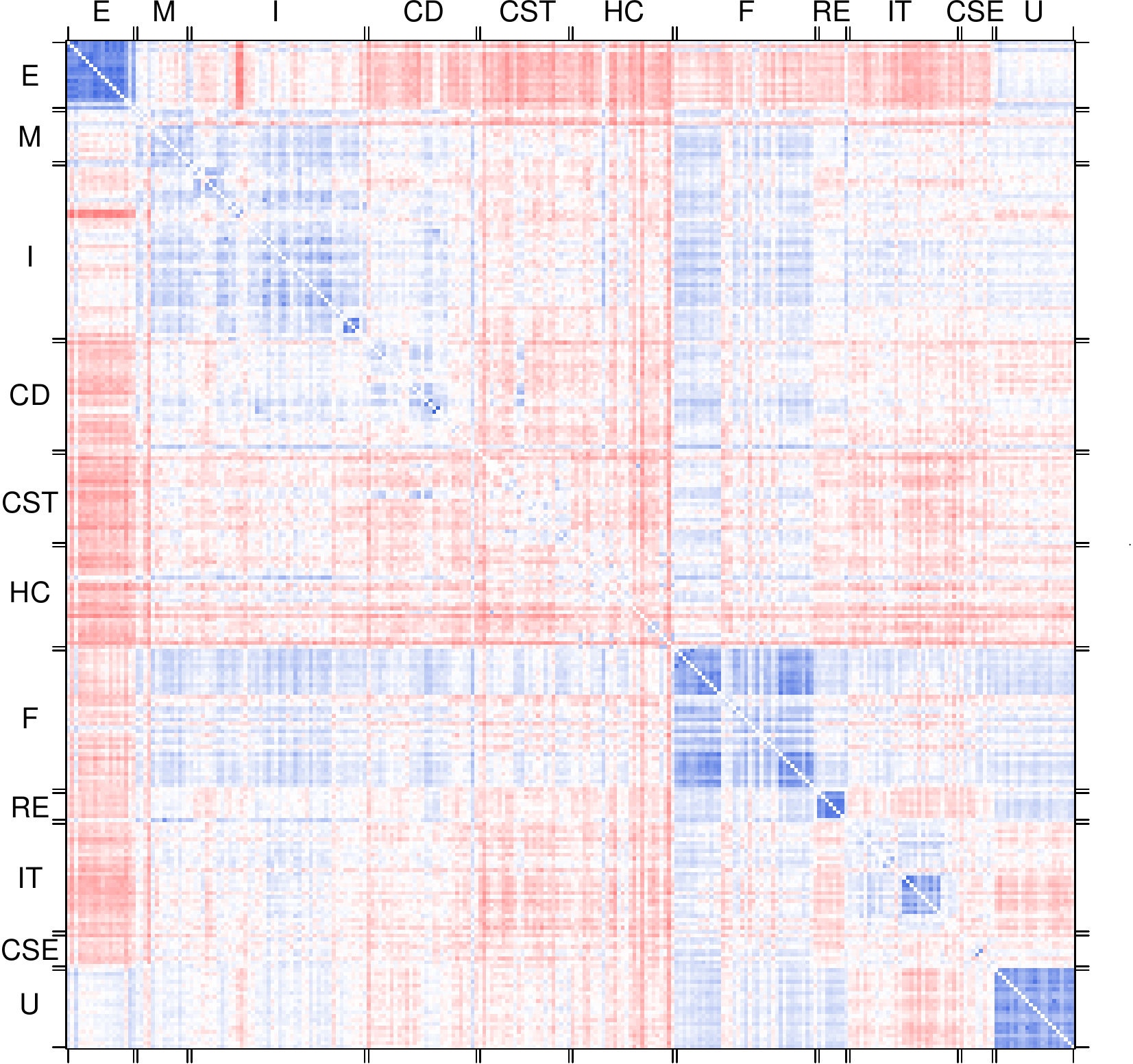}
	}
	\subfloat[\label{subfig:SubtractMean-5:e}state 5]{
		\includegraphics[width=0.33\textwidth]{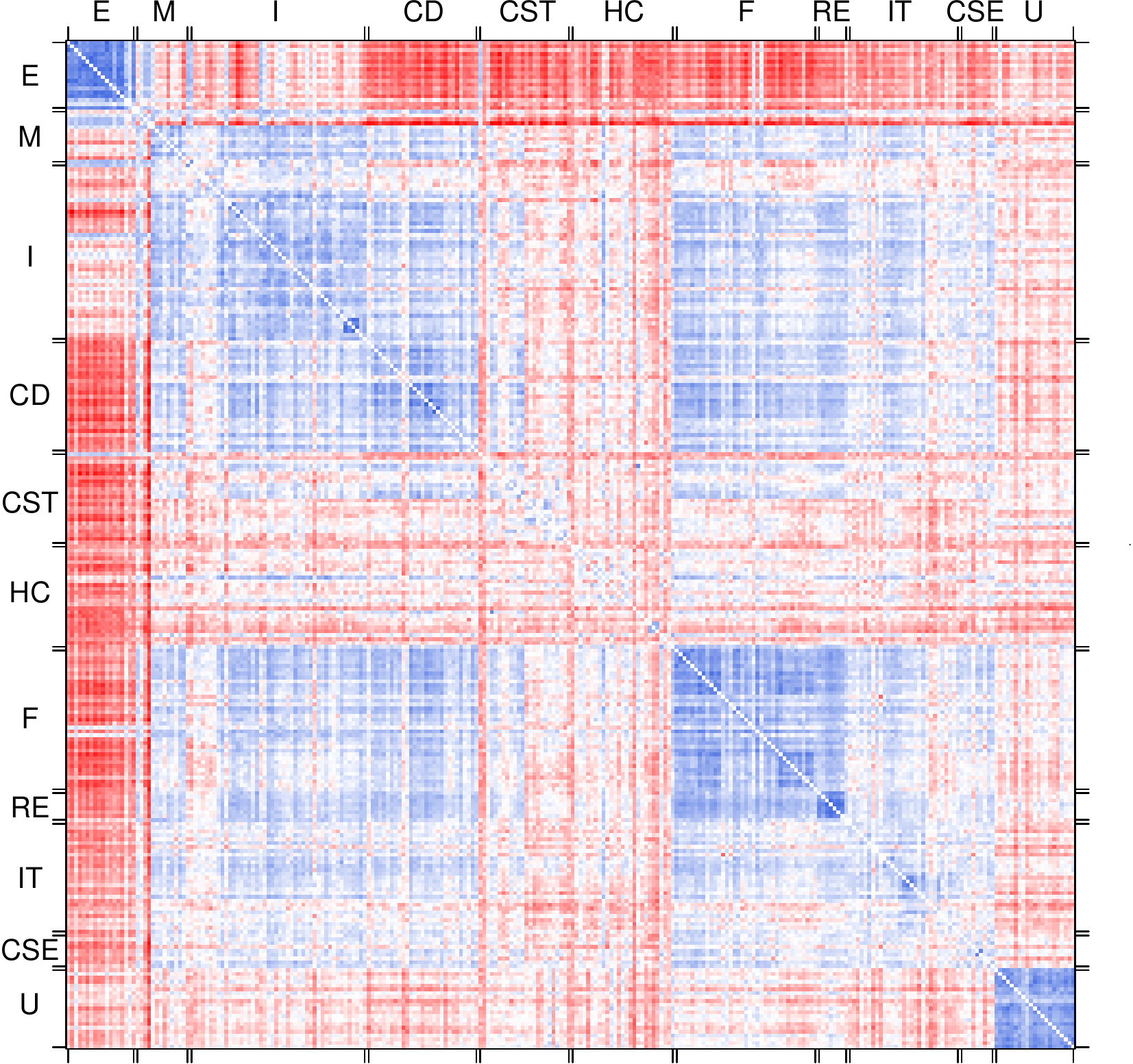}
	}\\
	\subfloat[\label{subfig:SubtractMean-6:f}overall average matrix (averaged over all 87 de-meaned matrices).]{
		\includegraphics[width=0.58\textwidth]{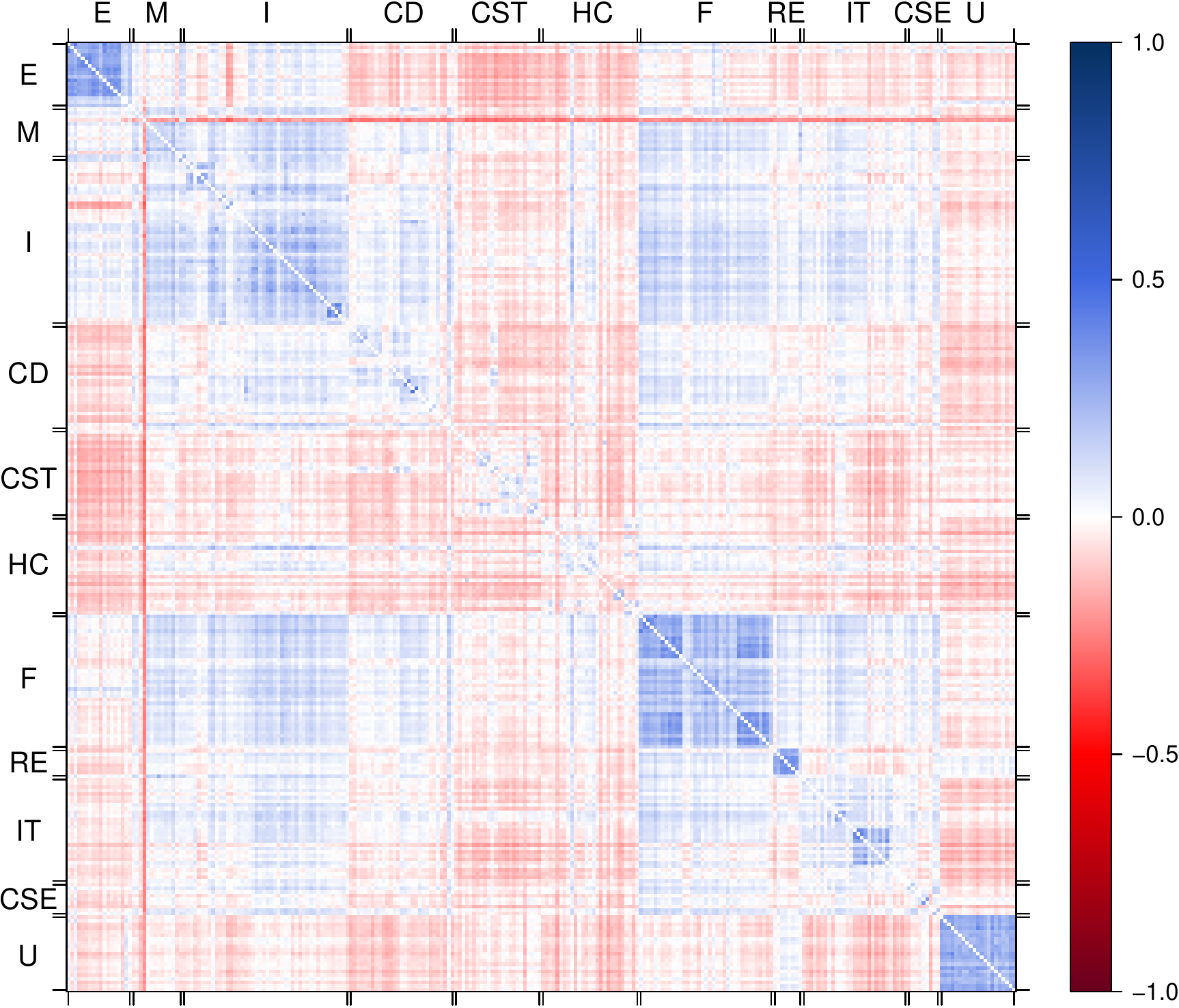}
	}
	\caption{\label{subfig:SubtractMean:main}Typical market states of the de-meaned matrix. The main diagonals are set to zero since the de-meaned matrix is not a correlation matrix (\href{https://www.quandl.com/}{Data 
			from QuoteMedia via Quandl}).}
\end{figure}
We perform a de-meaning procedure by taking the lower triangle of the standard correlation matrix in Sec.~\ref{sec:ResultsClusteringStandCorrMat} 
and subtracting its mean value.
The goal is to study the changes in the matrix structure induced by the eigenvector 
corresponding to the largest eigenvalue.
The diagonal of the standard correlation matrix should stay untouched since it has by definition no temporal evolution.
Afterwards, we cluster the 87 de-meaned lower triangular matrices by using the bisecting $k$-means algorithm assuming $k=5$
to make a comparison to the reduced-rank correlation matrices of the correlation approach possible. 
In Fig.~\ref{fig:TimeEvolutionStandSubstrMean-k5}, we observe a quasi-stationary temporal behavior of the de-meaned matrices. 
The temporal evolution differs from the one of the correlation approach in Fig.~\ref{fig:TimeEvolutionRedRankCorrCorrAnsatz}. 
The turning points are different and all market states also appear
at the beginning of the analyzed period. In Fig.~\ref{subfig:SubtractMean:main}, the typical market states of the de-meaned matrices show 
a different sector structure than the reduced-rank correlation matrices of Fig.~\ref{subfig:CorrCorrAnsatzTypClustCent:main}. 
Roughly speaking, there are four blocks which are very likely dominating the temporal
behavior of the de-meand matrices as well as stronger negative contributions in the typical market states 3 and 5. 
Interestingly, market state 5 can mainly be observed during the period of the ``crisis state"
detected by clustering the reduced-rank approaches (see Sec.~\ref{sec:ResultClusteringReducedRankCorrelationMatrices}).

\section{\label{sec:MarketStatesDependNumberStocks}Analyzing the market states depending on the choice and number of stocks}

We want to compare the original cluster solutions $Z^{(\text{opt})}$ or $Z^{\ast}$ (cf. Sec.~\ref{sec:ResultsClusteringStandCorrMat}) in Figs.~\ref{fig:TimeEvolutionStandCorr}, 
\ref{fig:TimeEvolutionRedRankCorrCovAnsatz} and~\ref{fig:TimeEvolutionRedRankCorrCorrAnsatz} with 
the cluster solutions $Z^{(K)}$ of randomly chosen companies from all $K = 262$ companies.
We perform the following procedure for the standard correlation matrices and for both reduced-rank approaches:
For each $K= 50, 100, 150, 200, 250$, 
we randomly select 50 times $K$ companies, 
calculate 50 times 87 correlation matrices of size $K \times K$ and cluster 50 times 87 correlation matrices using bisecting $k$-means (see Sec.~\ref{sec:DeterminationMarketStates})
assuming the number of clusters of  $Z^{(\text{opt})}$ or $Z^{\ast}$. Thus, we obtain 50 cluster solution $Z^{(K)}$ for each $K$. 
To compare the cluster solution of $Z^{(\text{opt})}$ or $Z^{\ast}$ with the corresponding $Z^{(K)}$ which belong to one 
of our three ``types" of correlation matrices, we choose the
adjusted Rand index as a similarity measure of two cluster solutions \cite{hubert1985comparing,wagner2007comparing,musmeci2015risk}.

In order to understand the adjusted Rand index, it is necessary to introduce the Rand index \cite{rand1971objective}
since it is a main ingredient in the definition of the adjusted Rand index (Eq.~(\ref{eqn:ARI})).
We calculate the Rand index between two cluster solutions, for example between 
$Z^{\ast} = \{ z_1^{(\ast)}, z_2^{(\ast)}, \ldots, z_l^{(\ast)}, \ldots z_k^{(\ast)} \} $ and 
$Z^{(K)} = \{ z_1^{(K)}, z_2^{(K)}, \ldots, z_{l}^{(K)}, \ldots z_{k}^{(K)}   \}$
as follows:
\begin{equation} \label{eqn:RI}
{R} \left( Z^{\ast}, Z^{(K)} \right) = \frac{ a+b } {a+b+c+d} = \frac{a+b}{\binom{N_{\text{ep}}}{ 2  }} \,.
\end{equation}
$k$ is the cluster number for which all correlation matrices are clustered. $z$ denotes the clusters (subsets) of the respective cluster solutions. $a$, $b$, $c$ and $d$ are
four ``sums of unordered pairs" (more precisely, the cardinalities of four sets consisting of unordered pairs) 
of two correlation matrices (from the $N_{\text{ep}} = 87$ correlation matrices) for which, in the case of
\begin{itemize}
	\item $a$, two correlation matrices belong to the same cluster of $Z^{\ast}$ and to the same cluster of $Z^{(K)}$ \,,  
	\item $b$, two correlation matrices belong to different clusters of $Z^{\ast}$ and to different clusters of $Z^{(K)}$ \,,
	\item $c$, two correlation matrices belong to the same cluster of $Z^{\ast}$ and to different clusters of $Z^{(K)}$ \,,
	\item $d$, two correlation matrices belong to different clusters of $Z^{\ast}$ and to the same cluster of $Z^{(K)}$ \,.
\end{itemize}
That means that in Eq.~(\ref{eqn:RI}), the numerator takes into account all agreeing pairs concerning the comparison 
of the two cluster solutions $Z^{\ast}$ and $Z^{(K)}$. In the denominator,
all possible pairs of two elements of the 87 correlation matrices are calculated.
The maximum value of the Rand index is ${R}^{(\text{max})}= 1$ which means that two cluster solutions are identical, 
the minimum value is ${R}^{(\text{min})} = 0$ which
means that there is no agreement between two cluster solutions.

\begin{figure}[!htb]
	\centering
	\subfloat[\label{subfig:TimeEvol-Stand:a}Standard correlation matrix; $\text{ARI} = 0.578$]{
		\includegraphics[width=.7\columnwidth]{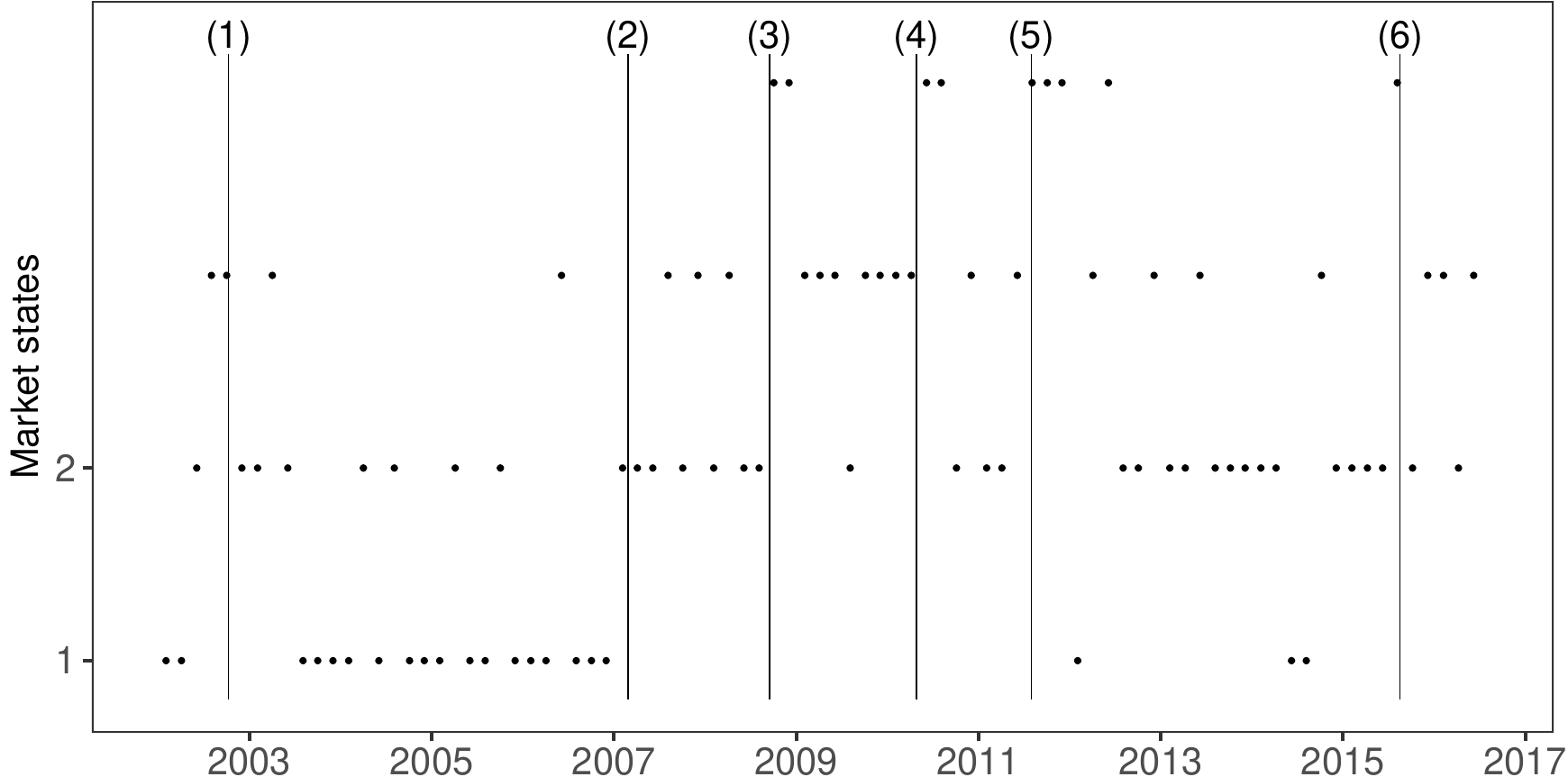}
	} \\
	\subfloat[\label{subfig:TimeEvol-CorrAppr:b}Reduced-rank correlation matrix (covariance approach); $\text{ARI} = 0.272$]{
		\includegraphics[width=.7\columnwidth]{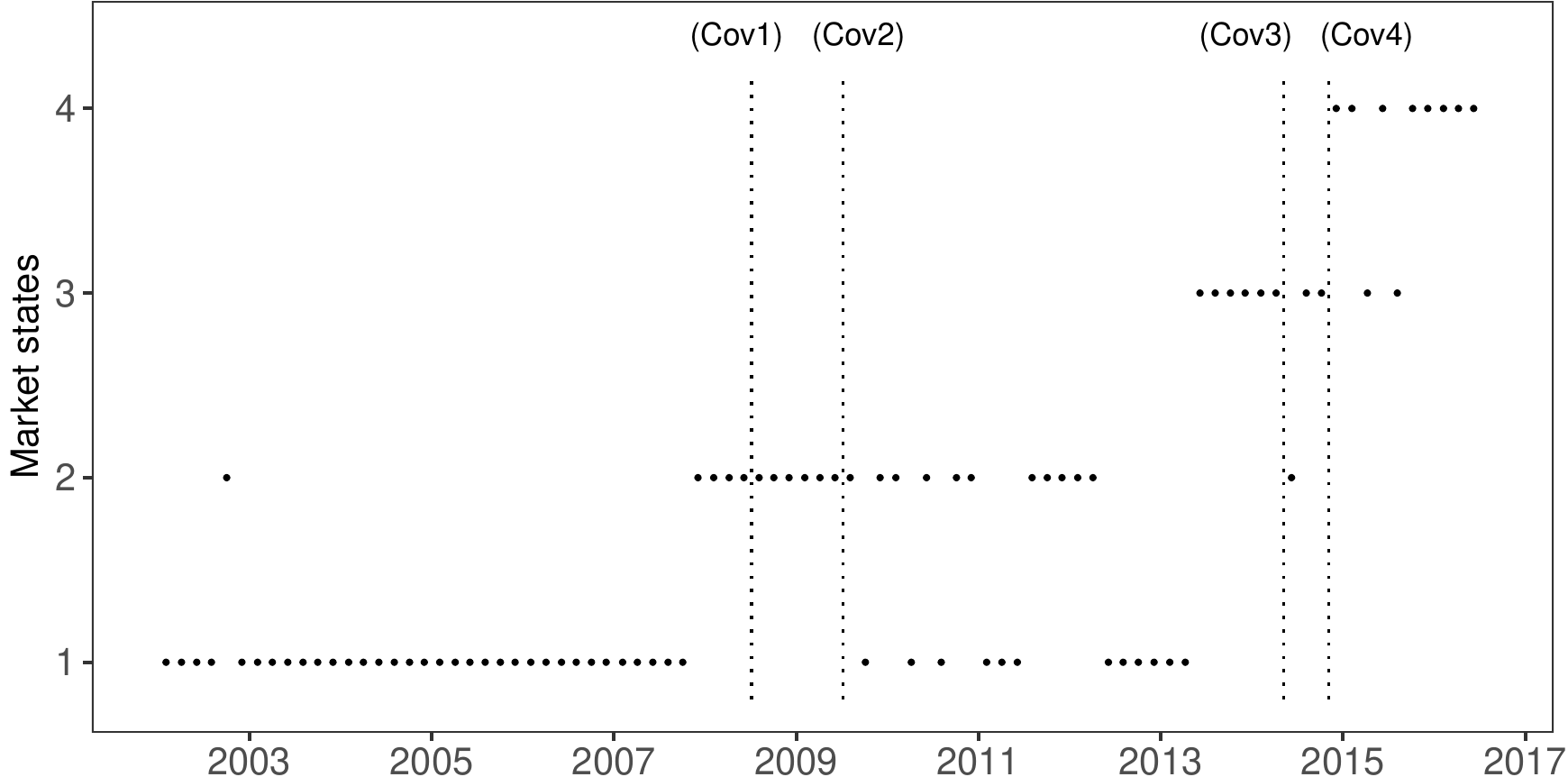}
	} \\
	\subfloat[\label{subfig:TimeEvol-CovAppr:c}Reduced-rank correlation matrix (correlation approach); $\text{ARI} = 0.457$]{
		\includegraphics[width=.7\columnwidth]{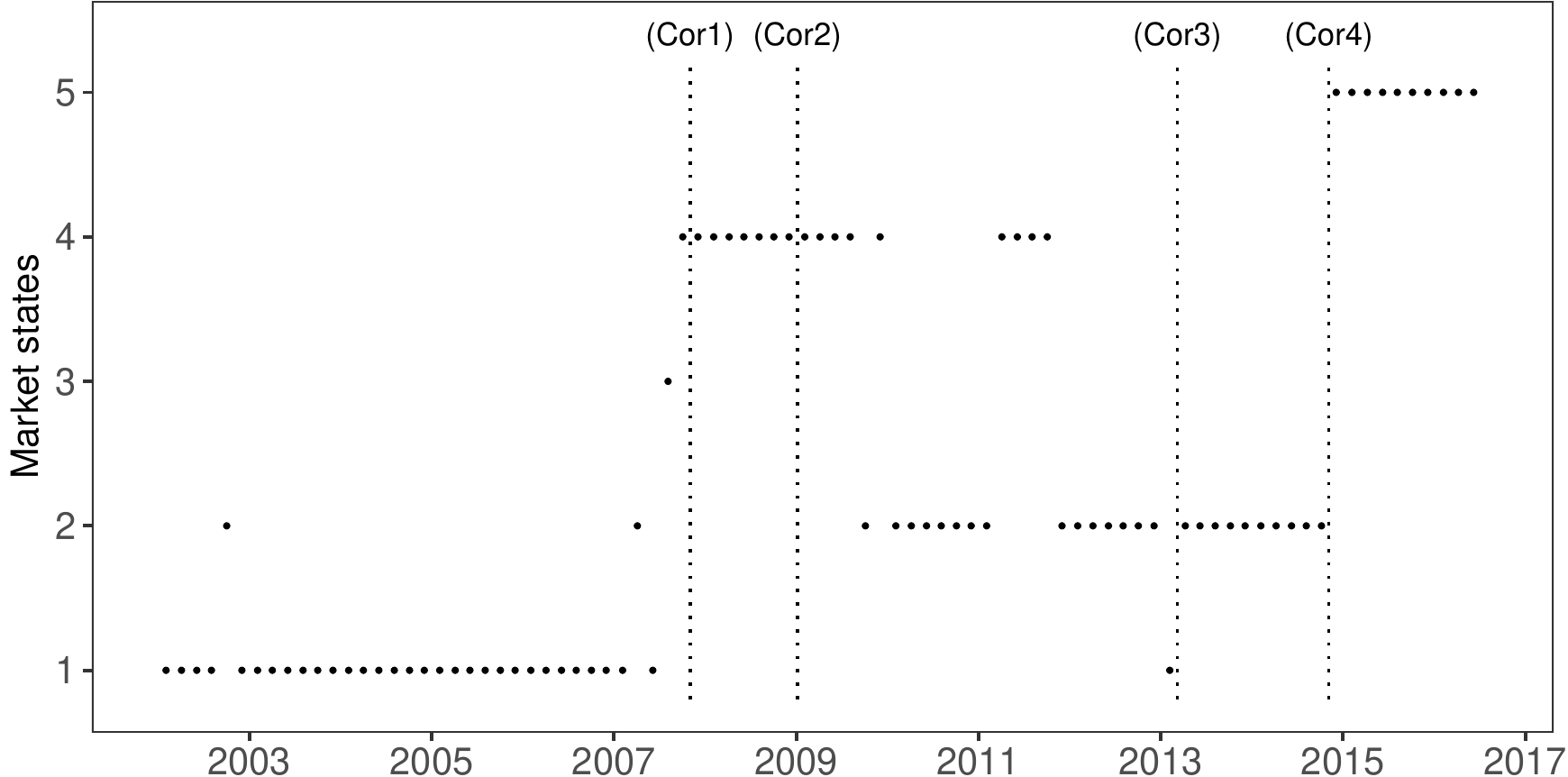}
	}
	\caption{\label{subfig:TimeEvol:main} Time evolution of randomly chosen companies ($K=50$). The adjusted Rand index was calculated from the here shown cluster solutions $Z^{(50)}$ and the cluster solutions $Z^{(\text{opt})}$ or $Z^{\ast}$ discussed in main part of the paper  (\href{https://www.quandl.com/}{Data 
			from QuoteMedia via Quandl}).}
\end{figure}

One disadvantage of the Rand index is that it depends on the number of clusters \cite{wagner2007comparing}.
To correct this a so-called permutation model \cite{gates2017impact} as a reference model 
is introduced which takes into account random overlaps between two cluster solutions.
The underlying null hypothesis for the reference model is that the correlation matrices are randomly shuffled between the clusters \cite{wagner2007comparing}
for a fixed number of clusters and for a fixed number of correlation matrices within a cluster.
$\text{PM}$ is the expected value of the Rand index under this null hypothesis.
In literature, the adjusted Rand index \cite{hubert1985comparing,gates2017impact} is introduced as
\begin{equation} \label{eqn:ARI}
\text{ARI} \left( Z^{\ast}, Z^{(K)} \right) = \frac{ {R} \left( Z^{\ast}, Z^{(K)} \right)  - \text{PM} \left( Z^{\ast}, Z^{(K)} \right) }{ {R}^{ \left(\text{max} \right)} - \text{PM} \left( Z^{\ast}, Z^{(K)} \right) } = \frac{ {R} \left( Z^{\ast}, Z^{(K)} \right)  - \text{PM} \left( Z^{\ast}, Z^{(K)} \right) }{ 1 - \text{PM} \left( Z^{\ast}, Z^{(K)} \right) } \,,
\end{equation}
where the Rand index is subtracted by Rand index $\text{PM}$ normalized by the maximum possible value of the numerator.
Due to the normalization, the maximum possible value of the adjusted Rand index is $\text{ARI}^{(\text{max})} = 1$ 
which is the case for two identical cluster solutions.
The adjusted Rand index can also take negative values.
The adjusted Rand index is a frequently used measure
for comparing two cluster solutions~\cite{vinh2010information}. In our analysis, we use the implementation of the R-package \texttt{mclust} 
\cite{fraley2020package}.

In Fig.~\ref{subfig:TimeEvol:main}, the time evolution of all three ``types" of correlation matrices are illustrated for $K=50$ randomly chosen companies.
The $\text{ARI}$-values are specified.
Three cluster solutions were chosen as representatives of the later explained Fig.~\ref{subfig:CompareClust:main} 
for which the adjusted Rand index comes the closest to the mean adjusted Rand index in Fig.~\ref{subfig:CompareClust:main}. 
In Fig.~\ref{subfig:TimeEvol:main}, the cluster solution for the standard correlation matrix possesses the largest $\text{ARI}$-value, followed by the correlation approach.
For the standard correlation matrix, crises events can be identified (Tab.~\ref{tab:FinancialCrises}). However, 
in the case of the reduced-rank correlation matrix approaches, the turning points are quite different (Tab.~\ref{tab:MeasuredTurningPoints}).
Exceptions are (Cor4) and (Cov4) and approximately (Cor1). The reduced-rank correlation matrices 
show again a higher quasi-stationarity than the standard correlation matrices.

Our goal is to compare the cluster solution $Z^{\text{(opt)}}$ or $Z^{\ast}$ with the cluster solutions of different number of companies $K$ more systematically.
In Fig.~\ref{subfig:CompareClust:main}, we plot for each $K$ the mean value of the computed 50 $\text{ARI}$-values, as well as the corresponding error bars.    
The following applies to all drawings in Fig.~\ref{subfig:CompareClust:main}: The larger the number of companies $K$, the more similar is the cluster
solution $Z^{(K)}$ to $Z^{(\text{opt})}$ or $Z^{\ast}$. 
In the case of the standard correlation matrix, we see the largest mean value of the $\text{ARI}$-values which reflects the
collective behavior of the stocks. Nonetheless, for $K=50$ the mean value of 
the adjusted Rand index is not 1 since the collective behavior also possesses a certain structure.        
The correlation approach shows higher similarities than the covariance approach. The latter has the largest error bars.

The market states strongly depend on the choice and number of stocks for the reduced-rank correlation matrices.
This is not surprising
because the market states of the randomly chosen companies show a different time evolution due to the different sector structures of the samples
and the different time evolution of the industrial sectors.

\begin{figure}[htbp]
	\centering
	\subfloat[\label{subfig:Stand-ErrorBar}Standard correlation matrix]{
		\includegraphics[width=0.5\textwidth]{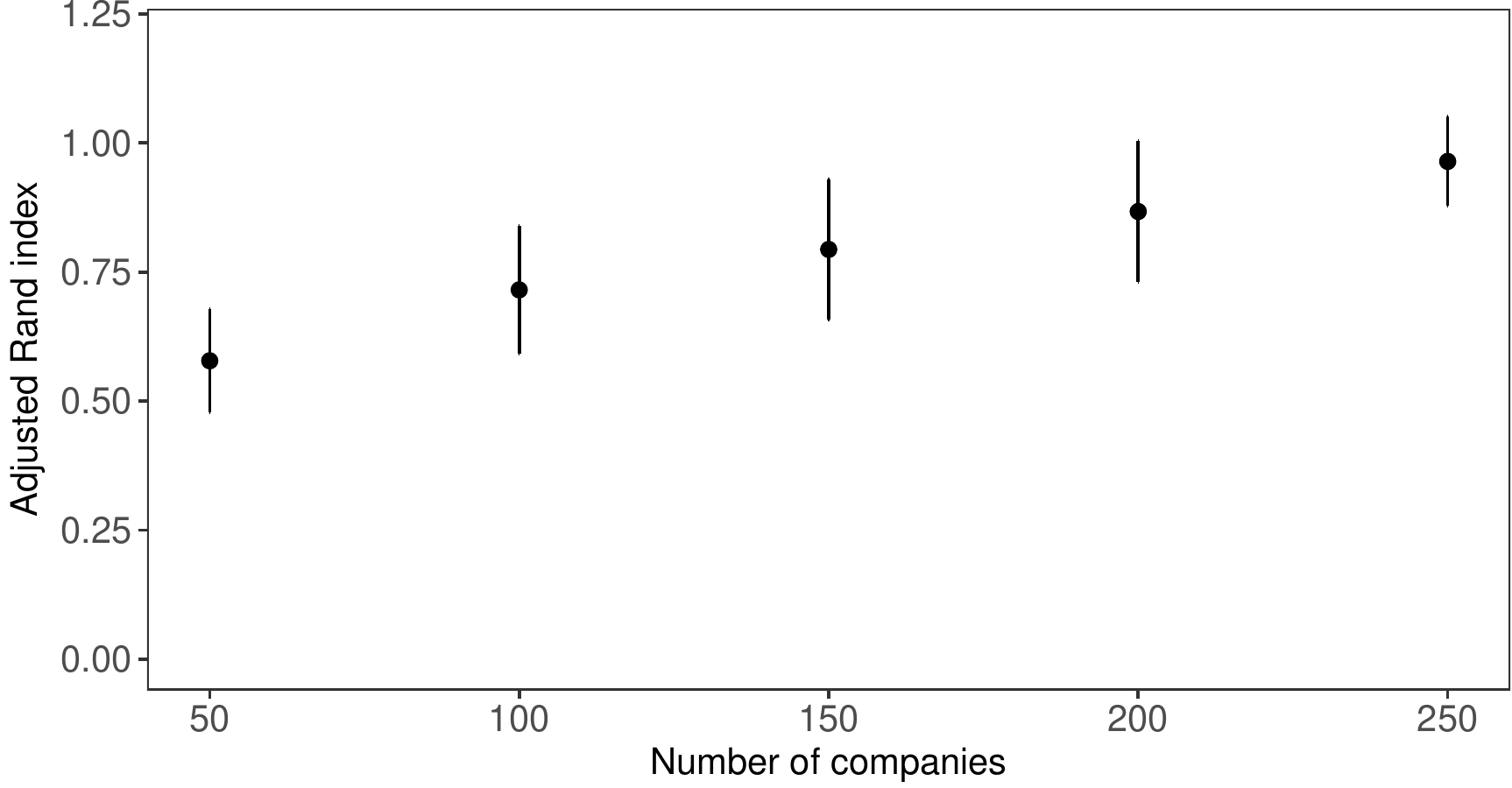}
	} \\
	\subfloat[\label{subfig:CovAppr-ErrorBar}Reduced-rank correlation matrix (covariance 
	approach)]{
		\includegraphics[width=0.5\textwidth]{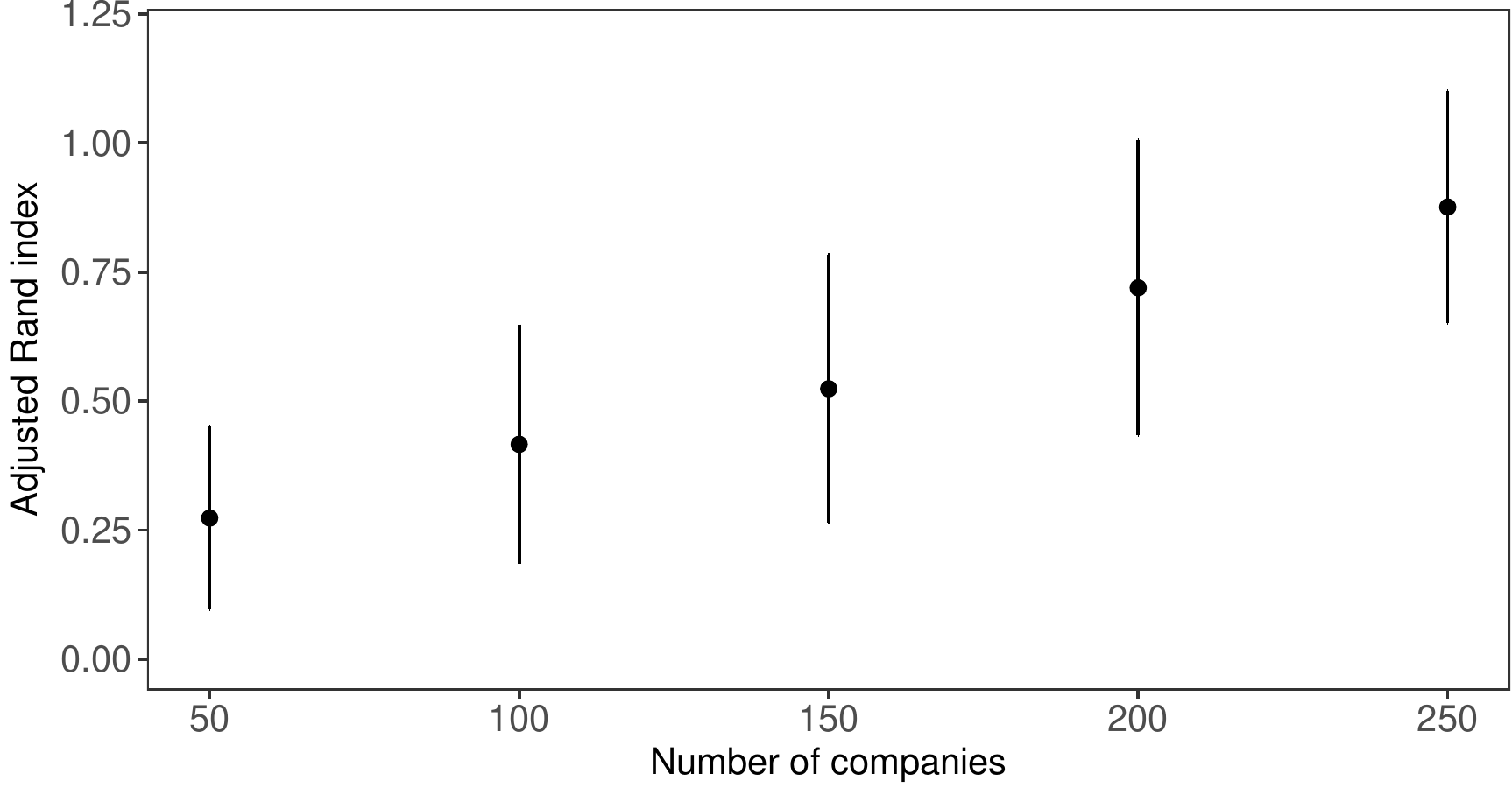}
	} \\
	\subfloat[\label{subfig:CorrAppr-ErrorBar}Reduced-rank correlation matrix (correlation 
	approach)]{
		\includegraphics[width=0.5\textwidth]{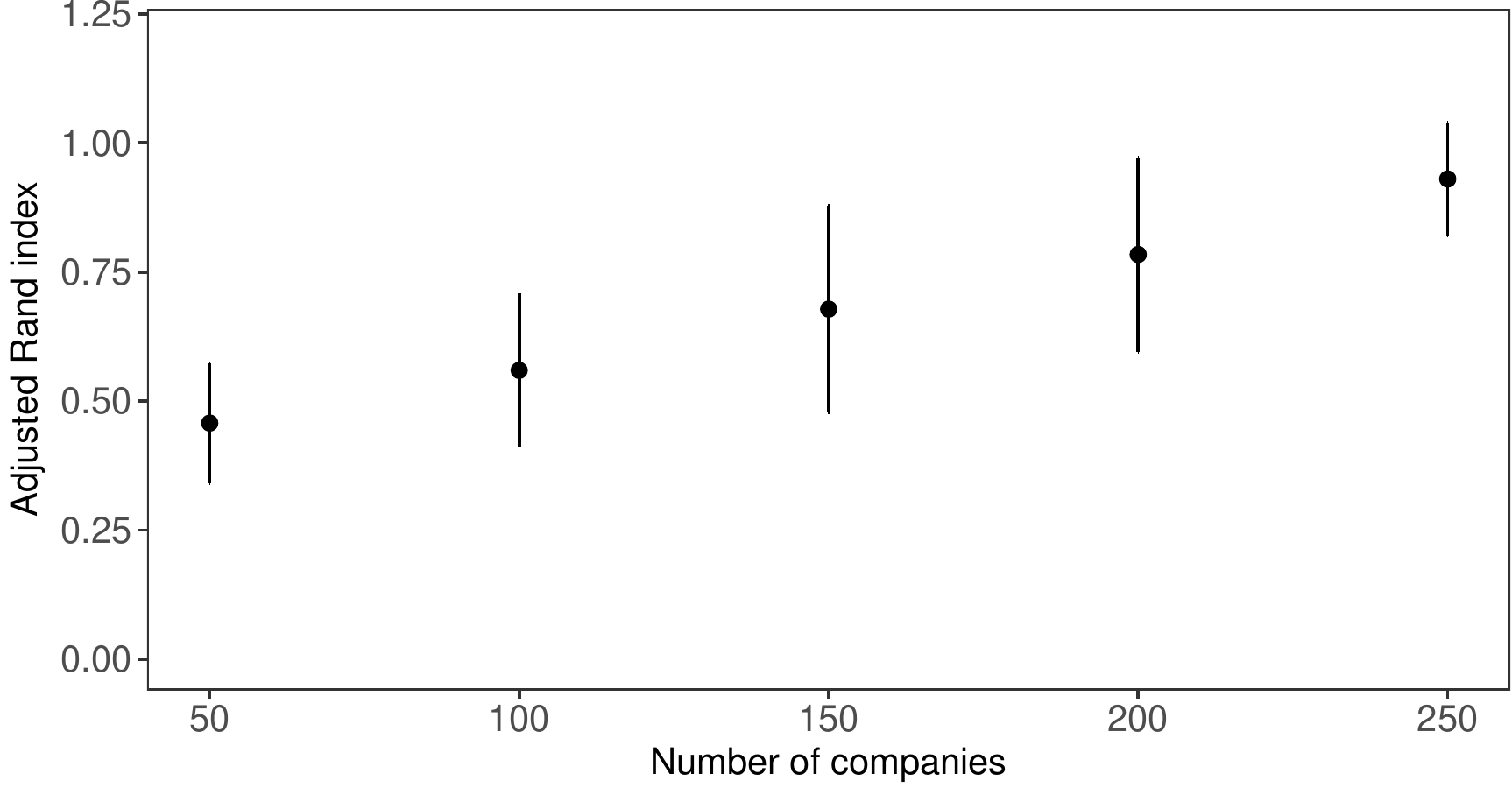}
	}
	\caption{\label{subfig:CompareClust:main}Comparing cluster solutions using the adjusted Rand index for the different correlation matrices depending on the number of companies $K$. The value range of the adjusted Rand index is indicated in error bars  (\href{https://www.quandl.com/}{Data from QuoteMedia 
			via Quandl}).}	
\end{figure}

\clearpage
\section{\label{sec:ListStocks}List of selected stocks}

{\tiny
	\begin{longtable}{rrllp{4cm}}
		\caption[]{Overview of the 262 selected companies of the S\&P 500 index (cf.~{\cite{wiki:2019:List500Companies}}).} 

		\label{tab:OverviewSP500} \\
		\toprule
		Number & Symbol & Security & Sector & Sub-Industry\\
		\midrule\endfirsthead
		\caption*{Continuation: Overview of the 262 selected companies of the S\&P 500 
index (cf.~{\cite{wiki:2019:List500Companies}}).}  \\
		Number & Symbol & Security & Sector & Sub-Industry\\
		\midrule\endhead
1 & CVX & Chevron Corp. & Energy & Integrated Oil \& Gas \\
2 & HES & Hess Corporation & Energy & Integrated Oil \& Gas \\
3 & XOM & Exxon Mobil Corp. & Energy & Integrated Oil \& Gas \\
4 & HP & Helmerich \& Payne & Energy & Oil \& Gas Drilling \\
5 & BHGE & Baker Hughes, a GE Company & Energy & Oil \& Gas Equipment \& 
Services \\
6 & HAL & Halliburton Co. & Energy & Oil \& Gas Equipment \& Services \\
7 & SLB & Schlumberger Ltd. & Energy & Oil \& Gas Equipment \& Services \\
8 & APA & Apache Corporation & Energy & Oil \& Gas Exploration \& Production \\
9 & APC & Anadarko Petroleum Corp & Energy & Oil \& Gas Exploration \& 
Production \\
10 & COG & Cabot Oil \& Gas & Energy & Oil \& Gas Exploration \& Production \\
11 & COP & ConocoPhillips & Energy & Oil \& Gas Exploration \& Production \\
12 & EOG & EOG Resources & Energy & Oil \& Gas Exploration \& Production \\
13 & MRO & Marathon Oil Corp. & Energy & Oil \& Gas Exploration \& Production \\
14 & NBL & Noble Energy Inc & Energy & Oil \& Gas Exploration \& Production \\
15 & OXY & Occidental Petroleum & Energy & Oil \& Gas Exploration \& Production 
\\
16 & VLO & Valero Energy & Energy & Oil \& Gas Refining \& Marketing \\
17 & OKE & ONEOK & Energy & Oil \& Gas Storage \& Transportation \\
18 & WMB & Williams Cos. & Energy & Oil \& Gas Storage \& Transportation \\
19 & VMC & Vulcan Materials & Materials & Construction Materials \\
20 & FMC & FMC Corporation & Materials & Fertilizers \& Agricultural Chemicals 
\\
21 & MOS & The Mosaic Company & Materials & Fertilizers \& Agricultural 
Chemicals \\
22 & NEM & Newmont Mining Corporation & Materials & Gold \\
23 & APD & Air Products \& Chemicals Inc & Materials & Industrial Gases \\
24 & BLL & Ball Corp & Materials & Metal \& Glass Containers \\
25 & AVY & Avery Dennison Corp & Materials & Paper Packaging \\
26 & IP & International Paper & Materials & Paper Packaging \\
27 & SEE & Sealed Air & Materials & Paper Packaging \\
28 & ECL & Ecolab Inc. & Materials & Specialty Chemicals \\
29 & IFF & Intl Flavors \& Fragrances & Materials & Specialty Chemicals \\
30 & PPG & PPG Industries & Materials & Specialty Chemicals \\
31 & SHW & Sherwin-Williams & Materials & Specialty Chemicals \\
32 & NUE & Nucor Corp. & Materials & Steel \\
33 & ARNC & Arconic Inc. & Industrials & Aerospace \& Defense \\
34 & BA & Boeing Company & Industrials & Aerospace \& Defense \\
35 & GD & General Dynamics & Industrials & Aerospace \& Defense \\
36 & HRS & Harris Corporation & Industrials & Aerospace \& Defense \\
37 & LMT & Lockheed Martin Corp. & Industrials & Aerospace \& Defense \\
38 & NOC & Northrop Grumman Corp. & Industrials & Aerospace \& Defense \\
39 & RTN & Raytheon Co. & Industrials & Aerospace \& Defense \\
40 & TXT & Textron Inc. & Industrials & Aerospace \& Defense \\
41 & UTX & United Technologies & Industrials & Aerospace \& Defense \\
42 & DE & Deere \& Co. & Industrials & Agricultural \& Farm Machinery \\
43 & EXPD & Expeditors & Industrials & Air Freight \& Logistics \\
44 & FDX & FedEx Corporation & Industrials & Air Freight \& Logistics \\
45 & ALK & Alaska Air Group Inc & Industrials & Airlines \\
46 & LUV & Southwest Airlines & Industrials & Airlines \\
47 & AOS & A.O. Smith Corp & Industrials & Building Products \\
48 & FAST & Fastenal Co & Industrials & Building Products \\
49 & JCI & Johnson Controls International & Industrials & Building Products \\
50 & MAS & Masco Corp. & Industrials & Building Products \\
51 & JEC & Jacobs Engineering Group & Industrials & Construction \& Engineering 
\\
52 & CAT & Caterpillar Inc. & Industrials & Construction Machinery \& Heavy 
Trucks \\
53 & PCAR & PACCAR Inc. & Industrials & Construction Machinery \& Heavy Trucks 
\\
54 & CTAS & Cintas Corporation & Industrials & Diversified Support Services \\
55 & AME & AMETEK Inc. & Industrials & Electrical Components \& Equipment \\
56 & EMR & Emerson Electric Company & Industrials & Electrical Components \& 
Equipment \\
57 & ETN & Eaton Corporation & Industrials & Electrical Components \& Equipment 
\\
58 & ROK & Rockwell Automation Inc. & Industrials & Electrical Components \& 
Equipment \\
59 & ROL & Rollins Inc. & Industrials & Environmental \& Facilities Services \\
60 & GE & General Electric & Industrials & Industrial Conglomerates \\
61 & HON & Honeywell Int'l Inc. & Industrials & Industrial Conglomerates \\
62 & MMM & 3M Company & Industrials & Industrial Conglomerates \\
63 & CMI & Cummins Inc. & Industrials & Industrial Machinery \\
64 & DOV & Dover Corp. & Industrials & Industrial Machinery \\
65 & FLS & Flowserve Corporation & Industrials & Industrial Machinery \\
66 & GWW & Grainger (W.W.) Inc. & Industrials & Industrial Machinery \\
67 & IR & Ingersoll-Rand PLC & Industrials & Industrial Machinery \\
68 & ITW & Illinois Tool Works & Industrials & Industrial Machinery \\
69 & PH & Parker-Hannifin & Industrials & Industrial Machinery \\
70 & PNR & Pentair plc & Industrials & Industrial Machinery \\
71 & SNA & Snap-on & Industrials & Industrial Machinery \\
72 & SWK & Stanley Black \& Decker & Industrials & Industrial Machinery \\
73 & CSX & CSX Corp. & Industrials & Railroads \\
74 & KSU & Kansas City Southern & Industrials & Railroads \\
75 & NSC & Norfolk Southern Corp. & Industrials & Railroads \\
76 & UNP & Union Pacific & Industrials & Railroads \\
77 & EFX & Equifax Inc. & Industrials & Research \& Consulting Services \\
78 & JBHT & J. B. Hunt Transport Services & Industrials & Trucking \\
79 & FL & Foot Locker Inc & Consumer Discretionary & Apparel Retail \\
80 & GPS & Gap Inc. & Consumer Discretionary & Apparel Retail \\
81 & LB & L Brands Inc. & Consumer Discretionary & Apparel Retail \\
82 & ROST & Ross Stores & Consumer Discretionary & Apparel Retail \\
83 & TJX & TJX Companies Inc. & Consumer Discretionary & Apparel Retail \\
84 & NKE & Nike & Consumer Discretionary & Apparel, Accessories \& Luxury Goods 
\\
85 & PVH & PVH Corp. & Consumer Discretionary & Apparel, Accessories \& Luxury 
Goods \\
86 & TIF & Tiffany \& Co. & Consumer Discretionary & Apparel, Accessories \& 
Luxury Goods \\
87 & VFC & V.F. Corp. & Consumer Discretionary & Apparel, Accessories \& Luxury 
Goods \\
88 & F & Ford Motor & Consumer Discretionary & Automobile Manufacturers \\
89 & MGM & MGM Resorts International & Consumer Discretionary & Casinos \& 
Gaming \\
90 & BBY & Best Buy Co. Inc. & Consumer Discretionary & Computer \& Electronics 
Retail \\
91 & JWN & Nordstrom & Consumer Discretionary & Department Stores \\
92 & TGT & Target Corp. & Consumer Discretionary & General Merchandise Stores \\
93 & LEG & Leggett \& Platt & Consumer Discretionary & Home Furnishings \\
94 & HD & Home Depot & Consumer Discretionary & Home Improvement Retail \\
95 & LOW & Lowe's Cos. & Consumer Discretionary & Home Improvement Retail \\
96 & LEN & Lennar Corp. & Consumer Discretionary & Homebuilding \\
97 & PHM & Pulte Homes Inc. & Consumer Discretionary & Homebuilding \\
98 & CCL & Carnival Corp. & Consumer Discretionary & Hotels, Resorts \& Cruise 
Lines \\
99 & WHR & Whirlpool Corp. & Consumer Discretionary & Household Appliances \\
100 & NWL & Newell Brands & Consumer Discretionary & Housewares \& Specialties 
\\
101 & HAS & Hasbro Inc. & Consumer Discretionary & Leisure Products \\
102 & MAT & Mattel Inc. & Consumer Discretionary & Leisure Products \\
103 & HOG & Harley-Davidson & Consumer Discretionary & Motorcycle Manufacturers 
\\
104 & MCD & McDonald's Corp. & Consumer Discretionary & Restaurants \\
105 & HRB & Block H\&R & Consumer Discretionary & Specialized Consumer Services 
\\
106 & GPC & Genuine Parts & Consumer Discretionary & Specialty Stores \\
107 & GT & Goodyear Tire \& Rubber & Consumer Discretionary & Tires \& Rubber \\
108 & ADM & Archer-Daniels-Midland Co & Consumer Staples & Agricultural 
Products \\
109 & TAP & Molson Coors Brewing Company & Consumer Staples & Brewers \\
110 & BF-B & Brown-Forman Corp. & Consumer Staples & Distillers \& Vintners \\
111 & WBA & Walgreens Boots Alliance & Consumer Staples & Drug Retail \\
112 & SYY & Sysco Corp. & Consumer Staples & Food Distributors \\
113 & KR & Kroger Co. & Consumer Staples & Food Retail \\
114 & CHD & Church \& Dwight & Consumer Staples & Household Products \\
115 & CL & Colgate-Palmolive & Consumer Staples & Household Products \\
116 & CLX & The Clorox Company & Consumer Staples & Household Products \\
117 & KMB & Kimberly-Clark & Consumer Staples & Household Products \\
118 & COST & Costco Wholesale Corp. & Consumer Staples & Hypermarkets \& Super 
Centers \\
119 & WMT & Walmart & Consumer Staples & Hypermarkets \& Super Centers \\
120 & CAG & Conagra Brands & Consumer Staples & Packaged Foods \& Meats \\
121 & CPB & Campbell Soup & Consumer Staples & Packaged Foods \& Meats \\
122 & GIS & General Mills & Consumer Staples & Packaged Foods \& Meats \\
123 & HRL & Hormel Foods Corp. & Consumer Staples & Packaged Foods \& Meats \\
124 & HSY & The Hershey Company & Consumer Staples & Packaged Foods \& Meats \\
125 & K & Kellogg Co. & Consumer Staples & Packaged Foods \& Meats \\
126 & MKC & McCormick \& Co. & Consumer Staples & Packaged Foods \& Meats \\
127 & TSN & Tyson Foods & Consumer Staples & Packaged Foods \& Meats \\
128 & PG & Procter \& Gamble & Consumer Staples & Personal Products \\
129 & KO & Coca-Cola Company (The) & Consumer Staples & Soft Drinks \\
130 & PEP & PepsiCo Inc. & Consumer Staples & Soft Drinks \\
131 & MO & Altria Group Inc & Consumer Staples & Tobacco \\
132 & AMGN & Amgen Inc. & Health Care & Biotechnology \\
133 & CELG & Celgene Corp. & Health Care & Biotechnology \\
134 & BMY & Bristol-Myers Squibb & Health Care & Health Care Distributors \\
135 & CAH & Cardinal Health Inc. & Health Care & Health Care Distributors \\
136 & ABMD & ABIOMED Inc & Health Care & Health Care Equipment \\
137 & ABT & Abbott Laboratories & Health Care & Health Care Equipment \\
138 & BAX & Baxter International Inc. & Health Care & Health Care Equipment \\
139 & BDX & Becton Dickinson & Health Care & Health Care Equipment \\
140 & DHR & Danaher Corp. & Health Care & Health Care Equipment \\
141 & HOLX & Hologic & Health Care & Health Care Equipment \\
142 & JNJ & Johnson \& Johnson & Health Care & Health Care Equipment \\
143 & MDT & Medtronic plc & Health Care & Health Care Equipment \\
144 & PKI & PerkinElmer & Health Care & Health Care Equipment \\
145 & SYK & Stryker Corp. & Health Care & Health Care Equipment \\
146 & TMO & Thermo Fisher Scientific & Health Care & Health Care Equipment \\
147 & VAR & Varian Medical Systems & Health Care & Health Care Equipment \\
148 & UHS & Universal Health Services, Inc. & Health Care & Health Care 
Facilities \\
149 & CVS & CVS Health & Health Care & Health Care Services \\
150 & COO & The Cooper Companies & Health Care & Health Care Supplies \\
151 & CERN & Cerner & Health Care & Health Care Technology \\
152 & CI & CIGNA Corp. & Health Care & Managed Health Care \\
153 & HUM & Humana Inc. & Health Care & Managed Health Care \\
154 & UNH & United Health Group Inc. & Health Care & Managed Health Care \\
155 & LLY & Lilly (Eli) \& Co. & Health Care & Pharmaceuticals \\
156 & MRK & Merck \& Co. & Health Care & Pharmaceuticals \\
157 & MYL & Mylan N.V. & Health Care & Pharmaceuticals \\
158 & PFE & Pfizer Inc. & Health Care & Pharmaceuticals \\
159 & BEN & Franklin Resources & Financials & Asset Management \& Custody Banks 
\\
160 & BK & The Bank of New York Mellon Corp. & Financials & Asset Management \& 
Custody Banks \\
161 & NTRS & Northern Trust Corp. & Financials & Asset Management \& Custody 
Banks \\
162 & STT & State Street Corp. & Financials & Asset Management \& Custody Banks 
\\
163 & TROW & T. Rowe Price Group & Financials & Asset Management \& Custody 
Banks \\
164 & AXP & American Express Co & Financials & Consumer Finance \\
165 & BAC & Bank of America Corp & Financials & Diversified Banks \\
166 & C & Citigroup Inc. & Financials & Diversified Banks \\
167 & CMA & Comerica Inc. & Financials & Diversified Banks \\
168 & JPM & JPMorgan Chase \& Co. & Financials & Diversified Banks \\
169 & USB & U.S. Bancorp & Financials & Diversified Banks \\
170 & WFC & Wells Fargo & Financials & Diversified Banks \\
171 & AJG & Arthur J. Gallagher \& Co. & Financials & Insurance Brokers \\
172 & AON & Aon plc & Financials & Insurance Brokers \\
173 & MMC & Marsh \& McLennan & Financials & Insurance Brokers \\
174 & RJF & Raymond James Financial Inc. & Financials & Investment Banking \& 
Brokerage \\
175 & SCHW & Charles Schwab Corporation & Financials & Investment Banking \& 
Brokerage \\
176 & AFL & AFLAC Inc & Financials & Life \& Health Insurance \\
177 & TMK & Torchmark Corp. & Financials & Life \& Health Insurance \\
178 & UNM & Unum Group & Financials & Life \& Health Insurance \\
179 & L & Loews Corp. & Financials & Multi-line Insurance \\
180 & LNC & Lincoln National & Financials & Multi-line Insurance \\
181 & JEF & Jefferies Financial Group & Financials & Multi-Sector Holdings \\
182 & AIG & American International Group, Inc. & Financials & Property \& 
Casualty Insurance \\
183 & CINF & Cincinnati Financial & Financials & Property \& Casualty Insurance 
\\
184 & PGR & Progressive Corp. & Financials & Property \& Casualty Insurance \\
185 & TRV & The Travelers Companies Inc. & Financials & Property \& Casualty 
Insurance \\
186 & BBT & BB\&T Corporation & Financials & Regional Banks \\
187 & FITB & Fifth Third Bancorp & Financials & Regional Banks \\
188 & HBAN & Huntington Bancshares & Financials & Regional Banks \\
189 & KEY & KeyCorp & Financials & Regional Banks \\
190 & PNC & PNC Financial Services & Financials & Regional Banks \\
191 & RF & Regions Financial Corp. & Financials & Regional Banks \\
192 & SIVB & SVB Financial & Financials & Regional Banks \\
193 & STI & SunTrust Banks & Financials & Regional Banks \\
194 & ZION & Zions Bancorp & Financials & Regional Banks \\
195 & PBCT & People's United Financial & Financials & Thrifts \& Mortgage 
Finance \\
196 & HCP & HCP Inc. & Real Estate & Health Care REITs \\
197 & HST & Host Hotels \& Resorts & Real Estate & Hotel \& Resort REITs \\
198 & DRE & Duke Realty Corp & Real Estate & Industrial REITs \\
199 & VNO & Vornado Realty Trust & Real Estate & Office REITs \\
200 & UDR & UDR Inc & Real Estate & Residential REITs \\
201 & FRT & Federal Realty Investment Trust & Real Estate & Retail REITs \\
202 & PSA & Public Storage & Real Estate & Specialized REITs \\
203 & WY & Weyerhaeuser & Real Estate & Specialized REITs \\
204 & ADBE & Adobe Systems Inc & Information Technology & Application Software 
\\
205 & ADSK & Autodesk Inc. & Information Technology & Application Software \\
206 & CDNS & Cadence Design Systems & Information Technology & Application 
Software \\
207 & ORCL & Oracle Corp. & Information Technology & Application Software \\
208 & SYMC & Symantec Corp. & Information Technology & Application Software \\
209 & CSCO & Cisco Systems & Information Technology & Communications Equipment 
\\
210 & MSI & Motorola Solutions Inc. & Information Technology & Communications 
Equipment \\
211 & JKHY & Jack Henry \& Associates Inc & Information Technology & Data 
Processing \& Outsourced Services \\
212 & GLW & Corning Inc. & Information Technology & Electronic Components \\
213 & ADP & Automatic Data Processing & Information Technology & Internet 
Software \& Services \\
214 & FISV & Fiserv Inc & Information Technology & Internet Software \& 
Services \\
215 & PAYX & Paychex Inc. & Information Technology & Internet Software \& 
Services \\
216 & TSS & Total System Services & Information Technology & Internet Software 
\& Services \\
217 & IBM & International Business Machines & Information Technology & IT 
Consulting \& Other Services \\
218 & AMAT & Applied Materials Inc. & Information Technology & Semiconductor 
Equipment \\
219 & KLAC & KLA-Tencor Corp. & Information Technology & Semiconductor 
Equipment \\
220 & LRCX & Lam Research & Information Technology & Semiconductor Equipment \\
221 & ADI & Analog Devices, Inc. & Information Technology & Semiconductors \\
222 & AMD & Advanced Micro Devices Inc & Information Technology & 
Semiconductors \\
223 & INTC & Intel Corp. & Information Technology & Semiconductors \\
224 & MU & Micron Technology & Information Technology & Semiconductors \\
225 & MXIM & Maxim Integrated Products Inc & Information Technology & 
Semiconductors \\
226 & SWKS & Skyworks Solutions & Information Technology & Semiconductors \\
227 & TXN & Texas Instruments & Information Technology & Semiconductors \\
228 & MSFT & Microsoft Corp. & Information Technology & Systems Software \\
229 & AAPL & Apple Inc. & Information Technology & Technology Hardware, Storage 
\& Peripherals \\
230 & HPQ & HP Inc. & Information Technology & Technology Hardware, Storage \& 
Peripherals \\
231 & WDC & Western Digital & Information Technology & Technology Hardware, 
Storage \& Peripherals \\
232 & XRX & Xerox & Information Technology & Technology Hardware, Storage \& 
Peripherals \\
233 & IPG & Interpublic Group & Communication Services & Advertising \\
234 & OMC & Omnicom Group & Communication Services & Advertising \\
235 & CMCSA & Comcast Corp. & Communication Services & Cable \& Satellite \\
236 & CTL & CenturyLink Inc & Communication Services & Integrated 
Telecommunication Services \\
237 & T & AT\&T Inc. & Communication Services & Integrated Telecommunication 
Services \\
238 & VZ & Verizon Communications & Communication Services & Integrated 
Telecommunication Services \\
239 & EA & Electronic Arts & Communication Services & Interactive Home 
Entertainment \\
240 & DIS & The Walt Disney Company & Communication Services & Movies \& 
Entertainment \\
241 & FOX & Twenty-First Century Fox Class B & Communication Services & Movies 
\& Entertainment \\
242 & AEP & American Electric Power & Utilities & Electric Utilities \\
243 & D & Dominion Energy & Utilities & Electric Utilities \\
244 & DUK & Duke Energy & Utilities & Electric Utilities \\
245 & ED & Consolidated Edison & Utilities & Electric Utilities \\
246 & EIX & Edison Int'l & Utilities & Electric Utilities \\
247 & ETR & Entergy Corp. & Utilities & Electric Utilities \\
248 & EVRG & Evergy & Utilities & Electric Utilities \\
249 & LNT & Alliant Energy Corp & Utilities & Electric Utilities \\
250 & PEG & Public Serv. Enterprise Inc. & Utilities & Electric Utilities \\
251 & PPL & PPL Corp. & Utilities & Electric Utilities \\
252 & SO & Southern Co. & Utilities & Electric Utilities \\
253 & WEC & Wec Energy Group Inc & Utilities & Electric Utilities \\
254 & CMS & CMS Energy & Utilities & Multi-Utilities \\
255 & CNP & CenterPoint Energy & Utilities & Multi-Utilities \\
256 & DTE & DTE Energy Co. & Utilities & Multi-Utilities \\
257 & EXC & Exelon Corp. & Utilities & Multi-Utilities \\
258 & NEE & NextEra Energy & Utilities & Multi-Utilities \\
259 & NI & NiSource Inc. & Utilities & Multi-Utilities \\
260 & PCG & PG\&E Corp. & Utilities & Multi-Utilities \\
261 & PNW & Pinnacle West Capital & Utilities & Multi-Utilities \\
262 & XEL & Xcel Energy Inc & Utilities & Multi-Utilities \\
		\bottomrule
	\end{longtable}
}

\end{document}